\documentclass[sts]{imsart}
\RequirePackage{amsthm,amsmath,amsfonts,amssymb,booktabs}
\RequirePackage[colorlinks,citecolor=blue,urlcolor=blue]{hyperref}
\usepackage{hyperref}
\hypersetup{
  colorlinks,
  citecolor=blue,
  linkcolor=blue}
\RequirePackage[authoryear]{natbib}
\RequirePackage{graphicx}
\usepackage{multirow}
\usepackage{amsfonts}
\usepackage{bbold}
\usepackage{multicol}
\usepackage{diagbox}
\usepackage{makecell}
\usepackage{enumerate}
\usepackage{lmodern}
\usepackage{sectsty}
\usepackage{placeins}
\usepackage{bbm}
\usepackage{fancyhdr}

\DeclareMathOperator{\E}{\mathbb{E}}
\DeclareMathOperator{\V}{\mathbb{V}}
\newtheorem{theoremP}{Proposition}
\startlocaldefs
\theoremstyle{plain}

\newtheorem{theorem}{Theorem}[section]

\usepackage{amsfonts,epsfig}
\theoremstyle{remark}
\newtheorem{definition}[theorem]{Definition}

\endlocaldefs
\begin{document}
\onecolumn

\begin{frontmatter}
\title{Analysis of Linked Files: A Missing Data Perspective}
\runtitle{}

\begin{aug}
\author[A]{\fnms{Gauri}~\snm{Kamat}\ead[label=e1]{gaurik1.edu@gmail.com}\orcid{0000-0001-9338-0820}}
\and
\author[B]{\fnms{Roee}~\snm{Gutman}\ead[label=e2]{rgutman@stat.brown.edu}}

\address[A]{Gauri Kamat is PhD candidate in Biostatistics, Brown University\printead[presep={\ }]{e1}.}

\address[B]{Roee Gutman is Associate Professor of Biostatistics, Brown University\printead[presep={\ }]{e2}.}

\end{aug}

\begin{abstract}
In many applications, researchers seek to identify overlapping entities across multiple data files. Record linkage algorithms facilitate this task, in the absence of unique identifiers. As these algorithms rely on semi-identifying information, they may miss records that represent the same entity, or incorrectly link records that do not represent the same entity.  Analysis of linked files commonly ignores such linkage errors, resulting in biased, or overly precise estimates of the associations of interest. We view record linkage as a missing data problem, and delineate the linkage mechanisms that underpin analysis methods with linked files. Following the missing data literature, we group these methods under three categories: likelihood and Bayesian methods, imputation methods, and weighting methods. We summarize the assumptions and limitations of the methods, and evaluate their performance in a wide range of simulation scenarios.
\end{abstract}


\end{frontmatter}

\section{Introduction} \label{sec:intro}

In many healthcare and social science applications, information about entities is dispersed
across multiple data files. To investigate the associations of interest, researchers seek to link records representing the same entity across these files. When unique identifiers, like social security numbers, are available, linking records is a relatively straightforward task. However, privacy regulations often limit access to such identifiers. In these situations, researchers rely on semi-identifying linking variables, such as names and addresses. Record linkage (data matching or entity resolution) is a statistical technique that identifies records representing the same entity across multiple files, in the absence of unique identifiers. Record linkage techniques have found applications in epidemiology and healthcare \citep{stern1986,brenner1997,mirel2021}, official statistics \citep{winkler1991,dasylva2014}, economics \citep{mittag2019,mittag2021}, and studies of human rights violations \citep{sadinle2014}. \par

Record linkage methods can be broadly categorized as being deterministic or probabilistic. Deterministic linkage methods are based on non-random similarity functions defined on the linking variables. These methods can link records with high accuracy, when the linking variables are informative and accurately reported \citep{gomatam2002}. However, when these variables are subject to variations in spelling, data entry errors, or incompleteness, deterministic methods may perform inadequately \citep{Campbell2008,newman2009,zhu2015}. Probabilistic linkage is based on the likelihood that a given pair of records represents the same entity. This likelihood can be computed in several ways, such as using mixture models \citep{Fellegi1969}, microclustering models \citep{zanellabetancourt}, or classification algorithms \citep{cochinwala2001}. \par

A widely-used approach to probabilistic record linkage is the mixture model proposed by \citet{Fellegi1969}. For computational convenience, this model relies on several tacit independence assumptions. In addition, identification of linked records is based on subjective thresholds \citep{Winkler1995,binette22}. These features have often been criticized \citep{tancrediliseo2011}, motivating the development of parametric Bayesian approaches for probabilistic linkage \citep{tancrediliseo2011,gutman2013,Steorts2015,Sadinle2017}.  \par

Probabilistic record linkage has also been viewed as a prediction problem \citep{lee2022}. In this approach, classification algorithms, like decision trees, are used to determine if a pair of records represents the same entity \citep{cochinwala2001,bilenko2003}. Such algorithms rely on the availability of training data, i.e., a known set of links, to estimate parameters in the classification algorithm. These estimates are then used to predict the unknown linkage status of record pairs. In the absence of training data, unsupervised classification algorithms, such as $k$-means clustering, can be employed to link records \citep{liu,gu2006,christen07,christen08,christen082}. The performance of both supervised and unsupervised classification algorithms can depend on the distance functions used to compare the records \citep{asher}. When reliable training data is available, supervised algorithms can outperform unsupervised algorithms in terms of linkage accuracy \citep{elfeky2003,liu}.

In this article, we exclusively focus on probabilistic record linkage when no training data is available. We also do not focus on predictive classification approaches to probabilistic linking, which have been comprehensively reviewed in \citet{binette22}.

When the linking variables contain limited information, probabilistic linking algorithms may miss links between records representing the same entity, or falsely link records representing different entities. When analyzing linked files, such errors can lead to substantial biases and a loss in efficiency \citep{Neter1965}. In this article, we review statistical methods to propagate the uncertainty stemming from linkage errors into the post-linkage analysis. We do so by formulating record linkage within a missing data framework \citep{rubin1976}, and formalizing \textit{linkage mechanisms} that underlie inferential methods with linked files. In addition, we characterize and compare the performance of selected linkage and inference methods under a wide variety of scenarios. 

This article is organized as follows. The remainder of Section \ref{sec:intro} provides notation and background on probabilistic record linkage. Sections \ref{sec:primary} and \ref{sec:secondary} identify and discuss inferential methods with linked files. Section \ref{sec:sim} describes simulations to compare the performance of selected methods from Sections \ref{sec:primary} and \ref{sec:secondary}. Section \ref{sec:discuss} concludes with a discussion.

\subsection{Notation and background} \label{sec:PRL}

Let $\mathbf{A}$ and $\mathbf{B}$ represent two files comprising $n_A$ and $n_B$ records, respectively. Unless mentioned otherwise, assume that there are no duplicates within each file. Let $n_{AB}$ be the number of records common to both the files. Without loss of generality, assume that $n_{AB} \leq n_A\leq n_B$. Further, let $\mathbf{L}_{Ai}=(L_{Ai1}, \dots, L_{AiK})$ represent $K$ linking variables for record $i \in \mathbf{A}$, where $i=1, \dots, n_A$. Similarly, let $\mathbf{L}_{Bj}=(L_{Bj1}, \dots, L_{BjK})$ represent the same $K$ linking variables for record $j \in \mathbf{B}$, where $j = 1,\dots,n_B$. Let $\mathbf{Y}_{Ai}=(Y_{Ai1}, \dots, Y_{AiP_A})$ denote $P_{A}$ variables that are exclusive to record $i\in \mathbf{A}$, and let $\mathbf{X}_{Bj}=(X_{Bj1}, \dots, X_{BjP_B})$ denote $P_B$ variables that are exclusive to record $j \in \mathbf{B}$. 

 Probabilistic record linkage seeks to identify the $n_{AB}$ overlapping records, using $\mathbf{L}_A = \{\mathbf{L}_{Ai}\}$ \text{ and } $\mathbf{L}_B=\{\mathbf{L}_{Bj}\}$, and potentially $\mathbf{Y}_A = \{\mathbf{Y}_{Ai}\}$ and $\mathbf{X}_B=\{\mathbf{X}_{Bj}\}$. Alternatively, each pair in the set $\mathbf{A} \times \mathbf{B} = \{(i,j); i \in \mathbf{A}$, $j\in \mathbf{B}\}$ is classified as a pair that represents the same entity (link), or a pair that does not represent the same entity (non-link). Formally, the parameter of interest is the \textit{linkage structure} between files $\mathbf{A}$ and $\mathbf{B}$.
 
\subsubsection{The linkage structure} \label{sec:LinkStr}
The linkage structure $\boldsymbol{\Delta}= [\Delta_{ij}]$ is a binary \textit{configuration matrix} \citep{Green2006} of order $n_A \times n_B$, where $\Delta_{ij} = 1$, if records $i \in \mathbf{A}$ and $j \in \mathbf{B}$ are linked, and $\Delta_{ij} = 0$, otherwise \citep{fortini2001,larsen2002,larsen2005,tancrediliseo2011}. The elements of $\boldsymbol{\Delta}$ can also be stacked to form an $n_An_B$-dimensional vector of binary indicator variables \citep{fienbergmanrique-vallier2008,hofzwinderman2015,Hof2017}.  

The linkage is said to be \textit{bipartite}, when (i) there are no duplicates within each of the two files; and (ii) a record in file $\mathbf{A}$ can be linked to at most one record in file $\mathbf{B}$, and vice versa. Formally, $\sum_{i =1}^{n_A} \Delta_{ij} \leq 1 \text{  }\forall j =1, \dots, n_B$, and $\sum_{j =1}^{n_B}  \Delta_{ij} \leq 1 \text{  } \forall i=1, \dots, n_A$. Under bipartite linkage, a computationally efficient and compact linkage structure is the \textit{matching labeling} \citep{Sadinle2017}, $\mathbf{Z} = (Z_1,\dots,Z_{n_A})$, where
\begin{align}
   Z_i = \begin{cases}
      j,  &\text{if records } i \in \mathbf{A} \text{ and } j \in \mathbf{B} \text{ are linked;} \\
      n_B + i,  &\text{ if no record in file } \mathbf{B}  \text{ can be linked to record }i \in \mathbf{A}.
  \end{cases}
\end{align}
A one-to-one map between $\mathbf{Z}$ and $\boldsymbol{\Delta}$ exists, in that $Z_i = j$ whenever $\Delta_{ij} = 1$, and $Z_i = n_B+i$ whenever $\Delta_{ij} = 0$ $\forall j =1, \dots, n_B$. If we link all $n_A$ records from file $\mathbf{A}$, and leave $n_B - n_A$ records from file $\mathbf{B}$ unlinked, then $\mathbf{Z}$ represents a \textit{matching permutation} \citep{gutman2013}. In this context, $\mathbf{Z}$ indicates the order in which a subset of $n_A$ records from file $\mathbf{B}$ are linked to the records in file $\mathbf{A}$. \par

An alternate form of the linkage structure is based on a group of latent population entities. Each latent entity represents the ``true" population unit, whose characteristics are distorted to generate a record in one of the files. Formally, let $\omega_{Ai}$ be an arbitrary, but fixed, integer representing the latent entity for record $i \in \mathbf{A}$. Similarly, let $\omega_{Bj}$ be an integer representing the latent entity for record $j \in \mathbf{B}$. If $\omega_{Ai} = \omega_{Bj}$, the pair $(i \in \mathbf{A},j \in \mathbf{B})$ is said to be a link. The set $ \boldsymbol{\Omega} = \{\omega_{Ai}, i \in \mathbf{A}\} \cup \{\omega_{Bj}, j \in \mathbf{B}\}$
represents the complete linkage structure between files $\mathbf{A}$ and $\mathbf{B}$.
This construct enables identification of duplicates within the same file, as well as multi-file linkage. It has been adopted by \citet{steorts16}, \citet{Steorts2015}, and \citet{tancredi2020}, among others. In the case of two files with no duplicates within a file, a one-to-one map between $\boldsymbol{\Delta}$ and $\boldsymbol{\Omega}$ exists, in that $\Delta_{ij} = 1$ whenever $\omega_{Ai} = \omega_{Bj}$, and $\Delta_{ij} = 0$ whenever $\omega_{Ai} \neq \omega_{Bj}$. \par

For ease of exposition, we use the matrix $\boldsymbol{\Delta}$ as a representation of the linkage structure. Given the direct correspondence between the various representations, adapting to other forms can be straightforward.

\subsubsection{The Fellegi-Sunter model} \label{sec:FS}
A commonly used model to estimate the linkage structure is that of \citet{Fellegi1969} (FS), which builds on the work of \citet{newcombe59}. This model considers the set of record pairs $\{(i,j): i\in \mathbf{A}, j\in \mathbf{B}\}$ to be the union of the set of links, $\mathbf{M} = \{(i,j): i \in \mathbf{A}, j \in \mathbf{B}, \Delta_{ij}=1\}$, and the set of non-links, $\mathbf{U} = \{(i,j): i \in \mathbf{A}, j \in \mathbf{B}, \Delta_{ij}=0\}$. The sets $\mathbf{M}$ and $\mathbf{U}$ are identified using the linking variables $\mathbf{L}_A$ and $\mathbf{L}_B$. 

To assess the similarity between records $i \in \mathbf{A}$ and $j \in \mathbf{B}$, comparison vectors $\gamma_{ij}=(\gamma_{ij1}, \dots, \gamma_{ijK})$ are constructed. The similarity on the $k^{th}$ linking variable is defined using $l_k$ levels of ordinal agreement, where the lowest level indicates the strongest agreement, and $l_k$ indicates the strongest disagreement \citep{Winkler1990}. Formally, $\gamma_{ijk}$ is represented by a set of indicator variables $\{\gamma_{ijkl} :k=1,\dots,K, l=1, \dots, l_k\}$, where
\begin{align}
    \gamma_{ijkl}=
     \begin{cases} 
      1, & \text{if $L_{Aik}$ and $L_{Bjk}$ have the $l^{th}$ level of agreement;} \\
      0, &\text{otherwise. }
   \end{cases} 
\end{align}

The comparison vectors follow a two-class mixture distribution of the form $
\text{Pr}(\gamma_{ij})=\nu\text{}m(\gamma_{ij}) +(1-\nu)\text{}u(\gamma_{ij})$, where
$m(\gamma_{ij})=\text{Pr}(\gamma_{ij}\text{  }|\text{  } (i,j) \in \mathbf{M}, \boldsymbol{\theta_M} )$, and $u(\gamma_{ij})=\text{Pr}(\gamma_{ij}\text{  }|\text{  } (i,j) \in \mathbf{U},\boldsymbol{\theta_U})$, and $\nu$ is the marginal probability that a pair of records is a link. The parameters $\boldsymbol{\theta_M}$ and $\boldsymbol{\theta_U}$ govern the distributions $m(\gamma_{ij})$ and $u(\gamma_{ij})$ among the links and non-links, respectively.

Here, $m(\gamma_{ij})$ and $u(\gamma_{ij})$ represent general probability distributions on the comparison space. \citet{Fellegi1969} specify them assuming conditional independence of the components of the comparison vectors, given the linkage status of a record pair. Other specifications allow for dependence between the components of $\gamma_{ij}$, using log-linear models \citep{larsenrubin2001} or multilevel normal models \citep{daggy2014evaluating}.

The probabilities $m(\gamma_{ij})$ and $u(\gamma_{ij})$ are  unknown, and are estimated using the Expectation-Maximization (EM) algorithm \citep{dempster1977,belinrubin1995}.  Let $\hat m(\gamma_{ij})$ and $\hat u(\gamma_{ij})$ be the estimates of $m(\gamma_{ij})$ and $u(\gamma_{ij})$, respectively. The FS model assigns an estimated weight to record pair $(i, j)$ as $\hat w_{ij} = \text{log } \frac{\hat m(\gamma_{ij})}{\hat u(\gamma_{ij})}$. 

Each record pair is classified as a link, non-link, or a possible link, by comparing $\hat{w}_{ij}$ to two fixed thresholds \citep{tepping68,binette22}. Choosing optimal thresholds is a subjective process, based on manual review of record pairs \citep{dusetzina2014}, or on the evaluation of linkage quality measures at different threshold values \citep{christen2005,christengoisier2007}. \par
The original FS model classifies the linkage status of each record pair independently, often resulting in many-to-one links. \citet{Jaro1989} addressed this, by introducing a linear sum optimization as a precursor to the FS model. \par
Multiple extensions to the FS model have been proposed. These include incorporating information from clerical reviews \citep{larsenrubin2001}, modeling the dependence between linking variables \citep{Winkler1993,schurle,xu2019}, handling continuous-valued linking variables \citep{vo2022}, handling missing data \citep{li2022,Sadinle2017}, testing the conditional independence assumption \citep{xuC2022},  improved scalability \citep{imaifastlink}, and extensions to multi-file linkage \citep{sadinlefienberg2013,aleshing,taylor24}.

\subsubsection{Linkage errors} 
Given that probabilistic linkage relies on semi-identifying information, the estimated $\mathbf{\Delta}$ is seldom error-free.
In particular, a linked file may contain two kinds of linkage errors: (i) false links between records representing different entities; and (ii) false non-links between records representing the same entity. 

False links can lead to erroneous estimates of the associations of interest \citep{Neter1965}. For instance, in linear regression models, false links can induce attenuation bias when estimating the regression coefficients \citep{wang2022, patki23}. False links can also have implications when estimating causal effects \citep{Wortman2018}, mortality rates \citep{krewski2005}, and population sizes \citep{isaki,consiglio1,consiglio2}. False non-links reduce the number of records available for analysis, leading to a loss in statistical power and increased variability in estimates \citep{haas1994,harron2017}. False non-links can also induce selection bias in the post-linkage inferences. Selection bias occurs when particular groups of records are less likely to link, and are thus excluded from the analysis \citep{lariscy,bohenskybook,reiter2021}. 

The extent of errors in a linked file is determined by the degree of information contained in the linking variables. This information can be summarized in terms of two metrics: the reliability and the discriminatory power \citep{shlomo19}. 

The \textit{reliability} is the probability that a linking variable is similar, given that a record pair is a link. Reliability reflects the quality of the linking variable, and is diminished when there is measurement error or missingness. For records $i \in \mathbf{A}$ and $j \in \mathbf{B}$, the aggregate reliability across all linking variables is quantified by $m(\gamma_{ij})$ \citep{shlomo19}. Low reliability of the linking variables is typically associated with a higher proportion of false non-links \citep{tromp2006,tromp11}. The \textit{discriminatory power} is the probability that a linking variable is similar, given that a record pair is a non-link \citep{doidge2018}. The theoretical number of possible values of the linking variable, and their empirical distribution, determines this probability. The combined discriminatory power of all linking variables is quantified by $u(\gamma_{ij})$ \citep{shlomo19}. If the combined discriminatory power is low, the record linkage algorithm may identify a high proportion of false links \citep{tromp2006,tromp11}. 

Both these metrics have been used to assess the feasibility and performance of the FS model \citep{roos1992,dean01}. Modifications to the FS weights based on the discriminatory power have also been proposed \citep{yancey2000,winkler2000,zhu2009,xu2021,vo2022}.

\subsubsection{Blocking} \label{sec:Block}
 With large file sizes, comparing all record pairs across the two files becomes computationally
cumbersome. Large files also prompt a higher chance of occurrence of linkage errors \citep{wang2022}. To improve the scalability and accuracy of the linkage
process, \textit{deterministic blocking} is a commonly implemented pre-processing technique. In this technique, only those records that agree on predefined, highly accurate blocking variables are compared. Records that do not agree on these variables are considered non-links. \par
When utilizing deterministic blocking, it is important to ascertain an optimal block size, which often involves a cost-benefit trade-off \citep{Baxter2003}. Large blocks generate a correspondingly large record comparison space, and may not provide substantial gains in computational efficiency or accuracy. When the blocks are too small, the record linking algorithm may miss many true links \citep{Baxter2003}. Methods that utilize training data to determine optimum blocking schemes have been proposed as possible solutions \citep{bilenkoicdm,michelson,mirel}.\par
When blocking variables are error-prone, records may be placed in the wrong block, leading to linkage errors. Possible solutions include using multiple blocking passes \citep{Baxter2003}, complex indexing schemes \citep{Murray2015}, or a combination of deterministic and data-driven blocking \citep{mcveigh2020}. Other alternatives include \textit{probabilistic blocking}, which simultaneously infers the blocking scheme and the linkage parameters \citep{Steorts2014}. Probabilistic blocking can effectively propagate blocking uncertainty into the downstream statistical analysis \citep{marchant2021}. \par 

When files are partitioned into deterministic blocks, typical practice is to
implement the linkage algorithm independently within each block \citep{tancrediliseo2011,Murray2015,Sadinle2017}. However, it is also possible to allow the linkage algorithm to borrow information across blocks \citep{gutman2013,shan2022}. \par

\subsubsection{Post-linkage analysis}
Linking two files is commonly not the end goal in a study; rather, researchers may seek to estimate the
population association between $\mathbf{Y}_A = \{\mathbf{Y}_{Ai}\}$ and $\mathbf{X}_B=\{\mathbf{X}_{Bj}\}$ among the $n_{AB}$ overlapping records.
One approach to summarizing this association is describing the conditional mean  $\E(\mathbf{Y}_A|\mathbf{X}_B) = \mu(\mathbf{X}_B;\boldsymbol{\beta})$, where $\mu(.)$ denotes an inverse link function of known form, and $\boldsymbol{\beta} = (\beta_1,\dots,\beta_{P{_B}})$ is a set of parameters that includes the estimand of interest. Alternatively, the association can be summarized using correlation coefficients, principal components analysis, or other multivariate analysis techniques \citep{rencher}.

\subsection{A missing data formulation}
Inference on $\boldsymbol{\beta}$ using the $n_{AB}$ overlapping records involves analysis of ``complete data'', given by $\mathcal{D} =   \big[\mathbf{L}_{Ai},\mathbf{L}_{Bj},\mathbf{Y}_{Ai},\mathbf{X}_{Bj},\Delta_{ij}\big]_{(i \in \mathbf{A},j\in \mathbf{B})}$. Assuming that the variables $(\mathbf{L}_A,\mathbf{L}_B,\mathbf{Y}_A,\mathbf{X}_B)$ are fully observed, the observed data are $\mathcal{D}^{obs} = \big[\mathbf{L}_{Ai},\mathbf{L}_{Bj},\mathbf{Y}_{Ai},\mathbf{X}_{Bj}\big]_{(i \in \mathbf{A},j\in \mathbf{B})}$, and the missing data are $\mathcal{D}^{mis} = \{\Delta_{ij}\}_{(i \in \mathbf{A},j\in \mathbf{B})}$. The linkage status $\Delta_{ij}$ can be viewed as a discrete latent variable that is never observed, but explains the observed data \citep{fienbergmanrique-vallier2008}. This view motivates inference using likelihood-based and Bayesian methods commonly employed in latent class modeling, such as the EM \citep{dempster1977} and data augmentation (DA) \citep{tannerwong1987} algorithms.

Alternatively, one can impute the missing $\boldsymbol{\Delta}$, using a stochastic model based on $\mathcal{D}^{obs}$. The resultant ``complete data'' comprises a set of identified links, which can be analyzed using standard methods. Uncertainty around the imputed $\boldsymbol{\Delta}$ can be accounted for using multiple imputation (MI) \citep{rubin1987,littlerubin2002}. Inference can also be accomplished by weighting the post-linkage analysis model, with weights estimated from models based on $\mathcal{D}^{obs}$.

\subsubsection{A taxonomy of linkage mechanisms} \label{sec:linkmech}
A \textit{linkage mechanism} characterizes the relationship between the linkage structure and the variables in the individual files. There are broad parallels between missing data mechanisms \citep{rubin1976}, and linkage mechanisms.  For a given prior on $\boldsymbol{\Delta}$, we define four mechanisms that describe $f(\boldsymbol{\Delta}|\mathbf{L}_A,\mathbf{L}_B,\mathbf{Y}_A,\mathbf{X}_B,\boldsymbol{\theta})$, where $\boldsymbol{\theta}$ is a vector parameter governing the distribution. The definitions assume that $\mathbf{L}_A,\mathbf{L}_B,\mathbf{Y}_A, \text{ and }\mathbf{X}_B$ are fully observed.

\begin{definition}
Strongly non-informative linkage (SNL): The linkage mechanism is said to be \textit{strongly non-informative}, if the conditional distribution of $\boldsymbol{\Delta}$ given $\mathbf{L}_A,\mathbf{L}_B,\mathbf{Y}_A,\mathbf{X}_B$ depends exclusively on $\mathbf{L}_A \text{ and }\mathbf{L}_B$. Formally, $f(\boldsymbol{\Delta}|\mathbf{L}_A,\mathbf{L}_B,\mathbf{Y}_A,\mathbf{X}_B,\boldsymbol{\theta})=f(\boldsymbol{\Delta}|\mathbf{L}_A,\mathbf{L}_B,\boldsymbol{\theta})$ for all $\mathbf{Y}_A,\mathbf{X}_B,$ and $\boldsymbol{\theta}$.
\end{definition}
The SNL mechanism asserts that variables exclusive to each of the files do not inform the linkage. This has analogies to when the data are missing at random (MAR) \citep{rubin1976}. Specifically, SNL can be viewed as a special case of MAR, where the missingness is determined by a set of auxiliary variables $\mathbf{L}_A$ and $\mathbf{L}_B$. These variables may not inherently be of interest in the analysis, but predict the missingness \citep{schafer1997,collins2001}. 

A more restrictive version of SNL states that the linkage is completely independent of $\mathbf{L}_A,\mathbf{L}_B,\mathbf{Y}_A,\mathbf{X}_B$. Thus, $f(\boldsymbol{\Delta}|\mathbf{L}_A,\mathbf{L}_B,\mathbf{Y}_A,\mathbf{X}_B,\boldsymbol{\theta})=f(\boldsymbol{\Delta}|\boldsymbol{\theta})$ for all $\mathbf{L}_A,\mathbf{L}_B,\mathbf{Y}_A$, $\mathbf{X}_B,$ and $\boldsymbol{\theta}$. \citet{hanlahiri2019} refer to this as \textit{linkage completely at random (LCAR)}, similar to when the data are missing completely at random (MCAR) \citep{rubin1976}. A special case of the LCAR mechanism is when $f(\boldsymbol{\Delta}|\boldsymbol{\theta})$ is a constant. This mechanism is seldom plausible in practice, as it entails that every record pair has an equal probability of being a link.

\begin{definition}
Non-informative linkage (NL): The linkage mechanism is said to be \textit{non-informative}, if the conditional distribution of $\boldsymbol{\Delta}$ given $\mathbf{L}_A,\mathbf{L}_B,\mathbf{Y}_A,\mathbf{X}_B$ depends on $\mathbf{L}_A,\mathbf{L}_B$ and $\mathbf{X}_B$, but not the outcome $\mathbf{Y}_A$. Formally,  $f(\boldsymbol{\Delta}|\mathbf{L}_A,\mathbf{L}_B,\mathbf{Y}_A,\mathbf{X}_B,\boldsymbol{\theta})=f(\boldsymbol{\Delta}|\mathbf{L}_A,\mathbf{L}_B,\mathbf{X}_B,\boldsymbol{\theta})$ for all $\mathbf{Y}_A$ and $\boldsymbol{\theta}$. 
\end{definition}

Post-linkage analysis methods with limited access to the linkage process \citep[for e.g.,][]{lahirilarsen05} typically rely on this assumption for inference. \citet{hanlahiri2019} refer to the NL mechanism as \textit{linkage at random (LAR)}. \par
To exemplify an NL scenario, suppose that file $\mathbf{B}$ contains the covariate age, and file $\mathbf{A}$ is a subset of units in file $\mathbf{B}$ that are less than 60 years old. While age is not a linking variable common to both files, it implicitly informs the linkage. The NL mechanism is also a special case of the MAR mechanism, where auxiliary variables $\mathbf{L}_{A}$ and $\mathbf{L}_{B}$, and substantively important variables $\mathbf{X}_{B}$, inform the linkage. 

\begin{definition}
Weakly non-informative linkage (WNL): The linkage mechanism is said to be \textit{weakly non-informative}, if $f(\boldsymbol{\Delta}|\mathbf{L}_A,\mathbf{L}_B,\mathbf{Y}_A,\mathbf{X}_B,\boldsymbol{\theta})$ depends on $\mathbf{L}_A,\mathbf{L}_B,\mathbf{Y}_A, \text{ and }\mathbf{X}_B$, for all $\boldsymbol{\theta}$.
\end{definition}

The WNL mechanism is also comparable to the MAR mechanism, as only fully observed variables inform the linkage. 

\begin{definition}
Informative linkage (IL): The linkage mechanism is said to be \textit{informative}, if none of the above linkage mechanisms hold. For instance, the conditional distribution of $\boldsymbol{\Delta}$ may depend on extraneous variables not observed in either file.
\end{definition}
IL may induce bias in the post-linkage inferences, particularly if  
such extraneous variables are associated with the substantively important variables. The IL mechanism is comparable to the missing not at random (MNAR) mechanism \citep{rubin1976}, as the linkage status depends on unobserved quantities.

In what follows, we discuss linked-data inferential methods that rely on one or more of the linkage mechanisms as their underlying assumptions.  

\section{Primary analysis of linked files} \label{sec:primary}
Primary analysis refers to the situation when the record linkage and the post-linkage analysis are performed by the same individual. Alternatively, the analyst has access to files $\mathbf{A}$ and $\mathbf{B}$, or to the likelihood of every record pair $(i \in \mathbf{A},j \in \mathbf{B})$ being a link. Following the missing data literature \citep{molenberghs2014,littlerubin2002}, we classify primary analysis methods into three broad categories:
likelihood and Bayesian methods, imputation methods, and weighting methods. We note that this classification can be overlapping. For instance, some methods can be classified as being both Bayesian and imputation-based, as proper MI methods commonly rely on sampling from (approximate) Bayesian posterior distributions \citep{Rubin1996}.

\subsection{Likelihood and Bayesian methods} \label{sec:primaryMLBayes}
Likelihood-based and Bayesian methods rely on the specification of the complete-data
likelihood, namely $\mathcal{L}_c(\mathbf{L}_{A},\mathbf{L}_{B},\mathbf{Y}_{A},\mathbf{X}_{B},\boldsymbol{\Delta}| \boldsymbol{\beta},\boldsymbol{\theta})$, where $\boldsymbol{\theta}$ denotes a set of parameters governing the linkage process. It is often reasonable to assume that $\boldsymbol{\theta}$ and $\boldsymbol{\beta}$ have distinct parameter spaces \citep{rubin1976,littlerubin2002}. This allows for $\mathcal{L}_c$ to be factored as 
\begin{align} \label{jointLikelihood}
\begin{split}
\mathcal{L}_c(\mathbf{L}_{A},\mathbf{L}_{B},\mathbf{Y}_{A},\mathbf{X}_{B},\boldsymbol{\Delta}|\boldsymbol{\beta},\boldsymbol{\theta}) &=
\mathcal{L}_l(\mathbf{L}_{A},\mathbf{L}_{B}|\boldsymbol{\Delta},\boldsymbol{\theta})\text{ }\mathcal{L}_a(\mathbf{Y}_{A},\mathbf{X}_{B}|\mathbf{L}_{A},\mathbf{L}_{B},\boldsymbol{\Delta},\boldsymbol{\beta}),
\end{split}
\end{align}
where $\mathcal{L}_l$ denotes the record linkage likelihood, and $\mathcal{L}_a$ denotes the likelihood for associations between $\mathbf{Y}_{A} \text{ and }\mathbf{X}_{B}$. Equation \eqref{jointLikelihood} treats the missing $\boldsymbol{\Delta}$ as a parameter in the likelihood. The joint posterior distribution of $(\boldsymbol{\Delta},\boldsymbol{\beta},\boldsymbol{\theta})$ is  
\begin{align} \label{jointBayes1}
\begin{split}
\mathcal{P}(\boldsymbol{\Delta},\boldsymbol{\beta},\boldsymbol{\theta}|\mathbf{L}_{A},\mathbf{L}_{B},\mathbf{Y}_{A},\mathbf{X}_{B})  &\propto \mathcal{L}_l(\mathbf{L}_{A},\mathbf{L}_{B}|\boldsymbol{\Delta},\boldsymbol{\theta})\text{ }\mathcal{L}_a(\mathbf{Y}_{A},\mathbf{X}_{B}|\mathbf{L}_{A},\mathbf{L}_{B},\boldsymbol{\Delta},\boldsymbol{\beta})  \text{ }\pi_{\Delta}(\boldsymbol{\Delta}) \text{ }\pi_{\theta,\beta}(\boldsymbol{\theta},\boldsymbol{\beta}), 
\end{split}
\end{align}
where $\pi_{\Delta}$ denotes a prior on $\boldsymbol{\Delta}$, and $\pi_{\theta,\beta}$ denotes a joint prior on $(\boldsymbol{\theta},\boldsymbol{\beta})$.

For Bayesian inference, distinctness of $\boldsymbol{\theta}$ and $\boldsymbol{\beta}$ requires that these parameters are \textit{a priori} independent \citep{rubin1976,van2018}. Hence, $\pi_{\theta,\beta}(\boldsymbol{\theta},\boldsymbol{\beta}) = \pi_{\theta}(\boldsymbol{\theta}) \text{ }\pi_\beta(\boldsymbol{\beta})$, where $\pi_{\theta}$ and $\pi_\beta$ represent prior distributions on $\boldsymbol{\theta}$ and $\boldsymbol{\beta}$, respectively. Thus,
\begin{align} \label{jointBayes}
\begin{split}
\mathcal{P}(\boldsymbol{\Delta},\boldsymbol{\beta},\boldsymbol{\theta}|\mathbf{L}_{A},\mathbf{L}_{B},\mathbf{Y}_{A},\mathbf{X}_{B})  &\propto  \mathcal{L}_l(\mathbf{L}_{A},\mathbf{L}_{B}|\boldsymbol{\Delta},\boldsymbol{\theta})\text{ }\mathcal{L}_a(\mathbf{Y}_{A},\mathbf{X}_{B}|\mathbf{L}_{A},\mathbf{L}_{B},\boldsymbol{\Delta},\boldsymbol{\beta})  \text{ }\pi_{\Delta}(\boldsymbol{\Delta}) \text{ } \pi_{\theta}(\boldsymbol{\theta})\text{ }\pi_\beta(\boldsymbol{\beta}).
\end{split}
\end{align}

We note that the distinctness assumption is more plausible under the SNL and NL mechanisms. It may not hold under WNL, when some components of $\boldsymbol{\beta}$ may inform the linkage, and affect $\boldsymbol{\theta}$. Moreover, $\boldsymbol{\theta}$ and $\boldsymbol{\beta}$ may be correlated \textit{a priori}; for example, when the linking and substantive variables represent longitudinal versions of the same characteristic. Here and throughout, we assume that the $\boldsymbol{\beta}$ represent the parameters of scientific interest, and the $\boldsymbol{\theta}$ are nuisance parameters in the likelihood. For notational simplicity, we henceforth suppress the dependence of Equation \eqref{jointBayes} on $\boldsymbol{\theta}$.

\subsubsection{Likelihood and Bayesian inference under SNL} \label{sec:snl}
Under the SNL assumption, Bayesian inference on $\boldsymbol{\beta}$ can be accomplished using an approximation to Equation \eqref{jointBayes}, called \textit{linkage averaging} \citep{sadinle2018}. Formally, under SNL, the posterior distribution of $\boldsymbol{\beta}$ is
\begin{align} \label{eq:linkavg}
\begin{split}
    \mathcal{P}(\boldsymbol{\beta}|\mathbf{L}_{A},\mathbf{L}_{B},\mathbf{Y}_{A},\mathbf{X}_{B}) &= \sum_{\boldsymbol{\Delta}} \mathcal{P}(\boldsymbol{\beta}|\mathbf{L}_{A},\mathbf{L}_{B},\mathbf{Y}_{A},\mathbf{X}_{B},\boldsymbol{\Delta}) \text{ } \mathcal{P}(\boldsymbol{\Delta}|\mathbf{L}_{A},\mathbf{L}_{B}).
\end{split}
\end{align}

Summing over all possible values of $\boldsymbol{\Delta}$ can be computationally prohibitive. One solution is to approximate Equation \eqref{eq:linkavg} using a method akin to Markov Chain Monte Carlo model composition ($MC^3$) \citep{madiganyork1995,madigan1996}. Specifically, let $\boldsymbol{\Delta}^{(1)},\dots,\boldsymbol{\Delta}^{(M)}$ be $M$ samples of the linkage structure drawn from $\mathcal{P}(\boldsymbol{\Delta}|\mathbf{L}_{A},\mathbf{L}_{B})$. For a well-behaved function $g(.)$ defined on the space $\{0,1\}^{n_A\times n_B}$, the average
\begin{equation} \label{mc3}
{G} = \frac{1}{M}\sum_{m=1}^{M} g(\boldsymbol{\Delta}^{(m)})
\end{equation}
converges almost surely to  $\E(g(\boldsymbol{\Delta}))$, as $M\rightarrow \infty$ \citep{smithroberts1993}. Thus, Equation \eqref{eq:linkavg} can be approximated by setting $g(\boldsymbol{\Delta}^{(m)})\equiv\mathcal{P}(\boldsymbol{\beta}|\mathbf{L}_{A},\mathbf{L}_{B},\mathbf{Y}_{A},\mathbf{X}_{B},\boldsymbol{\Delta}^{(m)})$ in Equation \eqref{mc3}.

For large $M$, point and interval estimates for $\boldsymbol{\beta}$ can be obtained using $\hat{\boldsymbol{\beta}}^{(1)},\dots, \hat{\boldsymbol{\beta}}^{(M)}$, where $\hat{\boldsymbol{\beta}}^{(m)}$ is a draw from $\mathcal{P}(\boldsymbol{\beta}|\mathbf{L}_{A},\mathbf{L}_{B},\mathbf{Y}_{A},\mathbf{X}_{B},\boldsymbol{\Delta}^{(m)})$, $m=1,\ldots,M$. Alternatively, each $\boldsymbol{\hat\beta}^{(m)}$ can represent a set of samples from the posterior $\mathcal{P}(\boldsymbol{\beta}|\mathbf{L}_{A},\mathbf{L}_{B},\mathbf{Y}_{A},\mathbf{X}_{B},\boldsymbol{\Delta}^{(m)})$; i.e., $\hat{\boldsymbol{\beta}}^{(m)} = \{\hat{\boldsymbol{\beta}}_s^{(m)}\}$, $s=1,\ldots,S$. Stacking these samples across the $M$ linkage structures enables us to obtain point and interval estimates for $\boldsymbol{\beta}$ \citep{zhoureiter}. In some cases, it is possible to compute the posterior mean and variance of $\boldsymbol{\hat\beta}^{(m)}$ analytically. Estimation can then follow from the identities in \citet{sadinle2018}, which we summarize in the supplementary materials. 

 %The linkage averaging approach will yield valid estimates, if $\mathcal{P}(\boldsymbol{\beta}|\mathbf{L}_{A},\mathbf{L}_{B},\mathbf{Y}_{A},\mathbf{X}_{B})$ in Equation \eqref{eq:linkavg} represents a proper posterior distribution. This in turn depends 

The posterior distribution of the linkage structure, $\mathcal{P}(\boldsymbol{\Delta}|\mathbf{L}_{A},\mathbf{L}_{B}) \propto \mathcal{L}_l(\mathbf{L}_{A},\mathbf{L}_{B}|\boldsymbol{\Delta}) \pi_{\Delta}(\boldsymbol{\Delta})$, can be specified using one of two approaches:  direct modeling of the information in $\mathbf{L}_{A} \text{ and }\mathbf{L}_{B}$, or using comparison vectors. 

The direct modeling approach for $\mathcal{P}(\boldsymbol{\Delta}|\mathbf{L}_{A},\mathbf{L}_{B})$ assumes that the linking variables $\mathbf{L}_{A}$ and $\mathbf{L}_{B}$ are erroneous versions of their true latent values. Under the assumption of bipartite linkage, \citet{tancrediliseo2011} utilize a hierarchical generative model to specify $\mathcal{L}_l(\mathbf{L}_{A},\mathbf{L}_{B}|\boldsymbol{\Delta})$.
The model posits that $\mathbf{L}_{Ai}$ and  $\mathbf{L}_{Bj}$ are noisy realizations of the variables $\boldsymbol{\eta}_A$ and $\boldsymbol{\eta}_B$, which are independent realizations of a latent variable $\boldsymbol{\eta}$, with a discrete and finite support. Conditional on $\boldsymbol{\eta}_A$ and $\boldsymbol{\eta}_B$, $\mathbf{L}_{Ai}$ and  $\mathbf{L}_{Bj}$ follow an identifiable variant of the hit-and-miss model \citep{copashilton,schwartz}. This is a mixture distribution with two components: a point mass at the true values of $\mathbf{L}_{Ai}$ or $\mathbf{L}_{Bj}$, and a uniform distribution over the remaining points in the support of $\boldsymbol{\eta}$. At the top level of the hierarchy, a distribution for $\boldsymbol{\eta}_A$ and $\boldsymbol{\eta}_B$ given $\boldsymbol{\Delta}$ is postulated. The number of links $n_{AB} = \sum_i \sum_j \Delta_{ij}$ is treated as a random quantity. Under diffuse prior distributions on $n_{AB}$ and $\boldsymbol{\Delta}$, the joint posterior distribution is explored via Metropolis-within-Gibbs algorithms \citep{robertcasella}.

 \citet{Steorts2015} build on the measurement error model of \citet{hallfienberg}, and posit a likelihood for a bipartite graph that maps every record to a latent population entity. Akin to \citet{tancrediliseo2011}, the likelihood postulates that the linking variables are noisy representations of latent, multinomial random variables. The prior distribution on the linkage structure is assumed to be uniform, and posterior sampling is accomplished via Metropolis-within-Gibbs algorithms with split-merge proposal moves \citep{smered}. Linking records to latent entities, rather than to other records, allows the model to handle linkage of multiple files, and duplicates within a file. 
 
 The original methodologies of \citet{tancrediliseo2011} and \citet{Steorts2015} are restricted to categorical linking variables. Extensions to incorporate multivariate normal linking variables \citep{liseo2011}, and string-valued linking variables \citep{steorts16} have also been proposed.

In some applications, it may be difficult to directly model complex linking variables, like addresses or telephone numbers. A solution is to define the likelihood  $\mathcal{L}_l(\mathbf{L}_{A},\mathbf{L}_{B}|\boldsymbol{\Delta})$ using comparison vectors $\gamma_{ij}$ (Section \ref{sec:FS}). Akin to the FS model, the comparison vectors are assumed to follow a mixture distribution over the set of links and non-links. However, bipartite linkage is explicitly enforced via a prior distribution on $\boldsymbol{\Delta}$. This approach was developed by \citet{fortini2001} and \citet{larsen2002,larsen2005} for the case of binary comparison vectors. \citet{Sadinle2017} extended it to include partial agreements on complex linking variables, and to accommodate ignorable missingness \citep{rubin1976} in the linking variables. Formally, the model is 
\begin{eqnarray} \label{sadinle}
\begin{split}
    \gamma_{ij}|\Delta_{ij}=1 \sim m(\gamma_{ij}) \\
\gamma_{ij}|\Delta_{ij}=0 \sim u(\gamma_{ij}) \\
\boldsymbol{\Delta} \sim \pi_\Delta(\boldsymbol{\Delta}),
\end{split}
\end{eqnarray}
where $m(\gamma_{ij})$ and $u(\gamma_{ij})$ are as defined in Section \ref{sec:FS}. The prior $\pi_\Delta(\boldsymbol{\Delta})$ is defined over the space of all bipartite linkage structures. 

One possible joint prior distribution is $\pi_{\Delta}(\boldsymbol{\Delta},n_{AB})=\pi_{1}(n_{AB})\text{ }\pi_{2}(\boldsymbol{\Delta}|n_{AB})$, where $\pi_{1}$ and $\pi_{2}$ are prior distributions on the number of links, and the space of bipartite linkage structures with $n_{AB}$ links, respectively \citep{larsen2002,Sadinle2017}. The joint posterior distribution of $(n_{AB},\boldsymbol{\Delta})$ can be explored using a Gibbs sampler based on an adaptive multinomial distribution \citep{Sadinle2017}, or a Metropolis-Hastings algorithm with proposals akin to Fisher-Yates shuffling \citep{fisheryates,larsen2005}.

\subsubsection{Likelihood and Bayesian inference under WNL} \label{sec:wnl}
The linkage averaging approach in Equation \eqref{eq:linkavg} can propagate linkage uncertainty to the downstream analysis. However, information from the analysis models is not utilized in the linkage process. An alternative is to allow the analysis models to inform the linkage. In a Bayesian framework, this can be accomplished by sampling $\boldsymbol{\Delta}$ and $\boldsymbol{\beta}$ simultaneously from their joint posterior distribution (Equation \eqref{jointBayes}). 

It has been shown analytically, and via simulations, that incorporating substantive relationships into the linkage process can facilitate correct identification of links and non-links \citep{brlvof}. These improvements are commonly called the ``feedback effect'' of the post-linkage analysis model \citep{steorts2018}. The magnitude of the improvements can depend on (i) the fit of the specified likelihood $\mathcal{L}_{a}(\mathbf{Y}_{A},\mathbf{X}_{B}|\mathbf{L}_{A},\mathbf{L}_{B},\boldsymbol{\Delta},\boldsymbol{\beta})$; and (ii) the strength of the association between $\mathbf{Y}_{A}$ and $\mathbf{X}_{B}$.

Sampling from the joint posterior distribution in Equation \eqref{jointBayes} can be accomplished via the DA algorithm. At the $m^{th}$ iteration, this algorithm iterates between two steps: (i) I-Step: Impute $\boldsymbol{\Delta}^{(m)}$ from the posterior full conditional distribution $\mathcal{P}(\boldsymbol{\Delta}|\mathbf{L}_{A},\mathbf{L}_{B},\mathbf{Y}_{A},\mathbf{X}_{B},\boldsymbol{\beta}^{(m-1)})$; and (ii) P-step: Sample $\boldsymbol{\beta}^{(m)}$ from the posterior full conditional distribution $\mathcal{P}(\boldsymbol{\beta}|\mathbf{L}_{A},\mathbf{L}_{B},\mathbf{Y}_{A},\mathbf{X}_{B},\boldsymbol{\Delta}^{(m)})$. The resultant samples of $\boldsymbol{\beta}$ incorporate the uncertainty around $\boldsymbol{\Delta}$, and can be used for point and interval estimation.

\citet{gutman2013} model the associations between variables common to both files, and those exclusive to either file. High quality linking variables are utilized in an informative blocking step, which reduces the computational complexity. Within each block, the linkage is assumed to be bipartite and complete, so that every record in the smaller block can be linked to a unique record in the larger block. The linkage likelihood $\mathcal{L}_{l}(\mathbf{L}_{A},\mathbf{L}_{B}|\boldsymbol{\Delta})$ is defined using a joint distribution that describes inconsistently recorded linking variables. The distributional form of $\mathcal{L}_{a}(\mathbf{Y}_{A},\mathbf{X}_{B}|\mathbf{L}_{A},\mathbf{L}_{B},\boldsymbol{\Delta},\boldsymbol{\beta})$ is specific to the application, and can include models of scientific interest, and models for
relationships that only inform the linkage. For unlinked records in the larger block, variables in the smaller block are treated as missing data, and imputed as part of an overarching DA scheme. Under a uniform prior on $\boldsymbol{\Delta}$, posterior sampling for $\boldsymbol{\Delta}$ is accomplished via Metropolis algorithms. \citet{dalzell2018} extend the model of \citet{gutman2013} to incorporate erroneous blocking variables, and to allow individual records to shift between blocks.  \par
A limitation of the approach of \citet{gutman2013} is the assumption of complete linkage, which can be violated in practice. An alternative approach augments the Bayesian FS likelihood (Equation \eqref{sadinle}) with relationships between variables exclusive to either file \citep{tang2020,brlvof}. Bayesian generative models have also been applied in a joint modeling framework, in the context of linear regression \citep{tancrediliseo2015,steorts2018}, and small area estimation \citep{briscolini}.

\citet{hofzwinderman2015} propose to estimate $\boldsymbol\beta$ by maximizing the complete-data likelihood $\mathcal{L}_c$, over all possible values of $\boldsymbol\Delta$. Since the parameter space of $\boldsymbol\Delta$ is fairly large, maximizing $\mathcal{L}_c$ can be computationally prohibitive. As a solution, \citet{hofzwinderman2015} propose a two-step pseudo-likelihood approach to estimate $\boldsymbol\beta$. In the first step, estimates of the linking probabilities for all record pairs are computed under the FS model. These estimates are treated as fixed constants in the pseudo-likelihood. In the second step, the pseudo-likelihood is maximized using the EM algorithm, to obtain an estimate $\boldsymbol{\hat\beta}$. The sampling variance of $\boldsymbol{\hat\beta}$, and confidence intervals for $\boldsymbol{\beta}$, can be computed using the Bootstrap, or an asymptotic normal approximation for the sampling distribution of $\boldsymbol{\hat\beta}$. A similar inferential approach for time-to-event data has been proposed in \citet{Hof2017}.

%The validity of the pseudo-likelihood approach depends on the assumption that ($\mathbf{Y}_A,\mathbf{X}_B) \independent (\mathbf{L}_A,\mathbf{L}_B)|\boldsymbol\Delta$. This assumption may be violated under WNL and IL, when linkage errors are associated with $\mathbf{Y}_A, \mathbf{X}_B$, or other unobserved variables. 

%\subsubsection{Point estimation for $\boldsymbol{\Delta}$ and canonicalization}

\subsection{Imputation}\label{sec:primaryMI} 
Fully Bayesian approaches to estimating $\boldsymbol{\beta}$ can propagate linkage uncertainty in a principled manner, but are computationally intensive. Moreover, in a joint modeling framework, the models used in the linkage, namely $\mathcal{L}_a(\mathbf{Y}_{A},\mathbf{X}_{B}|\mathbf{L}_{A},\mathbf{L}_{B},\boldsymbol{\Delta},\boldsymbol{\beta})$, may differ from the scientific model of interest. In such situations, an alternative is to approximate a fully Bayesian analysis using multiple imputation (MI) \citep{rubin1987}. 

In MI, $M>1$ imputations of the linkage structure are generated from its posterior distribution. Each imputation is analyzed separately, and estimates are pooled across the imputations using common combination rules \citep{rubin1987,littlerubin2002}. This procedure can propagate linkage uncertainty into the downstream analysis, while simplifying the computation. Application of MI to linked-data inference has been exemplified in \citet{gutman2013}, \citet{abowd2022}, and \citet{shan2021}, among others. \par

Formally, in each of the $M$ posterior samples of the linkage structure, $\boldsymbol{\Delta}^{(1)},\dots,\boldsymbol{\Delta}^{(M)}$, we obtain a point estimate, $\boldsymbol{\hat\beta}^{(m)}$, and its sampling variance $U^{(m)}$, for $m = 1,\dots,M$. A point estimate for $\boldsymbol{\beta}$ across the $M$ linked files is  $\boldsymbol{\hat\beta}=M^{-1}\sum_{m=1}^{M}\boldsymbol{\hat\beta}^{(m)}$. Its estimated variance is $T_M=\bar{U}_M + (1+M^{-1})B_M$, where $\bar{U}_M=M^{-1}\sum_{m=1}^{M} U^{(m)}$ and $B_M=(M-1)^{-1}\sum_{m=1}^{M}(\boldsymbol{\hat\beta}^{(m)}-\boldsymbol{\hat\beta})^2$. A confidence interval for $\boldsymbol{\hat\beta}$ can be constructed using a Student's-$t$ approximation, $(\boldsymbol{\hat\beta}-\boldsymbol{\beta})/ \sqrt{T_M} \sim t_{\nu}$, where $\nu=(M-1)(1+\bar{U}_M/( 
(1+M^{-1})B_M))^2$ \citep{rubin1987}.

A limitation of MI is that the imputation model for $\boldsymbol{\Delta}$ may be uncongenial \citep{meng1994,xiemeng2017} to the post-linkage analysis model. The analysis model is \textit{congenial} to the imputation model, if there exists a unifying Bayesian model for the complete data that embeds both of them \citep{murray2018}. An uncongenial imputation model may lead to biased point estimates, and interval estimates with invalid coverages \citep{van2018}. Since congeniality is hard to ascertain in practice, a general solution is to use an imputation model that is less restrictive than the analysis model. Thus, it may be advisable to include as many relationships of substantive interest as possible in the imputation model for $\boldsymbol{\Delta}$ \citep{Rubin1996,xiemeng2017}. The joint modeling approaches in Section \ref{sec:wnl} can provide a way to accomplish this.

\citet{goldsteinetal12} propose to cast linking two files in a statistical matching framework \citep{dorazio2006}. In this framework, all records in file $\mathbf{A}$ have $\mathbf{X}_B$ missing. For record $i \in \mathbf{A}$, all records in file $\mathbf{B}$ serve as a pool of candidate donors. Assuming that a set of known links is available, \citet{goldsteinetal12} construct a posterior distribution, $p^{*}_{ij} \propto p_{ij} \text{ }p_{B|A}(\mathbf{X}_{B}|\mathbf{Y}_{A},\mathbf{L}_{A},\mathbf{L}_{B})$, where $p_{ij} \propto \hat w_{ij}$ (Section \ref{sec:FS}) are prior linking probabilities, and $p_{B|A}$ denotes the imputation model. For record $i \in \mathbf{A}$, records $j \in \mathbf{B}$ with $p^{*}_{ij}$ below a predefined threshold are considered non-links. If multiple records exceed the threshold, the value of $\mathbf{X}_{Bj}$ such that $ p^{*}_{ij}>p^{*}_{ij'} \forall j' \neq j$ is used for imputing the missing $\mathbf{X}_{Bi}$. If no $p^{*}_{ij}$ reaches the threshold, the missing $\mathbf{X}_{Bi}$ is multiply imputed by sampling from $p_{B|A}(\mathbf{X}_{B}|\mathbf{Y}_{A},\mathbf{L}_{A},\mathbf{L}_{B})$. Each imputed file is analyzed separately, and estimates are pooled using MI combining rules. 

The imputation model $p_{B|A}$ assumes joint multivariate normality of $(\mathbf{Y}_{A},\mathbf{X}_{B}, \mathbf{L}_{A},\mathbf{L}_{B})$, when all variables are continuous. When the data comprise a mix of continuous and categorical variables, $p_{B|A}$ is derived using latent normal models \citep{goldsteinlatentnorm}. 

One limitation of the algorithm is that a set of known links is required to estimate the parameters in $p_{B|A}(\mathbf{X}_{B}|\mathbf{Y}_{A},\mathbf{L}_{A},\mathbf{L}_{B})$. Such a set is not always available in practice. Moreover, the imputation procedure produces statistically valid estimates, as long as the linkage errors are independent of $\mathbf{X}_{B}$ given $\mathbf{Y}_{A},\mathbf{L}_{A}, \text{ and }\mathbf{L}_{B}$. This is equivalent to $\mathbf{X}_{B}$ being missing at random \citep{goldsteinetal12}. Violation of this assumption can lead to substantial bias in point estimates \citep{harron2014}. A possible mitigation strategy is to make the MAR assumption more plausible, by specifying rich imputation models that accommodate variables from auxiliary data sources \citep{goldsteinharonbook}.

\subsection{Weighting} \label{sec:primaryWeight}
Weighting methods aim to obtain unbiased parameter estimates of generalized linear models fit with linked data.  These methods are confined to scenarios where a univariate
response $\mathbf{Y}_A$ is regressed on a set of covariates $\mathbf{X}_B$. Weighting methods rely on two assumptions. The first assumption is that the linkage is complete, i.e., the smaller file $\mathbf{A}$ is a subset of the larger file $\mathbf{B}$, so that $n_{AB}=n_A$.  The second assumption is that the linkage mechanism is NL.

Let $\mathbf{Y}^R_A = \{\mathbf{Y}^R_{Aj}:j=1,\ldots,n_B\}$ be the correct (but unobserved) permutation of records in file $\mathbf{A}$ to those in file $\mathbf{B}$. Then, a probabilistically linked file is of the form $(\mathbf{Y}^*_A,\mathbf{X}_B)$, where $\mathbf{Y}^*_A = \boldsymbol{\Delta}\mathbf{Y}_A^{R}$. 
Under the NL assumption, 
\begin{align}\label{eqmn}
    \begin{split}
        \E(\mathbf{Y}_A^*|\mathbf{L}_A,\mathbf{L}_B,\mathbf{X}_B)&= \E(\E(\mathbf{Y}^*_{A}|\mathbf{Y}^R_{A},\mathbf{L}_A,\mathbf{L}_B,\mathbf{X}_B)|\mathbf{L}_A,\mathbf{L}_B,\mathbf{X}_B) \\
        &= \E(\mathbf{Q}\mathbf{Y}^R_A|\mathbf{L}_A,\mathbf{L}_B,\mathbf{X}_B)
        = \mathbf{Q}\mu(\mathbf{X}_B;\boldsymbol{\beta}),
    \end{split}
\end{align}
where $\mathbf{Q}=[q_{ij}]=[\text{Pr}(\Delta_{ij}=1|\mathbf{L}_{Ai},\mathbf{L}_{Bj},\mathbf{X}_{Bj})]_{(i \in \mathbf{A}, j \in \mathbf{B})}$ is either known, or can be estimated during the linkage process. Under the one-to-one linkage constraint, the row sums of $\mathbf{Q}$, $\sum_{ j \in \mathbf{B}} q_{ij} = 1$.

A common model to describe the association between $\mathbf{Y}^R_A$ and $\mathbf{X}_B$ is the estimating equation,
\begin{equation} \label{eqw1}
 \kappa(\mathbf{X}_B;\boldsymbol{\beta}) \equiv 	\upsilon (\mathbf{X}_B;\boldsymbol{\beta}) \Big[\mathbf{Y}^R_A - \mu(\mathbf{X}_B;\boldsymbol{\beta})\Big] = 0,
\end{equation} 
where $\upsilon(\mathbf{X}_B;\boldsymbol{\beta})$ is the \textit{weighting matrix}, which is independent of $\mathbf{Y}^R_A$. 
Ignoring linkage errors and solving the estimating equation
\begin{equation} \label{eqw2}
\kappa^*(\mathbf{X}_B;\boldsymbol{\beta}) \equiv \upsilon(\mathbf{X}_B;\boldsymbol{\beta})\Big[\mathbf{Y}^*_A - \mu(\mathbf{X}_B;\boldsymbol{\beta})\Big] = 0
\end{equation} 
yields the naive coefficient estimate $\hat{\boldsymbol{\beta}}^*$. 
\begin{theoremP}  \label{prop1}
Under the NL assumption, the following holds:
\begin{enumerate} [(i)]
    \item If $\tilde{\boldsymbol{\beta}}$ is the true value of $\boldsymbol{\beta}$, then $\E(\kappa^*(\mathbf{X}_B;\tilde{\boldsymbol{\beta}})|\mathbf{X}_B) \neq 0$, and 
    \item $\E(\hat{\boldsymbol{\beta}}^*-\boldsymbol{\beta}|\mathbf{X}_{B}) \approx  - \Big[\frac{\partial \kappa^*(\mathbf{X}_{B};\boldsymbol{\beta})}{\partial \boldsymbol{\beta}}\Big]^{-1} \upsilon(\mathbf{X}_{B};\boldsymbol{\beta})\text{ }(\mathbf{Q}-\mathbf{I})\text{ }\mu(\mathbf{X}_{B};\boldsymbol{\beta})$.
\end{enumerate}
\end{theoremP}
We defer the proof to the supplementary materials. Proposition \ref{prop1} indicates that $\kappa^*(\mathbf{X}_B;\boldsymbol{\beta})$ is a biased estimating equation, and $\hat{\boldsymbol{\beta}}^*$ is a biased estimator for $\boldsymbol{\beta}$. 

\citet{scheurenwinkler93} describe an approach to mitigate the bias in $\hat{\boldsymbol{\beta}}^*$ for linear regression, when $n_A=n_B$, $\mu(\mathbf{X}_B;{\boldsymbol{\beta}})=\mathbf{X}_B{\boldsymbol{\beta}}$ and $\upsilon(\mathbf{X}_B;{\boldsymbol{\beta}})=\mathbf{X}_B^{T}$. Their bias-adjusted estimator is $\hat{{\boldsymbol{\beta}}}_{SW} = \hat{{\boldsymbol{\beta}}}^* - (\mathbf{X}_B^{T} \mathbf{X}_B)^{-1}\mathbf{X}_{B}^{T}(\mathbf{Q} - \boldsymbol{\Delta})\mathbf{Y}^*_A$. As most record pairs from files $\mathbf{A}$ and $\mathbf{B}$ do not represent links, \citet{scheurenwinkler93} truncate the $i^{th}$ row of $\mathbf{Q}$, so that all entries except the two highest $q_{ij}$ are set to null. \citet{scheurenwinkler97} utilize $\hat{{\boldsymbol{\beta}}}_{SW}$ in a procedure that recursively iterates between identifying non-links, and improving the estimates of ${\boldsymbol{\beta}}$.

Although $\boldsymbol{\hat{{\boldsymbol{\beta}}}}_{SW}$ has a lower bias compared to the naive estimator, it is not unbiased for ${\boldsymbol{\beta}}$ \citep{lahirilarsen05}. 
An alternative is to solve the adjusted estimating equation \citep{hanlahiri2019}
\begin{equation} \label{eqw4}
\Tilde{\kappa}(\mathbf{X}_B;{\boldsymbol{\beta}}) \equiv \upsilon(\mathbf{X}_B;{\boldsymbol{\beta}})\Big[\mathbf{Y}^*_A - \mathbf{Q}\mu(\mathbf{X}_B;{\boldsymbol{\beta}})\Big] = 0.
\end{equation} 
For the identity link function, $\mu(\mathbf{X}_B;{\boldsymbol{\beta}})=\mathbf{X}_B{\boldsymbol{\beta}}$, and equally sized files, setting $\upsilon(\mathbf{X}_B;{\boldsymbol{\beta}})=\mathbf{X}_B^T\mathbf{Q}^{T}$, yields the estimator $\hat{{\boldsymbol{\beta}}}^{LL}=(\mathbf{X}_B^T\mathbf{Q}^T\mathbf{Q}\mathbf{X}_B)^{-1}\mathbf{X}_B^T\mathbf{Q}^T\mathbf{Y}_A^*$ \citep{lahirilarsen05}. Alternatively, $\hat{\boldsymbol{\beta}}^{LL}$ can be obtained using a least squares regression of $\mathbf{Y}_A^*$ on a weighted combination of the covariates, $\mathbf{Q}\mathbf{X}_B$. 

The point estimator $\hat{{\boldsymbol{\beta}}}^{LL}$ is unbiased when $\mathbf{Q}$ is known. However, in most applications, only an estimate of $\mathbf{Q}$ is available. When ${\mathbf{Q}}$ is estimated using the FS model, its rows may not sum to one, as the one-to-one linkage constraint is not enforced. A possible solution is to normalize each entry in $\mathbf{Q}$ as $q^{'}_{ij}= q_{ij}/\sum_{j\in \mathbf{B}} q_{ij}$.

\citet{chipperfield2015} argue that using $q^{'}_{ij}$ can lead to biased inferences under one-to-one linkage. This limitation can be overcome by estimating $\mathbf{Q}$ using a bootstrap for the entire linkage process \citep{winglee2005,chipperfield2015}. Other variations of the weighting approach have been proposed, including using weighted least squares when covariates are spread over multiple files \citep{Hof2012}, and relaxing the assumption of complete linkage \citep{Hof2012,chipperfield2015}. \par

%Weighting estimators can have poor finite sample performance, even if they are correctly specified \cite{basu1971,rodnitzky2014}. To address this issue, setting some record pairs to have zero linkage probability  can improve finite sample performance \cite{hanlahiri2019}.However, this procedure can result in biased estimates when the linking probabilities are estimated inaccurately. 

Computation of variances with estimated weights is generally not straightforward \citep[][]{littlerubin2002}. Valid variance estimators require elaborate theoretical calculations \citep{tsiatis2006}, or computationally-intensive resampling techniques \citep[][]{littlerubin2002}.
For particular choices of $\upsilon(\mathbf{X}_B;\boldsymbol{\beta})$, \citet{han2018} derives theoretical plug-in estimators for the sampling variance of $\hat{\boldsymbol\beta}$ under Equation \eqref{eqw4}. To simplify calculations, \citet{han2018} suggests truncating every row of $\mathbf{Q}$ to include the highest (or two highest) values. These 
estimators, however, do not account for the uncertainty arising from using an estimate of $\mathbf{Q}$. This can be addressed by using resampling methods that re-estimate $\mathbf{Q}$ at every iteration. Two such methods are based on the unified Jackknife theory \citep{jiang2002,hanlahiri2019}, and the Bootstrap \citep{lahirilarsen05}.

\subsection{Interlude: Estimating complex models with linked files}
A majority of the linked-data inferential methods in the literature center around estimating parameters of generalized linear models. Some work has also focused on inferring the size of a closed population, using capture-recapture \citep{fienberg72} or multiple-systems estimation \citep{birdking}. For example, Bayesian models for concurrently inferring the linkage and the population size have been developed, for the case of two files \citep{tancrediliseo2011}, and multiple files \citep{tancredi2020}.

The Bayesian and imputation-based approaches also
offer a flexible framework to estimate other complex models, such as hierarchical models \citep{gelman2013}, or multivariate analyses like finite mixture models \citep{peel2000}. As some multivariate analysis methods are exploratory in nature (for e.g, factor analysis), posterior sampling within a fully Bayesian framework can be computationally daunting \citep{conti2014}. In such cases, an MI approach can offer more feasibility. In Section \ref{sec:multivar}, we provide an example of using MI to perform model-based cluster analysis with a linked file.

\section{Secondary analysis of linked files} \label{sec:secondary}
\textit{Secondary analysis} refers to settings when the linkage and the data analysis are performed by different individuals. Specifically, the data analyst is provided a linked file, but does not have access to the individual files, the linking variables, or linking probabilities for record pairs.  This situation arises when the original data comprise sensitive information, and releasing individual files interferes with privacy regulations, or poses considerable disclosure risks \citep{gilbert2017,christen20}. 

In secondary analysis, the analyst does not have access to records that were left unlinked during the linkage process. Consequently, most analysis methods can only propagate the uncertainty due to false links. Inferences from the linked file may not be generalizable to the entire population of entities, without assumptions on the relationship between the linked and unlinked records, or availability of additional information about the unlinked records. 

Formally, secondary analysis is based on the file $\mathbf{F} \equiv \{\mathbf{Y}_{Al},\mathbf{X}_{Bl}\}_{l=1}^{n_{AB}}$, obtained by linking files $\mathbf{A}$ and $\mathbf{B}$. The file $\mathbf{F}$ can be partitioned into true and false links, but this partitioning is unknown. In this case, we reformulate the linkage structure using a vector of indicators $\mathbf{C}= \{C_l\}$, where $C_l=1$ if record pair  $l$ is a true link, and $C_l=0$, otherwise, where $l=1, \dots n_{AB}$.

\subsection{Likelihood, Bayesian, and imputation methods}
Under the WNL assumption, \citet{gsamgreen16} postulate that record pairs in $\mathbf{F}$ are derived from a two-class mixture model. Formally, the likelihood is 
\begin{align}\label{gtmix}
\begin{split}
    \mathbf{Y}_{Al},\mathbf{X}_{Bl}| C_{l}=1 & \sim \mathcal{L}_M(\mathbf{Y}_{Al},\mathbf{X}_{Bl}|{\boldsymbol{\beta}}_M),  \\
    \mathbf{Y}_{Al},\mathbf{X}_{Bl}| C_{l}=0 & \sim \mathcal{L}_U(\mathbf{Y}_{Al},\mathbf{X}_{Bl}|{\boldsymbol{\beta}}_{U}), \\
    C_l &\overset{\mathrm{iid}}{\sim} \text{ Bernoulli }(\lambda),
\end{split}
\end{align}
where $\boldsymbol{{\boldsymbol{\beta}}}_M$ and $\boldsymbol{{\boldsymbol{\beta}}}_U$ are parameters governing the joint distribution of $(\mathbf{Y}_{A}, \mathbf{X}_{B})$ among the true and false links, respectively, and $\lambda$ is the marginal probability of a true link. The models $\mathcal{L}_M$ and $\mathcal{L}_U$ can be augmented to include additional variables present in the linked file.

To avoid label switching, \citet{gsamgreen16} rely on informative prior distributions for $({\boldsymbol{\beta}}_M,{\boldsymbol{\beta}}_{U},\lambda)$. These priors are constructed using a validation sample, and other available linkage paradata. The missing $\mathbf{C}$ is imputed using a DA scheme, that iterates between sampling $\mathbf{C}$, and the parameters $(\boldsymbol{\beta}_M,\boldsymbol{\beta}_U,\lambda)$. When $\mathcal{L}_M$ represents the scientific model of interest, inference based on posterior samples of ${\boldsymbol{\beta}}_M$ can account for the uncertainty around $\mathbf{C}$. Alternatively, each DA iteration provides a set of links, that can be used to fit the analysis model of interest. Point and interval estimates across posterior samples of $\mathbf{C}$ can be obtained using the MI combining rules (Section \ref{sec:primaryMI}). \par

\citet{slawski2021} consider the case of linear regression, and assume independence between $\mathbf{Y}_{Al}$ and $\mathbf{X}_{Bl}$ when postulating $\mathcal{L}_U$ in Equation \eqref{gtmix}.
An estimate $\hat{\boldsymbol{\beta}}_M$ is obtained by maximizing the likelihood in \eqref{gtmix} using the EM algorithm. The variance of $\hat{\boldsymbol{\beta}}_M$ is approximated using the Huber-Sandwich estimator \citep{huber}, and confidence intervals are constructed under an asymptotic normal approximation for the sampling distribution of $\hat{\boldsymbol{\beta}}_M$.

\subsection{Weighting} \label{sec:secondaryWeight}
In a secondary analysis setting, the matrix $\mathbf{Q}$ (Equation \eqref{eqmn}) can only be modeled using coarse summary statistics from the linkage process. One such statistic is the overall proportion of correct links, $\lambda$. This can be calculated, for example, by comparing a representative audit sample from the linked file to a gold-standard linked file \citep{chambersdasilva2020}.

Under the assumption that $n_A=n_B$, and that the linkage is complete and one-to-one, \citet{Chambers2009} imposes an ``exchangeable'' structure on $\mathbf{Q}$. Formally, the exchangeable linkage error (ELE) model is \citep{Chambers2009} 
\begin{align} \label{eq:ele}
\begin{split}
[\mathbf{Q}]_{(i \in \mathbf{A},j\in \mathbf{B})} &= [\text{Pr }(\Delta_{ij}=1|\mathbf{L}_{Ai},\mathbf{L}_{Bj},\mathbf{X}_{Bj})]_{(i \in \mathbf{A},j\in \mathbf{B})}  \\&= \begin{cases}
     \lambda, \text{ if } i=j \\
     \frac{1-\lambda}{n_A-1}, \text{ if } i\neq j.
    \end{cases} 
\end{split}
\end{align}
This construct implies that (i) every record $j \in \mathbf{B}$ has probability $\lambda$ of being the correct link for record $i\in \mathbf{A}$, and (ii) any two records that are not linked to $i\in \mathbf{A}$ are equally likely to be the correct links. 

Exchangeability implies that the linkage is independent of the linking and substantive variables, so that the linkage mechanism is LCAR. In many applications, this assumption may be overly stringent, as the probability of correct linkage can be heterogeneous across records \citep{winkler2021}. A possible solution is to assume that the linkage probabilities are homogeneous within disjoint subgroups (or blocks) of the linked file, rather than across all the linked records \citep{chambersdasilva2020}. \par

For inference on $\boldsymbol{\beta}$ in Equation \eqref{eqw4}, \citet{Chambers2009} proposes different types of weighting matrices $\upsilon(\mathbf{X}_B;\boldsymbol{\beta})$. In the case of linear regression, setting $\upsilon(\mathbf{X}_B;\boldsymbol{\beta})=\mathbf{X}_B ^\top$ yields an estimator with a ratio-type bias correction, similar in spirit to \citet{scheurenwinkler93}. Setting $\upsilon(\mathbf{X}_B;\boldsymbol{\beta})=\mathbf{X}_B^\top \mathbf{Q}^\top $, yields the Lahiri-Larsen estimator, with $\mathbf{Q}$ defined in Equation \eqref{eq:ele}. Setting $\upsilon(\mathbf{X}_B;\boldsymbol{\beta})=\mathbf{X}_B^\top \mathbf{Q}^\top (\sigma^2\mathbf{I}_{n_A}+\mathbf{\Sigma})^{-1}$, where $\sigma^2=\V(\mathbf{Y}_{A}|\mathbf{X}_{B})$ and $\mathbf{\Sigma}=\V(\boldsymbol{\Delta}\E(\mathbf{Y}_A|\mathbf{X}_B)|\mathbf{X}_B)$,  yields the asymptotically efficient best linear unbiased estimator (BLUE) of $\boldsymbol{\beta}$. Corresponding estimators for the case of logistic regression have also been proposed in \citet[][pp.22-23]{Chambers2009}. \citet{Chambers2009} derives Sandwich-type asymptotic sampling variances of these estimators, which account for the variability from using an estimate of $\lambda$. \par

 \citet{kimchambers2010} extend Equation \eqref{eqw4} to the case of incomplete linkage, when some records from both files may remain unlinked. For valid inference, \citet{kimchambers2010} assume  ignorable non-linkage, which implies that unlinked records in each file are a random sample of all the records in that file. Equivalently, the relationship between $\mathbf{Y}^*_A$ and $\mathbf{X}_B$ is the same among linked and unlinked records.  This assumption is equivalent to unit non-response in surveys occurring completely at random \citep{rubin1976}. When the non-linkage is not ignorable, \citet{kimchambers2010} propose to adjust regression coefficient estimates using inverse probability weighting \citep{seamanwhite2011}. 
 
 \citet{zhang2021} use the ELE model to develop pseudo-OLS estimators for linear regression, that accommodate incomplete linkage and heterogeneous linkage probabilities. Other extensions of the ELE model include its application to multi-file linkage \citep{kimchambers2012,kimchambers2015}, to estimation of  survey-weighted population totals \citep{chipperfield2019}, to survival analysis \citep{Vo23surv}, to robust regression in the presence of outliers \citep{chambers2022}, and to estimation using calibrated auxiliary information \citep{chambersdasilva2020}.
 Sensitivity of the linear regression estimators to the ELE assumption has also been demonstrated, theoretically \citep{zhang2021}, and via simulations \citep{chambersinf}.

\section{Simulation studies} \label{sec:sim}
We examine the performance of selected methods from Sections \ref{sec:primary} and \ref{sec:secondary} using extensive simulations. These methods encompass a wide range of linkage and inference techniques, and depict the main performance trends. We consider two simulation scenarios, corresponding to the primary and secondary analysis settings.

\subsection{Scenario 1: Primary analysis}
\subsubsection{Simulation design}
We generate files $\mathbf{A}$ and $\mathbf{B}$ of sizes $n_A=600$ and $n_B=900$, respectively. The simulations are designed as a factorial experiment with six factors classified into five main categories:
\begin{enumerate}
\item \textbf{Overlap}: The overlap represents the proportion of true links. It is defined as the proportion of records from file $\mathbf{A}$ that exist in file $\mathbf{B}$, as a fraction of the smaller file. We examine three levels of the overlap: 100\% (complete overlap), 50\% (medium overlap), and 20\% (low overlap).
\item \textbf{Blocking}: We consider two blocking scenarios. In the first scenario, both
files comprise 6 blocks. Consequently, there are 100 records per block in file
$\mathbf{A}$, and 150 records per block in file $\mathbf{B}$. In the second scenario, both files comprise 60 blocks, so that each block in file $\mathbf{A}$ contains 10 records, and each block in file $\mathbf{B}$
contains 15 records. 
\item \textbf{Linking variables}: We examine the effect of the discriminatory power (DP) of the linking variables under two scenarios. In the high DP scenario, we simulate four linking variables that represent an individual's zip code stratum (with 30 categories), last name (with $n_A+n_B-n_{AB}$ unique values), race (with 5 categories), and year of birth. In the low DP scenario, we simulate four variables to represent individual's gender (with 2 categories), last name (with $n_A$ unique values), race (with 5 categories), and year of birth. In both cases, we generate the year of birth and race from the same distributions, but vary how the other linking variables are generated.
\item \textbf{Variables for post-linkage analysis}: For record pairs that are links, we generate $\mathbf{Y}_A$ and $\mathbf{X}_B$ from a bivariate normal distribution 
\begin{align} \label{eq:datagen}
\begin{pmatrix}
Y_{Al} \\
X_{Bl}
\end{pmatrix}\overset{iid}\sim N_2\left(\begin{pmatrix}
0 \\
0
\end{pmatrix},\begin{pmatrix}
4 & 2\rho \\
2\rho & 1
\end{pmatrix}\right).
\end{align}
Under this setup, a linear regression of $\mathbf{Y}_{A}$ on $\mathbf{X}_{B}$ is of the form ${Y}_{Al} = \beta X_{Bl} + \epsilon_l$, where $\beta=2\rho$, and $\epsilon_l \overset{iid}\sim N(0,4(1-\rho^2))$. We vary the value of $\rho$, which controls the signal-to-noise ratio (SNR) of the regression model. In the high SNR scenario, we set $\rho =0.9$, and in the low SNR scenario, we set $\rho =0.4$. Among non-links, we generate $\mathbf{Y}_{A}$ and $\mathbf{X}_{B}$ independently as $Y_{Ai} \overset{iid}\sim N(0,1)$ and $X_{Bj} \overset{iid}\sim N(0,1)$.

\item \textbf{Measurement error}: We induce measurement errors in two out of the four linking variables: the last name and the year of birth. We vary the level and the mechanism of error generation. 
\begin{enumerate}
\item \textbf{Error level}: We consider two error levels: 10\% (low) and 40\% (high).
\item \textbf{Error mechanism}: Within the high and low error levels, we generate measurement errors according to the LCAR, SNL, NL, WNL, and IL mechanisms. Under LCAR, we randomly select records from the two files and introduce errors in the last name and the year of birth. Under SNL, we let the probability of error in a record to depend on an individual's race. Under NL and WNL, we let the probability of error in a record depend on $\mathbf{X}_B$, and $\mathbf{Y}_A$, respectively. Under IL, this probability depends on an unobserved variable which is associated with $\mathbf{Y}_A$ and $\mathbf{X}_B$.
\end{enumerate}
\end{enumerate}
In summary, we generate $2^4 \times 3 \times 5 = 240$ simulation configurations. For each configuration, we generate 100 simulation replications. We describe the data generation process in the supplementary materials. 

\subsubsection{Evaluation metrics}
We evaluate the performance of seven methods when estimating the regression slope, $\beta$.  For each method, in each simulation replication, we record the point estimate  $\hat{\beta}$, its estimated variance, and whether a 95\% interval estimate covers the true value of $\beta$. We summarize the results across all replications by calculating the mean bias of the point estimate, the average estimated standard error (SE), and the coverage rate of the 95\% interval estimate.

\subsubsection{Record linkage and inference methods}
We link files $\mathbf{A}$ and $\mathbf{B}$ using the FS method, implemented in the \texttt{fastlink} package \citep{imaifastlink} in \texttt{R}. Using the linked file, we obtain an estimate of the slope, $\hat\beta$, its sampling variance, and a 95\% confidence interval for $\beta$. 

We implement the weighted estimated equations approach of \citet{hanlahiri2019} in the complete overlap scenario. We estimate the matrix of posterior matching probabilities, $\mathbf{\hat Q}$, using the \texttt{fastLink} package in \texttt{R}. We consider three versions of the matrix $\mathbf{\hat Q}$: (i) the full matrix without any modifications (HLF); (ii) in every row, we set all entries except the two highest to zero (HL2); and (iii) in every row, we set all entries except the highest to zero (HL1). We obtain the estimate $\hat\beta$ using the blocked version of Equation \eqref{eqw4} \citep[][p.5]{hanlahiri2019}. To estimate the sampling error of $\hat\beta$, we follow the jacknife resampling approach in \citet{hanlahiri2019}. In each iteration of the jacknife, we leave out records from a block in both files, so that the number of jacknife replicates is equal to the number of blocks. We aggregate results across all jacknife replicates to obtain the sampling variance of $\hat\beta$, and a 95\% confidence interval for $\beta$. \par

We also examine the pseudo-likelihood method of \cite{Hof2012} (abbreviated as HZ). To construct the pseudo-likelihood for the distribution of $\mathbf{Y}_A$ given $\mathbf{X}_B$, we use Equation (8) from \citet[][p.79]{Hof2012}. We calculate the probabilities for the observed agreement patterns using the \texttt{fastLink} package in \texttt{R}. We obtain the point estimate $\hat\beta$ using the EM algorithm \citep[][Appendix B]{Hof2012}. To compute the sampling variance of $\hat\beta$, we invert the observed Fisher information matrix \citep[][p.80]{Hof2012}. We compute a 95\% confidence interval for $\beta$ under asymptotic normality of the sampling distribution of $\hat\beta$. 

We use four Bayesian methods to link files $\mathbf{A}$ and $\mathbf{B}$, and estimate the slope $\beta$. The first method is that of \citet{Sadinle2017} (abbreviated as SL), which we apply independently within each block. We assume a uniform prior distribution over the parameters of the record linkage likelihood in Equation \eqref{sadinle}. We use exact agreement to compare the zip stratum, gender, race, and the year of birth. We model the similarity in the last names using the Levenshtein edit $(LD)$ distance  \citep{levenshtein}. We standardize the $LD$ distance to obtain a final measure in the range of $[0, 1]$, where 0 represents total agreement and 1 represents total disagreement. Following \citet{Sadinle2017}, we categorize $LD$ into four levels of disagreement: perfect agreement ($LD$ = 0), mild disagreement ($0<LD\leq0.25$), low disagreement ($0.25 <LD \leq 0.5$), and extreme disagreement ($LD \geq 0.5$). We generate $1000$ samples from the posterior distribution of the linkage structure, using a Gibbs sampler implemented in the \texttt{R} package \texttt{BRL} \citep{brl}.\par

The second Bayesian method is that of \citet{steorts16} (abbreviated as ST), which we apply independently within each block. To implement the linking algorithm, we concatenate records from each block in the two files into a single file, and perform deduplication on the concatenated file. In the record linkage model, we assume that the year of birth, zip stratum, race, and gender are categorical variables, and use the $LD$ distance to model similarities on the last name. We set prior distributions and hyperparameters following the recommendations in \citet[][pp.9-17]{steorts16}. We provide specific details of the hyperparameter values in the supplementary materials. We generate 1000 samples from the posterior distribution of the linkage structure, using the Gibbs sampler implemented in the \texttt{R} package \texttt{blink} \citep{blink}.\par

The third Bayesian method is that of \citet{brlvof} (abbreviated as KSG), which augments the likelihood in Equation \eqref{sadinle} with associations between $\mathbf{Y}_A$ and $\mathbf{X}_B$. For record pair $(i,j)$, we assume $Y_{Ai}|X_{Bj} \sim N(\beta_{0M}+\beta_{1M} X_{Bj}, \sigma_M^2)$ if the pair is a link, and $Y_{Ai}|X_{Bj} \sim N(\beta_{0U}+\beta_{1U} X_{Bj}, \sigma_U^2)$ if the pair is a non-link. As in the SL algorithm, we assume uniform prior distributions on the parameters defining the similarity on the zip stratum, gender, race, year of birth and last name. To complete the Bayesian model, we assume the improper prior distributions $p(\boldsymbol{\beta}_M,\sigma_M) \propto \sigma_M^{-2}$ and $p(\boldsymbol{\beta}_U, \sigma_U) \propto \sigma_U^{-2}$. We adjust for blocking following the method described in \citet[][]{brlvof}. We generate 1000 samples from the posterior distribution of the linkage structure using the DA algorithm.\par

The fourth Bayesian method is that of \citet{gutman2013} (abbreviated as GT). We apply this method in the complete overlap scenario only. We extend the likelihood in \citet{gutman2013} to account for associations between variables common to both the files. We model the agreements on the year of birth, zip stratum, race, and gender as Bernoulli random variables. We specify Beta(1,1) prior distributions on the parameters of the Bernoulli distribution. In addition, we model the number of record pairs with LD distances in the ranges $(0,0.25)$, $(0.25, 0.5)$, and $(0.5,1)$ using a multinomial distribution with three parameters. We specify a Dirichlet (1,1,1) prior on each of the parameters.  We describe the complete model specification in the supplementary materials. We generate 1000 samples from the posterior distribution of the linkage structure, using the DA algorithm in \citet{gutman2013}. In each DA iteration, we perform $50,000$ record-pair-switches per block. \par

For all Bayesian methods, in each posterior sample of the linkage structure, we compute the estimated regression coefficient $\hat\beta$ among the identified links, and its sampling variance. We obtain point and interval estimates across all posterior samples using the MI combining rules described in Section \ref{sec:primaryMI}. \par

\subsubsection{Results}
Within each level of the overlap, we identify three simulation factors that have the largest effect on the bias, SE, and coverage, using ANOVA \citep{cangul2009}. The error mechanism, error level, and the discriminatory power are generally the most influential factors across the different 
evaluation metrics. We report results by averaging over levels of the remaining factors (the block size and the SNR).

Table \ref{tab:primComp} summarizes the performance for the complete overlap configuration, when the linking variables have high discriminatory power, and there is low measurement error. Across all linkage mechanisms, the bias of $\hat\beta$ is generally negative, as false links attenuate the estimates towards zero. The FS estimator of $\beta$ underestimates the true value, and has coverage below the nominal. When the mechanism is LCAR, SNL, or NL, the Bayesian, weighting, and likelihood-based methods have generally small biases. For these linkage mechanisms, the Bayesian methods have the smallest biases, with SL being the most efficient. Moreover, their interval estimates have coverages at or above the nominal. Weighting methods have relatively larger biases, but produce interval estimates with near-nominal coverage. Across the LCAR, SNL, and NL mechanisms, HZ exhibits small biases, but also has the largest SEs among all methods, which results in coverage rates well above 95\%.

\begin{table}[]
\caption[]{Bias, estimated SE, and coverage probabilities for estimates of $\beta$ in the primary analysis scenario with complete overlap. The measurement error level is low, the linking variables have high discriminatory power.\\} \label{tab:primComp}
\centering
\begin{tabular}{llccc}
\hline
\makecell{Linkage\\Mechanism}  & Method   & Bias      & SE      & Coverage \\ \hline
\multirow{6}{*}{LCAR} &FS &-0.099&0.039&0.64 \\ 
  &SL & -0.002 & 0.059 & 0.94 \\ 
  &ST & -0.007 & 0.065 & 0.94 \\ 
  &GT & 0.010 & 0.069 & 0.94 \\ 
  &KSG & 0.009 & 0.069 & 0.94 \\ 
  &HLF & -0.058 & 0.066 & 0.93 \\ 
  &HL2 & -0.058 & 0.066 & 0.92 \\ 
  &HL1 & -0.044 & 0.066 & 0.94 \\ 
  &HZ & 0.011 & 0.137 & 1.00 \\ 
   \hline
\multirow{6}{*}{SNL}  &FS &-0.094&0.039&0.64 \\ 
 &SL & -0.003 & 0.059 & 0.94 \\ 
 &ST & -0.007 & 0.065 & 0.94 \\ 
 &GT & 0.011 & 0.068 & 0.94 \\ 
 &KSG & 0.013 & 0.068 & 0.94 \\ 
 &HLF & -0.064 & 0.066 & 0.93 \\ 
 &HL2 & -0.064 & 0.066 & 0.93 \\ 
 &HL1 & -0.049 & 0.066 & 0.93 \\ 
 &HZ & 0.014 & 0.127 & 1.00 \\ \hline
\multirow{6}{*}{NL}  &FS &-0.097&0.039&0.54 \\ 
  &SL & -0.003 & 0.059 & 0.94 \\ 
  &ST & -0.006 & 0.069 & 0.95 \\ 
  &GT & 0.009 & 0.069 & 0.94 \\ 
  &KSG & 0.011 & 0.068 & 0.94 \\ 
  &HLF & -0.050 & 0.067 & 0.95 \\ 
  &HL2 & -0.050 & 0.067 & 0.95 \\ 
  &HL1 & -0.034 & 0.067 & 0.95 \\ 
  &HZ & -0.084 & 0.139 & 0.99 \\ \hline
\multirow{6}{*}{WNL} &FS & -0.125 & 0.061 & 0.42 \\ 
 &SL & -0.085 & 0.060 & 0.82 \\ 
 &ST & -0.143 & 0.061 & 0.31 \\ 
 &GT & -0.070 & 0.067 & 0.88 \\ 
 &KSG & -0.072 & 0.069 & 0.90 \\ 
 &HLF & -0.094 & 0.068 & 0.90 \\ 
 &HL2 & -0.095 & 0.068 & 0.89 \\ 
 &HL1 & -0.076 & 0.068 & 0.90 \\ 
 &HZ & -0.111 & 0.142 & 0.99 \\ 
   \hline
\multirow{6}{*}{IL} &FS &-0.172 & 0.059 & 0.16 \\ 
  &SL & -0.091 & 0.060 & 0.70 \\ 
  &ST & -0.204 & 0.058 & 0.04 \\ 
  &GT & -0.103 & 0.066 & 0.74 \\ 
  &KSG & -0.124 & 0.068 & 0.63 \\ 
  &HLF & -0.111 & 0.068 & 0.88 \\ 
  &HL2 & -0.112 & 0.068 & 0.86 \\ 
  &HL1 & -0.088 & 0.068 & 0.88 \\ 
  &HZ & -0.162 & 0.157 & 1.00 \\ 
   \hline
\end{tabular}
\end{table}

The bias under all the methods increases when the linkage mechanism is either WNL or IL. Under WNL, GT, KSG, and HL1 produce estimates with relatively smaller bias, but have coverage rates  below the nominal. Estimates under HZ have high bias, but inflated SEs result coverage probabilities above the nominal. Under IL, the bias under all the methods increases considerably. The coverage is below-nominal across all the Bayesian methods. The weighting methods have coverage near 88\%, while HZ has above-nominal coverage.

Increasing the measurement error has a noticeable impact on the performance of all the methods. When the linking variables are highly discriminatory, increase in the error level leads to larger biases under HZ and the weighting methods (Table S1 in the supplementary materials). In spite of the increased error, the Bayesian methods have relatively small biases under LCAR, SNL, and NL. Under WNL and IL, there is a substantial decrease in performance across all the methods. Under these linkage mechanisms, HZ provides valid coverage at the cost of higher SEs, and ST has the lowest coverage.
The results are qualitatively similar when the discriminatory power is low (Tables S2, S3 in the supplementary materials).

\begin{table}[t]
\caption[]{Bias, estimated SE, and coverage probabilities for estimates of $\beta$ in the primary analysis scenario with medium and low overlap. The measurement error level is low, the linking variables have high discriminatory power.\label{tab:prim} \\ }
\centering
\begin{tabular}{llcccccc}
\hline
&   &  \multicolumn{3}{c}{Medium Overlap} & \multicolumn{3}{c}{Low Overlap} \\ 
\cmidrule(r){3-5} \cmidrule(l){6-8}
\makecell{Linkage\\Mechanism}   &Method  & \makecell{Bias}        & SE     & Coverage       & \makecell{Bias}     & SE    & Coverage    \\ \hline
\multirow{2}{*}{LCAR} & FS & -0.158 & 0.098 & 0.51 & -0.314 & 0.132 &0.38 \\ 
 &SL & -0.109 & 0.115 & 0.90 & -0.280 & 0.207 & 0.79 \\ 
 &ST & 0.002 & 0.093 & 0.96 & 0.003 & 0.147 & 0.95 \\ 
 &KSG & -0.095 & 0.117 & 0.92 & -0.242 & 0.210 & 0.82 \\ 
 &HZ & 0.017 & 0.172 & 0.99 & 0.006 & 0.163 & 0.97 \\ 
   \hline
\multirow{2}{*}{SNL} & FS & -0.147 & 0.087 & 0.58 & -0.293 & 0.134 &0.44 \\ 
 &SL & -0.110 & 0.115 & 0.89 & -0.282 & 0.207 & 0.78 \\ 
 &ST & 0.003 & 0.092 & 0.95 & 0.010 & 0.148 & 0.94 \\ 
 &KSG & -0.103 & 0.118 & 0.91 & -0.236 & 0.209 & 0.80 \\ 
 &HZ & 0.012 & 0.178 & 1.00 & 0.079 & 0.185 & 0.95 \\ 
   \hline
                    
\multirow{2}{*}{NL} & FS & -0.208& 0.091 & 0.53 & -0.295& 0.139 &0.36 \\ 
 &SL & -0.125 & 0.120 & 0.89 & -0.325 & 0.216 & 0.70 \\ 
 &ST & -0.005 & 0.098 & 0.91 & -0.001 & 0.158 & 0.96 \\ 
 &KSG & -0.108 & 0.128 & 0.90 & -0.279 & 0.224 & 0.72 \\ 
 &HZ & 0.010 & 0.175 & 0.98 & -0.001 & 0.194 & 0.96 \\ 
   \hline  
\multirow{2}{*}{WNL} & FS & -0.239 & 0.085 & 0.29 & -0.472 & 0.134 &0.19 \\ 
 &SL & -0.137 & 0.118 & 0.85 & -0.335 & 0.213 & 0.66 \\ 
 &ST & -0.153 & 0.088 & 0.55 & -0.170 & 0.143 & 0.78 \\ 
 &KSG & -0.130 & 0.129 & 0.87 & -0.321 & 0.217 & 0.69 \\ 
 &HZ & -0.127 & 0.198 & 0.98 & -0.160 & 0.200 & 0.92 \\ 
   \hline            
\multirow{2}{*}{IL} & FS & -0.279& 0.083 & 0.15& -0.532 & 0.123& 0.17 \\ 
 &SL & -0.160 & 0.120 & 0.79 & -0.377 & 0.218 & 0.61 \\ 
 &ST & -0.254 & 0.085 & 0.14 & -0.280 & 0.141 & 0.46 \\ 
 &KSG & -0.154 & 0.126 & 0.80 & -0.351 & 0.226 & 0.64 \\ 
 &HZ & -0.204 & 0.187 & 0.94 & -0.246 & 0.207 & 0.84 \\ 
   \hline
\end{tabular}

\end{table}

Table \ref{tab:prim} depicts the results under the medium and low overlap scenarios, when the linking variables have high discriminatory power, and there is low measurement error. Under LCAR, SNL, and NL, estimates under ST have the lowest bias, near-nominal coverage rates, and are the most efficient. Estimates under HZ also have low bias and valid coverage, but with larger standard errors compared to ST. The performance of SL and KSG declines when the overlap is not complete, with KSG offering lower biases and better coverage rates than SL. Under WNL and IL, estimates under ST have larger biases, and below-nominal coverage. Under the medium overlap scenario, KSG and HZ have relatively low biases, with KSG exhibiting smaller SEs, but below-nominal coverage. The relative performance of the methods remains qualitatively unchanged when the error level increases (Table S4), or when the linking variables have low discriminatory power (Tables S5, S6).

We next examine the performance of the methods when the relative sizes of the two files change. Specifically, we let $n_A=600$ and $n_B=1800$, and consider the scenario when the linkage mechanism is IL, the error level is high, and the linking variables are highly discriminatory. We present results in Table S7 of the supplementary materials. Across all levels of the overlap, the biases under all methods generally persist at the same level as when $n_B=900$ (Tables S1 and S4). However, the estimated SEs are lower, leading to reduced coverage rates. 

We also evaluate the performance of GT and KSG, when the dependence between $\mathbf{Y}_A$ and the comparison vectors is incorporated into the models. For both methods, we let $Y_{Ai}|X_{Bj} \sim N(\beta_{0M}+\beta_{1M} X_{Bj} + \beta_{2M}\gamma_{ijl}, \sigma_M^2)$ among the links. Under KSG, we also model $Y_{Ai}|X_{Bj} \sim N(\beta_{0U}+\beta_{1U} X_{Bj}+\beta_{2U}\gamma_{ijl}, \sigma_U^2)$ among the non-links. Here, $\gamma_{ijl}$ denotes the comparison on the last name. We do not include comparisons on the other linking variables, to avoid singularity in the regression model. We consider the scenario when the overlap is complete, the error level is high, and the linking variables are highly discriminatory. Table S8 in the supplementary materials depicts the bias, estimated SEs and coverage probabilities. Under the LCAR and SNL mechanisms, the results are qualitatively similar to  Table S1. Under NL, WNL, and IL, both methods show improvement in the bias and the coverage.

\subsection{Example: Model-based clustering with linked data} \label{sec:multivar}
We illustrate the use of MI with linked data, when the downstream analysis involves complex modeling. We consider files $\mathbf{A}$ and $\mathbf{B}$, comprising $n_A=n_B=600$ records, with complete overlap. For record $i$ in both files, we generate a random vector from a mixture of four-dimensional Normal distributions. For mixture component $c \in \{1,\dots,4\}$, let $(Y_{A1i},Y_{A2i},X_{B1i},X_{B2i})^{'(c)}  \sim N_4(c\mathbf{u}_c, c\mathbf{\Sigma}_c)$, where $u_c$ is a unit vector with 1 at the $c^{th}$ position, and 0 elsewhere, and $\mathbf{\Sigma}_c = \bigg( \begin{smallmatrix}0.5 & 0.1 &0.1&0.1 \\ 0.1 & 0.5 &0.1&0.1\\ 0.1& 0.1 &0.5&0.1 \\ 0.1 & 0.1 &0.1&0.5\end{smallmatrix}\bigg)$. The proportions of records assigned to the four clusters are $\{0.4,0.3,0.2,0.1\}$. This design ensures that the clusters are relatively well-separated \citep{harelcluster}. We assign the variables $\mathbf{Y}_{A1}$ and $\mathbf{Y}_{A2}$ to file $\mathbf{A}$, and $\mathbf{X}_{B1}$ and $\mathbf{X}_{B2}$ to file $\mathbf{B}$. The goal of the analysis is to estimate (i) the population mean vector; and (ii) the correlation between $\mathbf{Y}_{A1}$ and $\mathbf{X}_{B1}$ within each cluster, after the files have been linked.

We let both files comprise three linking variables drawn from  a discrete uniform distribution over 5, 10, and 20 categories, respectively. We link the two files using the SL algorithm, using binary agreement  
to construct comparison vectors. Within each posterior sample of the linkage structure, we fit a four-component Gaussian mixture model via the EM algorithm implemented in the \texttt{mclust} package \citep{mclust} in \texttt{R}. We estimate the cluster means and correlations within each posterior sample of the linkage structure, and combine these estimates using the MI combining rules from Section \ref{sec:primaryMI}. Table \ref{tab:clus} depicts the results across 1000 posterior samples. The imputation procedure recovers the true cluster means, correlations,
and mixing proportions with reasonable accuracy. \par

In this example, the mixture components can be recovered well, as all components are well-separated. In more complex scenarios, label-invariance of the mixture model may require restrictions on the ordering of the cluster means. This issue is also relevant to other multivariate techniques like factor analysis \citep{nassiri}, where factor loadings may change order across posterior samples of the linkage structure.  With added computation, fully Bayesian estimation can also be performed using a version of the DA algorithm. As the weighting methods are restricted to generalized linear models, they cannot be directly applied to this example. 

\begin{table*}[t]
\centering

\caption[]{Cluster means, correlations, and mixing proportions across 1000 posterior samples of the linkage structure. The linkage structure is estimated using the SL algorithm.\\}  \label{tab:clus}
\begin{tabular}{ccccccc}
Cluster         & \makecell{Estimated \\mean vector} & \makecell{True \\mean vector} & \makecell{Estimated \\correlation} & \makecell{True \\correlation} & \makecell{Estimated mixing \\proportion} & \makecell{True mixing \\proportion}\\ \hline
1       & (1.191, 0.046, -0.021, -0.059)   &(1, 0, 0, 0)& 0.195  & 0.2  & 0.375  & 0.4\\
2       & (-0.019, 2.020,  0.031, 0.020)   &(0, 2, 0, 0)& 0.195  & 0.2 & 0.295  & 0.3   \\
3       & (0.021, -0.026, 2.930, -0.024)   &(0, 0, 3, 0)& 0.195  &0.2  & 0.213  &0.2 \\
4      & (0.058, -0.013, 0.068, 3.736)   &(0, 0, 0, 4)  & 0.195  &0.2  & 0.112  &0.1 \\ \hline

\end{tabular}

\end{table*}

\subsection{Scenario 2: Secondary analysis}
\subsubsection{Simulation design} 
In this scenario, we generate a linked file of size $n_{AB}=600$. For record $l$ in the linked file, we generate $(Y_{Al}, X_{Bl})$ based on Equation \eqref{eq:datagen}. We examine two levels of $\rho$: $\rho=0.9$ (high SNR) and $\rho=0.4$ (low SNR). \par

The $n_{AB}$ linked records are partitioned into three blocks of sizes $n_{1}=100$, $n_{2}=200$, and $n_{3}=300$. Following \citet{Chambers2009}, we constrain the linkage errors to be within blocks. In the low linkage error scenario, we let the error rates within the three blocks to be $\{5\%, 10\%, 15\%\}$. In the high linkage error scenario, we let the error rates to be $\{15\%, 20\%, 25\%\}$. We consider four mechanisms for generating the linkage errors: (i) the LCAR/ELE mechanism; (ii) the NL mechanism; (iii) the WNL mechanism, and (iv) the IL mechanism, where the linkage errors depend on an unobserved variable associated with $\mathbf{Y}_{A} \text{ and } \mathbf{X}_{B}$. Under each mechanism, we generate errors by shuffling the values of the outcome $\mathbf{Y}_{A}$. We provide a detailed description of the data generation process in the supplementary materials. \par
In summary, we generate $2^4=16$ simulation configurations, which are implemented as a full factorial design. In each configuration, we generate 100 simulation replications. 

\subsubsection{Inference methods} 
We compare five methods for inference on $\beta$, the slope of the regression of $Y_{A}$ on $X_{B}$. 
We first consider the naive estimator of $\beta$, that does not account for possible linkage errors. The second inferential method is the weighted estimating equations approach of \citet{Chambers2009}. We consider three weighting estimators:  the bias-adjusted ratio type estimator (ChR), the Lahiri-Larsen estimator under the ELE assumption (ChL), and the BLUE (ChB). These estimators require an estimate of the probability of correct linkage per block, $\hat\lambda_b, b =1,2,3$. To obtain $\hat\lambda_b$, we use a random audit sample of size $n_b/2$ in block $b$, and record the proportion of correct links $n_{bc}$. We set $\hat \lambda_b = \min\{n_b^{-1}(n_b-1),\max\{n_b^{-1},n_{bc}\}\}$.
This method avoids overly optimistic estimates of $\lambda_b$ when it is close to one \citep{Chambers2009}. 
We compute standard errors for each estimator using the Sandwich variance formulae in \citet[][pp.13-17]{Chambers2009}. \par

The third inferential method is that of \citet{gsamgreen16} (abbreviated as GT). We fit a mixture of linear regressions independently within each block. We assume diffuse proper prior distributions on the regression model parameters and the mixture class probabilities. We generate $1000$ samples from the posterior distribution of the mixture model parameters, using the DA algorithm. In each posterior sample, we estimate $\beta$, and its sampling variance among record pairs classified as links. We compute point and interval estimates using MI (Section \ref{sec:primaryMI}). \par

We also examine the likelihood-based method  of \citet{slawski2021} (abbreviated as SLW). Under this method, we fit a two-class mixture model using the EM algorithm, and obtain an estimate $\hat\beta$ as described in \citet[][pp. 996-997]{slawski2021}. We compute the covariance matrix as described in Appendix C of \citet{slawski2021}. We obtain confidence intervals for $\beta$ under an asymptotic normal approximation for the sampling distribution of $\hat\beta$.

We summarize results across all simulation replications in terms of the mean bias, the average estimated standard error (SE), and the coverage rate of the 95\% interval estimate.

\subsubsection{Results} 
Table \ref{tab:sec} depicts the results when the linkage error levels are low, and the SNR is high. The naive estimator underestimates $\beta$ across all the  linkage mechanisms, and has below-nominal coverage. All the other estimation methods reduce the linkage error bias. Under the ELE mechanism, all methods (except the naive) yield estimates with a small bias and above-nominal coverage, with GT and SLW having the smallest SEs. Under the NL mechanism, all methods yield coverage rates close to 95\%, with SLW being the most efficient. 

The bias under all methods increases when the linkage mechanism is WNL. The increased bias leads to interval estimates with coverages close to 70\% for ChR, ChL, and 50\% for ChB. GT and SLW have smaller biases and SEs, with coverage rates of 77\% and 73\%, respectively. The decrease in performance is more apparent under IL, where ChR, ChL, and ChB have biases that are larger than the naive estimator. The biases and SEs under GT and SLW are lower than the weighting methods, but coverage is below the nominal.

Table S9 in the supplementary materials displays the results when the linkage error level and the SNR are low. Across all the linkage mechanisms, the naive estimator continues to show large negative biases, with coverage rates below the nominal. Under the ELE mechanism,  the weighting methods have low biases with coverage rates near 95\%. Biases under the weighting methods increase under the NL, WNL, and IL mechanisms, with coverage rates near 95\% under NL, and near 85\% under WNL and IL. Across all linkage mechanisms, the performance of the weighting methods is better when the SNR is low compared to when it is high. Evidently, the linkage mechanism has a less noticeable impact on the inferences when the association between $Y_A \text{ and } X_B$ is low. Under all linkage mechanisms, the bias under GT and SLW is higher compared to the high SNR scenario. This is because both models rely on the association between $Y_A \text{ and } X_B$, which is attenuated in this scenario. Nonetheless, the coverage under GT and SLW is close to nominal.

\begin{table}[t]
\caption[]{Bias, estimated SE, and coverage probabilities for estimates of $\beta$ in the secondary analysis scenario, when linkage error levels are low, and the SNR is high.\\} \label{tab:sec}
\centering
\begin{tabular}{llccc}
\hline
Linkage Mechanism         & Method & Bias & SE & Coverage \\ \hline
                          & Naive  &-0.208       &0.050    & 0.00         \\
\multirow{5}{*}{ELE (LCAR)} & ChR  & 0.004       &0.077    & 0.97        \\
                          & ChL    & 0.003       &0.076    & 0.97         \\
                          & ChB    & -0.001       &0.070    & 0.99         \\
                          & GT     & -0.011       &0.047    & 0.97         \\ 
                          & SLW    & 0.000      &0.039    & 0.95        \\ \hline
                          & Naive  & -0.174       &0.048    & 0.06         \\
\multirow{5}{*}{NL}       & ChR    & 0.035       &0.076    & 0.96         \\
                          & ChL    & 0.033       &0.074    & 0.95         \\
                          & ChB    & 0.051       &0.067    & 0.91        \\
                          & GT     & -0.020       &0.048    & 0.92          \\ 
                          & SLW    & -0.007       &0.040    & 0.91         \\ \hline
                          & Naive  & -0.115       &0.044    & 0.22         \\
\multirow{5}{*}{WNL}      & ChR    & 0.107       &0.075    & 0.73         \\
                          & ChL    &0.104        &0.074    & 0.74        \\
                          & ChB    &0.130       &0.065    & 0.50       \\
                          & GT     & -0.076       &0.051    & 0.77       \\ 
                          & SLW    & -0.047       &0.041    & 0.73         \\ \hline
                          & Naive  & -0.098      &0.038    &0.47      \\
\multirow{5}{*}{IL}       & ChR    & 0.195       &0.076    & 0.11         \\
                          & ChL    & 0.191       &0.074    & 0.13       \\
                          & ChB    & 0.183       &0.063    & 0.13        \\
                          & GT     & -0.078      &0.061   & 0.80    \\ 
                          & SLW    & -0.049      &0.041    & 0.71    \\ \hline
\end{tabular}
\end{table}

Table S10 in the supplementary materials depicts the results when the error level is high and the SNR is high. Due to the higher error level, biases under all the methods increase, especially under the WNL and IL mechanisms. The weighting methods show the highest increase in bias, with coverage near 57\% under WNL, and 3\% under IL. SLW has the smallest bias across all the linkage mechanisms. GT and SLW have coverages near the nominal under the ELE and NL mechanisms, but close to 50\% when the linkage mechanism is WNL or IL. For the same error level, the relative performance of the methods is qualitatively similar when the SNR is low (Table S11).

\section{Concluding remarks} \label{sec:discuss}

In this article, we review inferential methods with linked data files. Following the missing data literature, we group these methods into three categories: likelihood and Bayesian methods, imputation methods, and weighting methods. We outline the assumptions underlying the methods, and discuss their possible benefits and shortcomings. \par
Using simulations, we evaluate the performance of selected methods for primary and secondary analysis of linked files. In the primary analysis setting, our simulations identify two factors that significantly affect performance: the amount of overlap between the files, and the linkage mechanism. With high overlap, and under LCAR, SNL, or NL, all methods yield generally valid inferences. Under these linkage mechanisms, the model of \citet{steorts16} holds up well with decreasing levels of overlap. Under WNL, the joint modeling approaches from Section \ref{sec:wnl} offer some advantages over the other methods.  When the linkage mechanism is IL, there is a noticeable decline in performance across most of the methods. The linkage mechanism also has a considerable impact on inferences in the secondary analysis setting.

We conclude by pointing out some important and open avenues for research. First, the area of post-linkage analysis has remained generally underdeveloped beyond generalized linear models. In Section \ref{sec:multivar}, we provide an illustrative example of applying MI to cluster analysis with linked files. However, other multivariate analysis techniques call for more sophisticated methodological developments. Second, as indicated in the simulations, linkage mechanisms can heavily influence post-linkage inferences. These mechanisms underpin every record linkage algorithm, and are untestable in practice. Thus, it is crucial to assess the sensitivity of inferences to alternative linkage mechanisms than those assumed in the linkage process. As there are parallels between the linkage and missing data mechanisms, sensitivity analysis methods for record linkage can borrow from the rich missing data literature. Finally, as demonstrated in the article, MI offers an appealing alternative to fully Bayesian inference with linked data. However, the validity of MI inferences under more elaborate scenarios, for example, with multiple analysis models and uncongeniality, are unstudied.

\begin{funding}
This work was supported by the National Institute on Aging (R21AG059120) and the Patient-Centered Outcomes Research Institute (ME-2017C3-10241). All statements in this report, including its findings and conclusions, are solely those of the authors and do not necessarily represent the views of the PCORI, its Board of Governors, or the Methodology Committee.
\end{funding}

\begin{acks}
We thank the Editor, Associate Editor, and two referees for their thoughtful comments that improved the quality of the article.
\end{acks}

%%%%%%%%%%%%%%%%%%%%%%%%%%%%%%%%%%%%%%%%%%%%%%
%% Supplementary Material, including data   %%
%% sets and code, should be provided in     %%
%% {supplement} environment with title      %%
%% and short description. It cannot be      %%
%% available exclusively as external link.  %%
%% All Supplementary Material must be       %%
%% available to the reader on Project       %%
%% Euclid with the published article.       %%
%%%%%%%%%%%%%%%%%%%%%%%%%%%%%%%%%%%%%%%%%%%%%%
\section*{Supplementary Material}
The supplementary material describes the data generation process for the simulation study, presents additional simulation results, and provides derivations referenced in the main text.

\bibliographystyle{apalike}
\bibliography{bibliography}       % Bibliography file (usually '*.bib')

\begin{thebibliography}{}

\bibitem[Abowd et~al., 2021]{abowd2022}
Abowd, J.~M., Abramowitz, J., Levenstein, M.~C., McCue, K., Patki, D., Raghunathan, T.~E., Rodgers, A.~M., Shapiro, M.~D., Wasi, N., and Zinsser, D. (2021).
\newblock {Finding needles in haystacks: Multiple-imputation record linkage using machine learning.}
\newblock Working Paper 21-35, Center for Economic Studies, U.S. Census Bureau.

\bibitem[Aleshin-Guendel and Sadinle, 2022]{aleshing}
Aleshin-Guendel, S. and Sadinle, M. (2022).
\newblock Multifile partitioning for record linkage and duplicate detection.
\newblock {\em Journal of the American Statistical Association}, 0(0):1--10.

\bibitem[Asher et~al., 2020]{asher}
Asher, J., Resnick, D., Brite, J., Brackbill, R., and Cone, J. (2020).
\newblock An introduction to probabilistic record linkage with a focus on linkage processing for wtc registries.
\newblock {\em International Journal of Environmental Research and Public Health}, 17(18):6937.

\bibitem[Baxter and Christen, 2003]{Baxter2003}
Baxter, R. and Christen, P. (2003).
\newblock A comparison of fast blocking methods for record linkage, cmis technical report 03/139.
\newblock In {\em Proceedings of ACM SIGKDD'03 Workshop on Data Cleaning, Record Linkage, and Object Consolidation}, pages 39--48.

\bibitem[Belin and Rubin, 1995]{belinrubin1995}
Belin, T.~R. and Rubin, D.~B. (1995).
\newblock A method for calibrating false-match rates in record linkage.
\newblock {\em Journal of the American Statistical Association}, 90(430):694–707.

\bibitem[Bilenko et~al., 2006]{bilenkoicdm}
Bilenko, M., Kamath, B., and Mooney, R. (2006).
\newblock Adaptive blocking: Learning to scale up record linkage.
\newblock In {\em Proceedings of the Sixth IEEE International Conference on Data Mining}, pages 87--96.

\bibitem[Bilenko and Mooney, 2003]{bilenko2003}
Bilenko, M. and Mooney, R.~J. (2003).
\newblock On evaluation and training-set construction for duplicate detection.
\newblock In {\em Proceedings of the KDD-2003 Workshop on Data Cleaning, Record Linkage, and Object Consolidation}, pages 7--12.

\bibitem[Binette and Steorts, 2022]{binette22}
Binette, O. and Steorts, R.~C. (2022).
\newblock (almost) all of entity resolution.
\newblock {\em Science Advances}, 8.

\bibitem[Bird and King, 2018]{birdking}
Bird, S.~M. and King, R. (2018).
\newblock Multiple systems estimation (or capture-recapture estimation) to inform public policy.
\newblock {\em Annual Review of Statistics and its Application}, 5:95--118.

\bibitem[Bohensky, 2015]{bohenskybook}
Bohensky, M. (2015).
\newblock {\em Bias in data linkage studies}, chapter~4, pages 63--82.
\newblock John Wiley and Sons, Ltd.

\bibitem[Brenner et~al., 1997]{brenner1997}
Brenner, H., Schmidtmann, I., and Stegmaier, C. (1997).
\newblock Effects of record linkage errors on registry-based follow-up studies.
\newblock {\em Statistics in Medicine}, 16(23):2633--2643.

\bibitem[Briscolini et~al., 2018]{briscolini}
Briscolini, D., Di~Consiglio, L., Liseo, B., Tancredi, A., and Tuoto, T. (2018).
\newblock New methods for small area estimation with linkage uncertainty.
\newblock {\em International Journal of Approximate Reasoning}, 94:30--42.

\bibitem[Campbell et~al., 2008]{Campbell2008}
Campbell, K., Deck, D., and Krupski, A. (2008).
\newblock Record linkage software in the public domain: A comparison of link plus, the link king, and a `basic' deterministic algorithm.
\newblock {\em Health Informatics Journal}, 14:5--15.

\bibitem[Campbell et~al., 2021]{mirel}
Campbell, S.~R., Resnick, D.~M., Cox, C.~S., and Mirel, L.~B. (2021).
\newblock Using supervised machine learning to identify efficient blocking schemes for record linkage.
\newblock {\em Statistical Journal of the IAOS}, 37(2):673--680.

\bibitem[Cangul et~al., 2009]{cangul2009}
Cangul, M.~Z., Chretien, Y.~R., Gutman, R., and Rubin, D.~B. (2009).
\newblock Testing treatment effects in unconfounded studies under model misspecification: Logistic regression, discretization, and their combination.
\newblock {\em Statistics in Medicine}, 28(20):2531--2551.

\bibitem[Chambers, 2009]{Chambers2009}
Chambers, R. (2009).
\newblock Regression analysis of probability-linked data.
\newblock {\em Statisphere Official Statistics}, 4.

\bibitem[Chambers and {Diniz da Silva}, 2020]{chambersdasilva2020}
Chambers, R. and {Diniz da Silva}, A. (2020).
\newblock {Improved secondary analysis of linked data: A framework and an illustration}.
\newblock {\em Journal of the Royal Statistical Society, Series A}, 183:37--59.

\bibitem[Chambers et~al., 2019]{chambersinf}
Chambers, R., Salvati, N., Fabrizi, E., and {Diniz da Silva}, A. (2019).
\newblock Domain estimation under informative linkage.
\newblock {\em Statistical Theory and Related Fields}, 3(2):90--102.

\bibitem[Chambers et~al., 2022]{chambers2022}
Chambers, R.~L., Fabrizi, E., Ranalli, M.~G., Salvati, N., and Wang, S. (2022).
\newblock Robust regression using probabilistically linked data.
\newblock {\em WIREs Computational Statistics}, page e1596.

\bibitem[Chipperfield, 2019]{chipperfield2019}
Chipperfield, J. (2019).
\newblock A weighting approach to making inference with probabilistically linked data.
\newblock {\em Statistica Neerlandica}, 73(3):333--350.

\bibitem[Chipperfield and Chambers, 2015]{chipperfield2015}
Chipperfield, J.~O. and Chambers, R.~L. (2015).
\newblock Using the bootstrap to account for linkage errors when analysing probabilistically linked categorical data.
\newblock {\em Journal of Official Statistics}, 31(3):397.

\bibitem[Christen, 2007]{christen07}
Christen, P. (2007).
\newblock A two-step classification approach to unsupervised record linkage.
\newblock In {\em Proceedings of the Sixth Australasian Conference on Data Mining and Analytics}, volume~70, page 111–119.

\bibitem[Christen, 2008a]{christen08}
Christen, P. (2008a).
\newblock Automatic record linkage using seeded nearest neighbour and support vector machine classification.
\newblock In {\em Proceedings of the 14th ACM SIGKDD International Conference on Knowledge Discovery and Data Mining}, pages 151--159.

\bibitem[Christen, 2008b]{christen082}
Christen, P. (2008b).
\newblock Automatic training example selection for scalable unsupervised record linkage.
\newblock In Washio, T., Suzuki, E., Ting, K.~M., and Inokuchi, A., editors, {\em Advances in Knowledge Discovery and Data Mining: 12th Pacific-Asia Conference, PAKDD}, pages 511--518.

\bibitem[Christen and Goiser, 2005]{christen2005}
Christen, P. and Goiser, K. (2005).
\newblock Assessing deduplication and data linkage quality: What to measure?
\newblock In {\em Proceedings of the Fourth Australasian Data Mining Conference}.

\bibitem[Christen and Goiser, 2007]{christengoisier2007}
Christen, P. and Goiser, K. (2007).
\newblock Quality and complexity measures for data linkage and deduplication.
\newblock In {\em Quality Measures in Data Mining}, pages 127--151. Springer.

\bibitem[Christen et~al., 2020]{christen20}
Christen, P., Ranbaduge, T., and Schnell, R. (2020).
\newblock {\em Linking Sensitive Data: Methods and Techniques for Practical Privacy-Preserving Information Sharing}.
\newblock Springer, Cham.

\bibitem[Cochinwala et~al., 2001]{cochinwala2001}
Cochinwala, M., Kurien, V., Lalk, G., and Shasha, D. (2001).
\newblock Efficient data reconciliation.
\newblock {\em Information Sciences}, 137(1):1--15.

\bibitem[Collins et~al., 2001]{collins2001}
Collins, L.~M., Schafer, J.~L., and Kam, C.~M. (2001).
\newblock {A comparison of inclusive and restrictive strategies in modern missing data procedures}.
\newblock {\em Psychological Methods}, 6:330--351.

\bibitem[Conti et~al., 2014]{conti2014}
Conti, G., Fr{\"u}hwirth-Schnatter, S., Heckman, J.~J., and Piatek, R. (2014).
\newblock Bayesian exploratory factor analysis.
\newblock {\em Journal of Econometrics}, 183(1):31--57.

\bibitem[Cook et~al., 2001]{dean01}
Cook, L.~J., Olson, L.~M., and Dean, J.~M. (2001).
\newblock {Probabilistic Record Linkage: Relationships between File Sizes, Identifiers, and Match Weights}.
\newblock {\em Methods of Information in Medicine}, 40:196--203.

\bibitem[Copas and Hilton, 1990]{copashilton}
Copas, J.~B. and Hilton, F.~J. (1990).
\newblock Record linkage: Statistical models for matching computer records.
\newblock {\em Journal of the Royal Statistical Society. Series A (Statistics in Society)}, 153(3):287--320.

\bibitem[Daggy et~al., 2014]{daggy2014evaluating}
Daggy, J., Xu, H., Hui, S., and Grannis, S. (2014).
\newblock Evaluating latent class models with conditional dependence in record linkage.
\newblock {\em Statistics in medicine}, 33(24):4250--4265.

\bibitem[Dalzell and Reiter, 2018]{dalzell2018}
Dalzell, N.~M. and Reiter, J.~P. (2018).
\newblock Regression modeling and file matching using possibly erroneous matching variables.
\newblock {\em Journal of Computational and Graphical Statistics}, 27(4):728--738.

\bibitem[Dasylva et~al., 2014]{dasylva2014}
Dasylva, A., Titus, R.-C., and Thibault, C. (2014).
\newblock Overcoverage in the 2011 canadian census.
\newblock In {\em Proceedings of Statistics Canada Symposium}.

\bibitem[Dempster et~al., 1977]{dempster1977}
Dempster, A.~P., Laird, N.~M., and Rubin, D.~B. (1977).
\newblock Maximum likelihood from incomplete data via the em algorithm.
\newblock {\em Journal of the Royal Statistical Society. Series B (Methodological)}, 39(1):1--38.

\bibitem[{Di Consiglio} and Tuoto, 2015]{consiglio1}
{Di Consiglio}, L. and Tuoto, T. (2015).
\newblock Coverage evaluation on probabilistically linked data.
\newblock {\em Journal of Official Statistics}, 31:415--429.

\bibitem[{Di Consiglio} and Tuoto, 2018]{consiglio2}
{Di Consiglio}, L. and Tuoto, T. (2018).
\newblock Population size estimation and linkage errors: The multiple lists case.
\newblock {\em Journal of Official Statistics}, 34:889--908.

\bibitem[Doidge and Harron, 2018]{doidge2018}
Doidge, J.~C. and Harron, K. (2018).
\newblock Demystifying probabilistic linkage: Common myths and misconceptions.
\newblock {\em International journal of population data science}, 3:410(1).

\bibitem[D'Orazio et~al., 2006]{dorazio2006}
D'Orazio, M., {Di Zio}, M., and Scanu, M. (2006).
\newblock {\em Statistical Matching: Theory and Practice}.
\newblock Hoboken, NJ: Wiley.

\bibitem[Dusetzina et~al., 2014]{dusetzina2014}
Dusetzina, S.~B., Tyree, S., Meyer, A.-M., Meyer, A., Green, L., and Carpenter, W.~R. (2014).
\newblock Linking data for health services research: A framework and instructional guide, rockville, md: Agency for healthcare research and quality (us).

\bibitem[Elfeky et~al., 2003]{elfeky2003}
Elfeky, M.~G., Verykios, V.~S., Elmagarmid, A.~K., Ghanem, T.~M., and Kuwait, W. A.~R. (2003).
\newblock Record linkage: A machine learning approach, a toolbox, and a digital government web service.
\newblock {\em Purdue e-Pubs. Purdue University, West Lafayette}.

\bibitem[Enamorado et~al., 2019]{imaifastlink}
Enamorado, T., Fifield, B., and Imai, K. (2019).
\newblock Using a probabilistic model to assist merging of large-scale administrative records.
\newblock {\em American Political Science Review,}, 113:353--371.

\bibitem[Fellegi and Sunter, 1969]{Fellegi1969}
Fellegi, I.~P. and Sunter, A.~B. (1969).
\newblock A theory for record linkage.
\newblock {\em Journal of the American Statistical Association}, (64):1183--1210.

\bibitem[Fienberg, 1972]{fienberg72}
Fienberg, S.~E. (1972).
\newblock The multiple recapture census for closed populations and incomplete 2k contingency tables.
\newblock {\em Biometrika}, 59:591--603.

\bibitem[Fienberg and Manrique-Vallier, 2009]{fienbergmanrique-vallier2008}
Fienberg, S.~E. and Manrique-Vallier, D. (2009).
\newblock Integrated methodology for multiple systems estimation and record linkage using a missing data formulation.
\newblock {\em AStA Advances in Statistical Analysis}, 93(1):49–60.

\bibitem[Fisher and Yates, 1963]{fisheryates}
Fisher, R.~A. and Yates, F. (1963).
\newblock {\em Statistical Tables for Biological, Agricultural and Medical Research}.
\newblock Oliver and Boyd: Edinburgh, UK.

\bibitem[Fortini et~al., 2001]{fortini2001}
Fortini, M., Liseo, B., Nuccitelli, A., and Scanu, M. (2001).
\newblock On bayesian record linkage.
\newblock {\em Research in Official Statistics}, 4:185--198.

\bibitem[Freedman, 2006]{huber}
Freedman, D.~A. (2006).
\newblock On the so-called “huber sandwich estimator” and “robust standard errors”.
\newblock {\em The American Statistician}, 60(4):299--302.

\bibitem[Gelman et~al., 2013]{gelman2013}
Gelman, A., Carlin, J.~B., Stern, H.~S., Dunson, D.~B., Vehtari, A., and Rubin, D.~B. (2013).
\newblock {\em Bayesian data analysis}.
\newblock CRC press.

\bibitem[Gilbert et~al., 2017]{gilbert2017}
Gilbert, R., Lafferty, R., Hagger-Johnson, G., Harron, K., Zhang, L.~C., Smith, P., Dibben, C., and Goldstein, H. (2017).
\newblock Guild: Guidance for information about linking data sets.
\newblock {\em Journal of Public Health}, 40:191--198.

\bibitem[Golden and Mirel, 2021]{mirel2021}
Golden, C. and Mirel, L.~B. (2021).
\newblock Enhancement of health surveys with data linkage.
\newblock In Chun, A.~Y., Larsen, M., Durrant, G., and Reiter, J.~P., editors, {\em Administrative Records for Survey Methodology}, pages 105--138. Wiley.

\bibitem[Goldstein et~al., 2009]{goldsteinlatentnorm}
Goldstein, H., Carpenter, J., Kenward, M.~G., and Levin, K.~A. (2009).
\newblock Multilevel models with multivariate mixed response types.
\newblock {\em Statistical Modelling}, 9(3):173--197.

\bibitem[Goldstein and Harron, 2015]{goldsteinharonbook}
Goldstein, H. and Harron, K. (2015).
\newblock Record linkage: A missing data problem.
\newblock In Harron, K., Goldstein, H., and Dibben, C., editors, {\em {Methodological Developments in Data Linkage}}, volume~1, pages 109--124. John Wiley \& Sons.

\bibitem[Goldstein et~al., 2012]{goldsteinetal12}
Goldstein, H., Harron, K., and Wade, A. (2012).
\newblock The analysis of record-linked data using multiple imputation with data value priors.
\newblock {\em Statistics in Medicine}, 31(28):3481--3493.

\bibitem[Gomatam et~al., 2002]{gomatam2002}
Gomatam, S., Carter, R., Ariet, M., and Mitchell, G. (2002).
\newblock An empirical comparison of record linkage procedures.
\newblock {\em Statistics in Medicine}, 21:1485--1496.

\bibitem[Green and Mardia, 2006]{Green2006}
Green, P.~J. and Mardia, K.~V. (2006).
\newblock Bayesian alignment using hierarchical models, with applications in protein bioinformatics.
\newblock {\em Biometrika}, 93(2):235--254.

\bibitem[Gu and Baxter, 2006]{gu2006}
Gu, L. and Baxter, R. (2006).
\newblock Decision models for record linkage.
\newblock In Williams, G. and Simoff, S., editors, {\em Data Mining, Lecture Notes in Computer Science}, pages 146--160. Springer, Berlin, Heidelberg.

\bibitem[Gutman et~al., 2013]{gutman2013}
Gutman, R., Afendulis, C.~C., and Zaslavsky, A.~M. (2013).
\newblock A bayesian procedure for file linking to analyze end-of-life medical costs.
\newblock {\em Journal of the American Statistical Association}, 108(501):34–47.

\bibitem[Gutman et~al., 2016]{gsamgreen16}
Gutman, R., Sammartino, C., Green, T., and Montague, B. (2016).
\newblock Error adjustments for file linking methods using encrypted unique client identifier (euci) with application to recently released prisoners who are hiv+.
\newblock {\em Statistics in Medicine}, 35(1):115--129.

\bibitem[Haas et~al., 1994]{haas1994}
Haas, J.~S., Brandenburg, J.~A., Udvarhelyi, I.~S., and Epstein, A.~M. (1994).
\newblock Creating a comprehensive database to evaluate health coverage for pregnant women: The completeness and validity of a computerized linkage algorithm.
\newblock {\em Medical Care}, 32(10):1053--1057.

\bibitem[Hall and Fienberg, 2012]{hallfienberg}
Hall, R. and Fienberg, S. (2012).
\newblock Valid statistical inference on automatically matched files.
\newblock In Domingo-Ferrer, J. and Muralidhar, K., editors, {\em Proceedings of the International Conference on Privacy in Statistical Databases}, pages 131--142.

\bibitem[Han, 2018]{han2018}
Han, Y. (2018).
\newblock {\em Statistical Inference Using Data From Multiple Files Combined Through Record Linkage.}
\newblock PhD thesis, University of Maryland.

\bibitem[Han and Lahiri, 2019]{hanlahiri2019}
Han, Y. and Lahiri, P. (2019).
\newblock {Statistical analysis with linked data}.
\newblock {\em International Statistical Review}, 87:1013 -- 1038.

\bibitem[Harron et~al., 2017]{harron2017}
Harron, K., Dibben, C., Boyd, J., Hjern, A., Azimaee, M., Barreto, M.~L., and Goldstein, H. (2017).
\newblock Challenges in administrative data linkage for research.
\newblock {\em Big Data \& Society}, 4(2).

\bibitem[Harron et~al., 2014]{harron2014}
Harron, K., Wade, A., Gilbert, R., Muller-Pebody, B., and Goldstein, H. (2014).
\newblock Evaluating bias due to data linkage error in electronic healthcare records.
\newblock {\em BMC Medical Research Methodology}, 14(1):1--10.

\bibitem[Hof et~al., 2017]{Hof2017}
Hof, M. H.~P., Ravelli, A.~C., and Zwinderman, A.~H. (2017).
\newblock A probabilistic record linkage model for survival data.
\newblock {\em Journal of the American Statistical Association}, 112(520):1504--1515.

\bibitem[Hof and Zwinderman, 2012]{Hof2012}
Hof, M. H.~P. and Zwinderman, A.~H. (2012).
\newblock Methods for analyzing data from probabilistic linkage strategies based on partially identifying variables.
\newblock {\em Statistics in Medicine}, 31(30):4231--4242.

\bibitem[Hof and Zwinderman, 2014]{hofzwinderman2015}
Hof, M. H.~P. and Zwinderman, A.~H. (2014).
\newblock A mixture model for the analysis of data derived from record linkage.
\newblock {\em Statistics in Medicine}, 34(1):74–92.

\bibitem[Isaki and Schultz, 1987]{isaki}
Isaki, C. and Schultz, L. (1987).
\newblock The effects of correlation and matching error on dual system estimation.
\newblock {\em Communications in Statistics - Theory and Methods}, 16:2405--2427.

\bibitem[Jaro, 1989]{Jaro1989}
Jaro, M. (1989).
\newblock Advances in record-linkage methodology as applied to matching the 1985 census of tampa, florida.
\newblock {\em Journal of the American Statistical Association}, (84):414--420.

\bibitem[Jiang et~al., 2002]{jiang2002}
Jiang, J., Lahiri, P., and Wan, S.-M. (2002).
\newblock A unified jackknife theory for empirical best prediction with m-estimation.
\newblock {\em The Annals of Statistics}, 30:1782--1810.

\bibitem[Kamat et~al., 2023]{brlvof}
Kamat, G., Shan, M., and Gutman, R. (2023).
\newblock Bayesian record linkage with variables in one file.
\newblock {\em Statistics in Medicine}, 42:4931--4951.

\bibitem[Kim and Chambers, 2012a]{kimchambers2010}
Kim, G. and Chambers, R. (2012a).
\newblock Regression analysis under incomplete linkage.
\newblock {\em Computational Statistics and Data Analysis}, 56(518):2756--2770.

\bibitem[Kim and Chambers, 2012b]{kimchambers2012}
Kim, G. and Chambers, R. (2012b).
\newblock Regression analysis under probabilistic multi-linkage.
\newblock {\em Statistica Neerlandica}, 66:64--79.

\bibitem[Kim and Chambers, 2015]{kimchambers2015}
Kim, G. and Chambers, R. (2015).
\newblock Unbiased regression estimation under correlated linkage errors.
\newblock {\em Stat}, 4(1):32--45.

\bibitem[Krewski et~al., 2005]{krewski2005}
Krewski, D., Dewanji, A., Wang, Y., Bartlett, S., Zielinski, J., and Mallick, R. (2005).
\newblock The effect of record linkage errors on risk estimates in cohort mortality studies.
\newblock {\em Survey Methodology}, 31:13–21.

\bibitem[Lahiri and Larsen, 2005]{lahirilarsen05}
Lahiri, P. and Larsen, M.~D. (2005).
\newblock Regression analysis with linked data.
\newblock {\em Journal of the American Statistical Association}, 100(469):222--230.

\bibitem[Lariscy, 2011]{lariscy}
Lariscy, J.~T. (2011).
\newblock Differential record linkage by hispanic ethnicity and age in linked mortality studies: Implications for the epidemiologic paradox.
\newblock {\em Journal of Aging and Health}, 23(8):1263--1284.

\bibitem[Larsen, 2002]{larsen2002}
Larsen, M.~D. (2002).
\newblock Comments on hierarchical bayesian record linkage.
\newblock In {\em Proceedings of the Survey Methods Section}, pages 1995--2000. American Statistical Association.

\bibitem[Larsen, 2005]{larsen2005}
Larsen, M.~D. (2005).
\newblock Advances in record linkage theory: Hierarchical bayesian record linkage theory.
\newblock In {\em JSM Proceedings of the Survey Research Methods Section}, pages 3277--3284. American Statistical Association.

\bibitem[Larsen and Rubin, 2001]{larsenrubin2001}
Larsen, M.~D. and Rubin, D.~B. (2001).
\newblock Iterative automated record linkage using mixture models.
\newblock {\em Journal of the American Statistical Association}, 96:32–41.

\bibitem[Lee et~al., 2022]{lee2022}
Lee, D., Zhang, L.-C., and Kim, J.-K. (2022).
\newblock Maximum entropy classification for record linkage.
\newblock {\em Survey Methodology}, 48.

\bibitem[Lee and Harel, 2022]{harelcluster}
Lee, J.~W. and Harel, O. (2022).
\newblock Incomplete clustering analysis via multiple imputation.
\newblock {\em Journal of Applied Statistics}, 0(0):1--18.

\bibitem[Levenshtein, 1966]{levenshtein}
Levenshtein, V.~I. (1966).
\newblock Binary codes capable of correcting deletions, insertions and reversals.
\newblock {\em Soviet Physics Doklady}, 10(3):707--710.

\bibitem[Li et~al., 2022]{li2022}
Li, X., Xu, H., and Grannis, S. (2022).
\newblock The data-adaptive fellegi-sunter model for probabilistic record linkage: Algorithm development and validation for incorporating missing data and field selection.
\newblock {\em Journal of Medical Internet Research}, 24:e33775.

\bibitem[Liseo and Tancredi, 2011]{liseo2011}
Liseo, B. and Tancredi, A. (2011).
\newblock Bayesian estimation of population size via linkage of multivariate normal data sets.
\newblock {\em Journal of official statistics}, 27(3):491--505.

\bibitem[Little and Rubin, 2019]{littlerubin2002}
Little, R. J.~A. and Rubin, D.~B. (2019).
\newblock {\em Statistical Analysis with Missing Data}.
\newblock Hoboken, NJ: John Wiley \& Sons.

\bibitem[Liu et~al., 2003]{liu}
Liu, B., Dai, Y., Li, X., Lee, W., and Yu, P. (2003).
\newblock Building text classifiers using positive and unlabeled examples.
\newblock In {\em Proceedings of Third IEEE International Conference on Data Mining}, pages 179--186.

\bibitem[Madigan et~al., 1996]{madigan1996}
Madigan, D., Raftery, A.~E., Volinsky, C., and Hoeting, J. (1996).
\newblock Bayesian model averaging.
\newblock In {\em Proceedings of the AAAI Workshop on Integrating Multiple Learned Models, Portland, OR}, pages 77--83.

\bibitem[Madigan et~al., 1995]{madiganyork1995}
Madigan, D., York, J., and Allard, D. (1995).
\newblock Bayesian graphical models for discrete data.
\newblock {\em International Statistical Review}, 63:215--232.

\bibitem[Marchant et~al., 2021]{marchant2021}
Marchant, N.~G., Kaplan, A., Elazar, D.~N., Rubinstein, B. I.~P., and Steorts, R.~C. (2021).
\newblock d-blink: Distributed end-to-end bayesian entity resolution.
\newblock {\em Journal of Computational and Graphical Statistics}, 30(2):406–421.

\bibitem[McVeigh et~al., 2020]{mcveigh2020}
McVeigh, B., Spahn, B.~T., and Murray, J.~S. (2020).
\newblock Scaling bayesian probabilistic record linkage with post-hoc blocking: An application to the california great registers.
\newblock {\em arXiv}.

\bibitem[Meng, 1994]{meng1994}
Meng, X.-L. (1994).
\newblock Multiple-imputation inferences with uncongenial sources of input.
\newblock {\em Statistical Science}, 9(4):538--558.

\bibitem[Meyer and Mittag, 2019]{mittag2019}
Meyer, B.~D. and Mittag, N. (2019).
\newblock Using linked survey and administrative data to better measure income: implications for poverty, program effectiveness and holes in the safety net.
\newblock {\em American Economic Journal: Applied Economics}, 11:176–204.

\bibitem[Meyer and Mittag, 2021]{mittag2021}
Meyer, B.~D. and Mittag, N. (2021).
\newblock Combining administrative and survey data to improve income measurement.
\newblock pages 105--138. Wiley.

\bibitem[Michelson and Knoblock, 2006]{michelson}
Michelson, M. and Knoblock, C.~A. (2006).
\newblock Learning blocking schemes for record linkage.
\newblock In {\em Proceedings of the 21st National Conference on Artificial Intelligence - Volume 1}, page 440–445.

\bibitem[Molenberghs et~al., 2014]{molenberghs2014}
Molenberghs, G., Fitzmaurice, G., Kenward, M.~G., Tsiatis, A., and Verbeke, G. (2014).
\newblock {\em Handbook of missing data methodology}.
\newblock CRC Press.

\bibitem[Murray, 2015]{Murray2015}
Murray, J. (2015).
\newblock Probabilistic record linkage and deduplication after indexing, blocking, and filtering.
\newblock {\em Journal of Privacy and Confidentiality}, 7:https://doi.org/10.29012/jpc.v7i1.643.

\bibitem[Murray, 2018]{murray2018}
Murray, J.~S. (2018).
\newblock Multiple imputation: A review of practical and theoretical findings.
\newblock {\em Statistical Science}, 33(2):142--159.

\bibitem[Nassiri et~al., 2018]{nassiri}
Nassiri, V., Lovik, A., Molenberghs, G., and Verbeke, G. (2018).
\newblock On using multiple imputation for exploratory factor analysis of incomplete data.
\newblock {\em Behavioral Research}, 50(0):501--517.

\bibitem[Neter et~al., 1965]{Neter1965}
Neter, J., Maynes, E.~S., and Ramanathan, R. (1965).
\newblock The effect of mismatching on the measurement of response error.
\newblock {\em Journal of the American Statistical Association}, 60(312):1005--1027.

\bibitem[Newcombe et~al., 1959]{newcombe59}
Newcombe, H., Kennedy, J.~M., Axford, S.~J., and James, A.~P. (1959).
\newblock Automatic linkage of vital records.
\newblock {\em Science}, 130:954--959.

\bibitem[Newman et~al., 2009]{newman2009}
Newman, L.~M., Samuel, M.~C., Stenger, M.~R., Gerber, T.~M., Macomber, K., Stover, J.~A., and Wise, W. (2009).
\newblock Practical considerations for matching std and hiv surveillance data with data from other sources.
\newblock {\em Public Health Reports}, pages 2--27.

\bibitem[Patki and Shapiro, 2023]{patki23}
Patki, D. and Shapiro, M.~D. (2023).
\newblock Implicates as instrumental variables: An approach for estimation and inference with probabilistically matched data.
\newblock {\em Journal of Survey Statistics and Methodology}.

\bibitem[Peel and MachLachlan, 2000]{peel2000}
Peel, D. and MachLachlan, G. (2000).
\newblock {\em Finite Mixture Models}.
\newblock John Wiley and Sons.

\bibitem[Reiter, 2021]{reiter2021}
Reiter, J.~P. (2021).
\newblock Assessing uncertainty when using linked administrative records.
\newblock In Chun, A.~Y., Larsen, M., Durrant, G., and Reiter, J.~P., editors, {\em Administrative Records for Survey Methodology}, pages 139--153. Wiley, Hoboken, New Jersey.

\bibitem[Rencher, 2002]{rencher}
Rencher, A.~C. (2002).
\newblock {\em Methods of Multivariate Analysis}.
\newblock John Wiley and Sons.

\bibitem[Robert and Casella, 1999]{robertcasella}
Robert, C.~P. and Casella, G. (1999).
\newblock {\em Monte Carlo Statistical Methods}.
\newblock Springer.

\bibitem[Roos et~al., 1992]{roos1992}
Roos, L.~L., Wajda, A., Nicol, J.~P., and Roberts, J. (1992).
\newblock {Record linkage: An overview.}
\newblock Technical report, Medical Effectiveness Research Data Methods, US Department of Health and Human Services.

\bibitem[Rubin, 1996]{Rubin1996}
Rubin, D. (1996).
\newblock Multiple imputation after 18+ years.
\newblock {\em Journal of the American Statistical Association}, 91:473–489.

\bibitem[Rubin, 1976]{rubin1976}
Rubin, D.~B. (1976).
\newblock Inference and missing data (with discussion).
\newblock {\em Biometrika}, 63:581--592.

\bibitem[Rubin, 1987]{rubin1987}
Rubin, D.~B. (1987).
\newblock {\em Multiple Imputation for Nonresponse in Surveys}.
\newblock New York, NY: John Wiley \& Sons.

\bibitem[Sadinle, 2014]{sadinle2014}
Sadinle, M. (2014).
\newblock Detecting duplicates in a homicide registry using a bayesian partitioning approach.
\newblock {\em The Annals of Applied Statistics}, 8:2404--2434.

\bibitem[Sadinle, 2017]{Sadinle2017}
Sadinle, M. (2017).
\newblock Bayesian estimation of bipartite matchings for record linkage.
\newblock {\em Journal of the American Statistical Association}, 112:600--612.

\bibitem[Sadinle, 2018]{sadinle2018}
Sadinle, M. (2018).
\newblock Bayesian propagation of record linkage uncertainty into population size estimation of human rights violations.
\newblock {\em The Annals of Applied Statistics}, 12(2):1013--1038.

\bibitem[Sadinle, 2020]{brl}
Sadinle, M. (2020).
\newblock {\em BRL: Beta Record Linkage}.
\newblock R package version 0.1.0.

\bibitem[Sadinle and Fienberg, 2013]{sadinlefienberg2013}
Sadinle, M. and Fienberg, S.~E. (2013).
\newblock A generalized fellegi–sunter framework for multiple record linkage with application to homicide record systems.
\newblock {\em Journal of the American Statistical Association}, 108(502):385–397.

\bibitem[Schafer, 1997]{schafer1997}
Schafer, J.~L. (1997).
\newblock {\em Analysis of incomplete multivariate data}.
\newblock CRC press.

\bibitem[Scheuren and Winkler, 1993]{scheurenwinkler93}
Scheuren, F. and Winkler, W.~E. (1993).
\newblock Regression analysis of data files that are computer matched - part i.
\newblock {\em Survey Methodology}, 19:106--125.

\bibitem[Scheuren and Winkler, 1997]{scheurenwinkler97}
Scheuren, F. and Winkler, W.~E. (1997).
\newblock Regression analysis of data files that are computer matched - part ii.
\newblock {\em Survey Methodology}, 19:106--125.

\bibitem[Schürle, 2005]{schurle}
Schürle, J. (2005).
\newblock A method for consideration of conditional dependencies in the fellegi and sunter model of record linkage.
\newblock {\em Statistical Papers}, 46(3):433--449.

\bibitem[Scrucca et~al., 2016]{mclust}
Scrucca, L., Fop, M., Murphy, T.~B., and Raftery, A.~E. (2016).
\newblock {mclust} 5: clustering, classification and density estimation using {G}aussian finite mixture models.
\newblock {\em The {R} Journal}, 8(1):289--317.

\bibitem[Seaman and White, 2011]{seamanwhite2011}
Seaman, S.~R. and White, I.~R. (2011).
\newblock Review of inverse probability weighting for dealing with missing data.
\newblock {\em Statistical Methods in Medical Research}, 22(3):278–295.

\bibitem[Shan et~al., 2022]{shan2022}
Shan, M., Thomas, K., and Gutman, R. (2022).
\newblock A bayesian multilayer record linkage procedure to analyze postacute care recovery of patients with traumatic brain injury.
\newblock {\em Biostatistics}.

\bibitem[Shan et~al., 2021]{shan2021}
Shan, M., Thomas, K.~S., and Gutman, R. (2021).
\newblock {A multiple imputation procedure for record linkage and causal inference to estimate the effects of home-delivered meals}.
\newblock {\em The Annals of Applied Statistics}, 15(1):412 -- 436.

\bibitem[Shlomo, 2019]{shlomo19}
Shlomo, N. (2019).
\newblock Overview of data linkage methods for policy design and evaluation.
\newblock In Crato, N. and Paruolo, P., editors, {\em {Data-Driven Policy Impact Evaluation: How Access to Microdata is Transforming Policy Design}}, pages 47--65. Springer.

\bibitem[Slawski et~al., 2021]{slawski2021}
Slawski, M., Diao, G., and Ben-David, E. (2021).
\newblock A pseudo-likelihood approach to linear regression with partially shuffled data.
\newblock {\em Journal of Computational and Graphical Statistics}, 30(4):991--1003.

\bibitem[Smith and Roberts, 1993]{smithroberts1993}
Smith, A.~F. and Roberts, G.~O. (1993).
\newblock Bayesian computation via the gibbs sampler and related markov chain monte carlo methods.
\newblock {\em Journal of the Royal Statistical Society: Series B (Methodological)}, 55:3--23.

\bibitem[Steorts, 2020]{blink}
Steorts, R. (2020).
\newblock {\em blink: Record Linkage for Empirically Motivated Priors}.
\newblock R package version 1.1.0.

\bibitem[Steorts et~al., 2014a]{Steorts2014}
Steorts, R., Ventura, S., Sadinle, M., and Fienberg, S. (2014a).
\newblock A comparison of blocking methods for record linkage.
\newblock In Domingo-Ferrer, J. and Muralidhar, K., editors, {\em Proceedings of the International Conference on Privacy in Statistical Databases}, pages 253--268.

\bibitem[Steorts, 2015]{steorts16}
Steorts, R.~C. (2015).
\newblock {Entity resolution with empirically motivated priors}.
\newblock {\em Bayesian Analysis}, 10(4):849 -- 875.

\bibitem[Steorts et~al., 2014b]{smered}
Steorts, R.~C., Hall, R., and Fienberg, S. (2014b).
\newblock Smered: A bayesian approach to graphical record linkage and de-duplication.
\newblock In {\em Artificial Intelligence and Statistics}, pages 922--930.

\bibitem[Steorts et~al., 2016]{Steorts2015}
Steorts, R.~C., Hall, R., and Fienberg, S.~E. (2016).
\newblock A bayesian approach to graphical record linkage and deduplication.
\newblock {\em Journal of the American Statistical Association}, 111:1660--1672.

\bibitem[Steorts et~al., 2018]{steorts2018}
Steorts, R.~C., Tancredi, A., and Liseo, B. (2018).
\newblock Generalized bayesian record linkage and regression with exact error propagation.
\newblock In Domingo-Ferrer, J. and Muralidhar, K., editors, {\em Proceedings of the International Conference on Privacy in Statistical Databases}, pages 297--313.

\bibitem[Stern, 1986]{stern1986}
Stern, R.~S. (1986).
\newblock {Record linkage: A powerful tool for epidemiologic analysis}.
\newblock {\em Archives of Dermatology}, 122:1383--1384.

\bibitem[Swartz et~al., 2004]{schwartz}
Swartz, T., Haitovsky, Y., Vexler, A., and Yang, T. (2004).
\newblock Bayesian identifiability and misclassification in multinomial data.
\newblock {\em The Canadian Journal of Statistics / La Revue Canadienne de Statistique}, 32(3):285--302.

\bibitem[Tancredi and Liseo, 2011]{tancrediliseo2011}
Tancredi, A. and Liseo, B. (2011).
\newblock A hierarchical bayesian approach to record linkage and population size problems.
\newblock {\em The Annals of Applied Statistics}, 5(2B):1553--1585.

\bibitem[Tancredi and Liseo, 2015]{tancrediliseo2015}
Tancredi, A. and Liseo, B. (2015).
\newblock Regression analysis with linked data: Problems and possible solutions.
\newblock {\em Statistica}, 75(1):19--35.

\bibitem[Tancredi et~al., 2020]{tancredi2020}
Tancredi, A., Steorts, R.~C., and Liseo, B. (2020).
\newblock {A unified framework for de-duplication and population size estimation (with Discussion)}.
\newblock {\em Bayesian Analysis}, 15:633 -- 682.

\bibitem[Tang et~al., 2020]{tang2020}
Tang, J., Reiter, J., and Steorts, R. (2020).
\newblock Bayesian modeling for simultaneous regression and record linkage.
\newblock In Domingo-Ferrer, J. and Muralidhar, K., editors, {\em Proceedings of the International Conference on Privacy in Statistical Databases}, pages 209--223.

\bibitem[Tanner and Wong, 1987]{tannerwong1987}
Tanner, M. and Wong, W. (1987).
\newblock The calculation of posterior distributions by data augmentation.
\newblock {\em Journal of the American Statistical Association}, 82:528--540.

\bibitem[Taylor et~al., 2024]{taylor24}
Taylor, I., Kaplan, A., and Betancourt, B. (2024).
\newblock Fast {Bayesian} record linkage for streaming data contexts.
\newblock {\em Journal of Computational and Graphical Statistics}, 0(0):1--12.

\bibitem[Tepping, 1968]{tepping68}
Tepping, B.~J. (1968).
\newblock A model for optimum linkage of records.
\newblock {\em Journal of the American Statistical Association}, 63:1321--1332.

\bibitem[Tromp et~al., 2011]{tromp11}
Tromp, M., Ravelli, A.~C., Bonsel, G., Hasman, A., and Reitsma, J.~B. (2011).
\newblock {Results from simulated data sets: Probabilistic record linkage outperforms deterministic record linkage}.
\newblock {\em Journal of Clinical Epidemiology}, 64:564--572.

\bibitem[Tromp et~al., 2006]{tromp2006}
Tromp, M., Reitsma, J.~B., Ravelli, A.~C., Méray, N., and Bonsel, G.~J. (2006).
\newblock {Record linkage: Making the most out of errors in linking variables}.
\newblock In {\em Proceedings of the AMIA Annual Symposium}, page 779–783.

\bibitem[Tsiatis, 2006]{tsiatis2006}
Tsiatis, A.~A. (2006).
\newblock {\em Semiparametric Theory and Missing Data}.
\newblock Springer.

\bibitem[van Buuren, 2018]{van2018}
van Buuren, S. (2018).
\newblock {\em Flexible Imputation of Missing Data}.
\newblock CRC press.

\bibitem[Vo et~al., 2023]{vo2022}
Vo, T.~H., Chauvet, G., Happe, A., Oger, E., Paquelet, S., and Gar{\`{e}}s, V. (2023).
\newblock Extending the fellegi-sunter record linkage model for mixed-type data with application to the french national health data system.
\newblock {\em Computational Statistics {\&} Data Analysis}, 179:107656.

\bibitem[Vo et~al., 2024]{Vo23surv}
Vo, T.~H., Garès, V., Zhang, L.-C., Happe, A., Oger, E., Paquelet, S., and Chauvet, G. (2024).
\newblock Cox regression with linked data.
\newblock {\em Statistics in Medicine}, 43(2):296--314.

\bibitem[Wang et~al., 2022]{wang2022}
Wang, Z., Ben-David, E., Diao, G., and Slawski, M. (2022).
\newblock Regression with linked datasets subject to linkage error.
\newblock {\em WIREs Computational Statistics}, 14:e1570.

\bibitem[Winglee et~al., 2005]{winglee2005}
Winglee, M., Valliant, R., and Scheuren, F. (2005).
\newblock A case study in record linkage.
\newblock {\em Survey Methodology}, 31:3--11.

\bibitem[Winkler, 1990]{Winkler1990}
Winkler, W. (1990).
\newblock String comparator metrics and enhanced decision rules in the fellegi-sunter model of record linkage.
\newblock In {\em Proceedings of the Section on Survey Research Methods, American Statistical Association}, page 354–359.

\bibitem[Winkler, 1993]{Winkler1993}
Winkler, W. (1993).
\newblock Improved decision rules in the fellegi–sunter model of record linkage.
\newblock In {\em Proceedings of the Section on Survey Research Methods, American Statistical Association}, pages 274--279.

\bibitem[Winkler, 2000]{winkler2000}
Winkler, W. (2000).
\newblock Frequency-based matching in fellegi-sunter model of record linkage.
\newblock In {\em Proceedings of the Section on Survey Research Methods, American Statistical Association}, pages 778--783.

\bibitem[Winkler, 2014]{Winkler1995}
Winkler, W. (2014).
\newblock Matching and record linkage.
\newblock {\em WIREs Computational Statistics}, 6:313--325.

\bibitem[Winkler, 2021]{winkler2021}
Winkler, W.~E. (2021).
\newblock Cleaning and using administrative lists: Enhanced practices and computational algorithms for record linkage and modeling/editing/imputation.
\newblock In Chun, A.~Y., Larsen, M., Durrant, G., and Reiter, J.~P., editors, {\em Administrative Records for Survey Methodology}, pages 271--296. Wiley.

\bibitem[Winkler and Thibaudeau, 1991]{winkler1991}
Winkler, W.~E. and Thibaudeau, Y. (1991).
\newblock {An application of the Fellegi-Sunter model of record linkage to the 1990 US decennial census.}
\newblock Technical report, US Bureau of the Census.

\bibitem[Wortman and Reiter, 2018]{Wortman2018}
Wortman, J.~H. and Reiter, J.~P. (2018).
\newblock Simultaneous record linkage and causal inference with propensity score subclassification.
\newblock {\em Statistics in Medicine}, 37(24):3533--3546.

\bibitem[Xie and Meng, 2017]{xiemeng2017}
Xie, X. and Meng, X.-L. (2017).
\newblock Dissecting multiple imputation from a multi-phase inference perspective: What happens when god’s, imputer’s and analyst’s models are uncongenial?
\newblock {\em Statistica Sinica}, 27:1485--1594.

\bibitem[Xu et~al., 2021]{xu2021}
Xu, H., Li, X., and Grannis, S. (2021).
\newblock A simple two-step procedure using the fellegi{\textendash}sunter model for frequency-based record linkage.
\newblock {\em Journal of Applied Statistics}, 49(11):2789--2804.

\bibitem[Xu et~al., 2019]{xu2019}
Xu, H., Li, X., Shen, C., Hui, S.~L., and Grannis, S. (2019).
\newblock {Incorporating conditional dependence in latent class models for probabilistic record linkage: Does it matter?}
\newblock {\em The Annals of Applied Statistics}, 13(3):1753 -- 1790.

\bibitem[Xu et~al., 2022]{xuC2022}
Xu, H., Li, X., Zhang, Z., and Grannis, S. (2022).
\newblock Score test for assessing the conditional dependence in latent class models and its application to record linkage.
\newblock {\em Journal of the Royal Statistical Society, Series C (Applied Statistics)}, 71(5):1663--1687.

\bibitem[Yancey, 2000]{yancey2000}
Yancey, W.~E. (2000).
\newblock Frequency-dependent probability measures for record linkage.
\newblock In {\em Proceedings of the Section on Survey Research Methods, American Statistical Association}, page 752–757.

\bibitem[Zanella et~al., 2016]{zanellabetancourt}
Zanella, G., Betancourt, B., Wallach, H., Miller, J., Zaidi, A., and Steorts, R.~C. (2016).
\newblock Flexible models for microclustering with application to entity resolution.
\newblock In {\em Proceedings of the 30th International Conference on Neural Information Processing Systems}, page 1425–1433, Red Hook, NY, USA.

\bibitem[Zhang and Tuoto, 2021]{zhang2021}
Zhang, L.-C. and Tuoto, T. (2021).
\newblock Linkage-data linear regression.
\newblock {\em Journal of the Royal Statistical Society Series A}, 184(2):522--547.

\bibitem[Zhou and Reiter, 2010]{zhoureiter}
Zhou, X. and Reiter, J.~P. (2010).
\newblock A note on {Bayesian} inference after multiple imputation.
\newblock {\em The American Statistician}, 64(2):159--163.

\bibitem[Zhu et~al., 2009]{zhu2009}
Zhu, V.~J., Overhage, M.~J., Egg, J., Downs, S.~M., and Grannis, S.~J. (2009).
\newblock An empiric modification to the probabilistic record linkage algorithm using frequency-based weight scaling.
\newblock {\em Journal of the American Medical Informatics Association}, 16(5):738--745.

\bibitem[Zhu et~al., 2015]{zhu2015}
Zhu, Y., Matsuyama, Y., Ohashi, Y., and Setoguchi, S. (2015).
\newblock When to conduct probabilistic linkage vs. deterministic linkage? {A} simulation study.
\newblock {\em Journal of Biomedical Informatics}, 56:80--86.

\end{thebibliography}

\newpage
\title{Supplementary Material}
\setcounter{table}{0}
\setcounter{section}{0}
\renewcommand{\thetable}{S\arabic{table}}

\vspace{0.5cm}
\section{Likelihood/Bayesian inference under SNL}

Let $\hat{\boldsymbol\beta}^{(m)}$ be the posterior mean of $\boldsymbol\beta$ obtained using the $m^{th}$ sample of the linkage structure, $\boldsymbol{\Delta}^{(m)}$, for $m=1,\ldots,M$. A point estimate $\hat{\boldsymbol\beta}$ across all $M$ samples can be computed as
\begin{align} 
   \hat{\boldsymbol\beta} &= \E(\boldsymbol\beta|\mathbf{Y}_{A},\mathbf{X}_{B},\mathbf{L}_{A},\mathbf{L}_{B})=\E(\E(\boldsymbol\beta|\mathbf{Y}_{A},\mathbf{X}_{B},\mathbf{L}_{A},\mathbf{L}_{B},\boldsymbol{\Delta})|\mathbf{Y}_{A},\mathbf{X}_{B},\mathbf{L}_{A},\mathbf{L}_{B}) \approx \frac{1}{M} \sum_{m=1}^{M} \hat{\boldsymbol\beta}^{(m)}. \label{eq:linkavg1} 
\end{align}
The variance of  $\hat{\boldsymbol\beta}$ can be obtained using 
\begin{align}
    \begin{split}
   \V(\hat{\boldsymbol\beta}|\mathbf{Y}_{A},\mathbf{X}_{B},\mathbf{L}_{A},\mathbf{L}_{B}) = &\E(\V(\boldsymbol\beta|\mathbf{Y}_{A},\mathbf{X}_{B},\mathbf{L}_{A},\mathbf{L}_{B},\boldsymbol{\Delta})|\mathbf{Y}_{A},\mathbf{X}_{B},\mathbf{L}_{A},\mathbf{L}_{B})  \\&+ \V(\E(\boldsymbol\beta|\mathbf{Y}_{A},\mathbf{X}_{B},\mathbf{L}_{A},\mathbf{L}_{B},\boldsymbol{\Delta})|\mathbf{Y}_{A},\mathbf{X}_{B},\mathbf{L}_{A},\mathbf{L}_{B}) \label{eq:linkavg2}.
\end{split}
\end{align}
The first term in Equation \eqref{eq:linkavg2} is the posterior variance of $\boldsymbol\beta$ averaged across all $M$ samples of $\boldsymbol{\Delta}$. The second term in Equation \eqref{eq:linkavg2} is given by $M^{-1}\sum_{m=1}^{M}(\hat{\boldsymbol\beta}^{(m)} - \hat{\boldsymbol\beta})^2$. This term represents the contribution of the linkage uncertainty, i.e., the added variability due to the missing $\boldsymbol\Delta$. It is possible to obtain a confidence interval for $\boldsymbol\beta$ analytically, using an asymtptotic normal or a Student's-t approximation from Section 2.2 of the main text.

\thispagestyle{empty}
\section{Proof of Proposition 1}
The proof of part (i) is trivial, since 
\begin{align}
    \E(\kappa^*(\mathbf{X}_B;\tilde{\boldsymbol{\beta}})|\mathbf{X}_B) =  \upsilon(\mathbf{X}_B;\tilde{\boldsymbol{\beta}}) [\E(\mathbf{Y}_A^*|\mathbf{X}_B) - \mu(\mathbf{X}_B,\tilde{\boldsymbol{\beta}})] = \upsilon(\mathbf{X}_B;\tilde{\boldsymbol{\beta}}) [(\mathbf{Q} - \mathbf{I})\mu(\mathbf{X}_B,\tilde{\boldsymbol{\beta}})] \neq 0.
\end{align}

To prove part (ii), we consider the first-order Taylor series expansion of $\kappa^*(\mathbf{X}_B;\hat{\boldsymbol{\beta}}^*)$ about ${\boldsymbol{\beta}}$, 
\begin{align} \label{taylor}
   &\kappa^*(\mathbf{X}_B;\hat{\boldsymbol{\beta}}^*) \approx \kappa^*(\mathbf{X}_B;{\boldsymbol{\beta}}) + \frac{\partial \kappa^*(\mathbf{X}_{B};{\boldsymbol{\beta}})}{\partial {\boldsymbol{\beta}}} (\hat{\boldsymbol{\beta}}^* - {\boldsymbol{\beta}}).
\end{align}

Equation \eqref{taylor} implies that 
\begin{align} 
\begin{split}
    \hat{\boldsymbol{\beta}}^* - {\boldsymbol{\beta}} &\approx \big[\frac{\partial \kappa^*(\mathbf{X}_{B};{\boldsymbol{\beta}})}{\partial {\boldsymbol{\beta}}}\big]^{-1}[\kappa^*(\mathbf{X}_B;\hat{\boldsymbol{\beta}}^*)- \kappa^*(\mathbf{X}_B;{\boldsymbol{\beta}})] \\
   &= \big[\frac{\partial \kappa^*(\mathbf{X}_{B};{\boldsymbol{\beta}})}{\partial {\boldsymbol{\beta}}}\big]^{-1}[- \kappa^*(\mathbf{X}_B;{\boldsymbol{\beta}})],   
\end{split}
\end{align}
as $\kappa^*(\mathbf{X}_B;\hat{\boldsymbol{\beta}}^*)=0$. The bias will be 
\begin{align} 
\begin{split}
    \E(\hat{\boldsymbol{\beta}}^*-{\boldsymbol{\beta}}|\mathbf{X}_{B}) &\approx -\big[\frac{\partial \kappa^*(\mathbf{X}_{B};{\boldsymbol{\beta}})}{\partial {\boldsymbol{\beta}}}\big]^{-1} \E(\kappa^*(\mathbf{X}_B;{\boldsymbol{\beta}})|\mathbf{X}_B) \\
    &= -\big[\frac{\partial \kappa^*(\mathbf{X}_{B};{\boldsymbol{\beta}})}{\partial {\boldsymbol{\beta}}}\big]^{-1} \upsilon(\mathbf{X}_B;{\boldsymbol{\beta}}) [(\mathbf{Q} - \mathbf{I})\mu(\mathbf{X}_B,{\boldsymbol{\beta}})].
\end{split}
\end{align}

\section{Details of simulation studies}
In this section, we describe the setup and implementation of the simulation studies in the main text.

\subsection{Scenario 1: Primary analysis}
We generate files $\mathbf{A}$ and $\mathbf{B}$ of sizes $n_A = 600$ and $n_B=900$, respectively. The number of true links $n_{AB}$ is determined by the overlap between the two files. For true links, we generate the linking variables in file $\mathbf{A}$, and replicate their values in file $\mathbf{B}$. For the non-linking record pairs, linking variables in both files are generated independently.

\subsubsection{Data generation}

\paragraph{Linking variables in the high discriminatory power setting:}
In this setting, we generate four linking variables to represent an individual's zip stratum, year of birth, race, and last name. We generate the zip stratum for each record from a discrete uniform distribution over $\{1,\dots,30\}$. 
We generate the year of birth by simulating age from a normal distribution, $N(30,10^2)$, and converting it to the year of birth. Specifically, for record $r$, we generate the age $a_r\overset{iid}\sim N(30,10^2)$,and compute the year of birth as $YOB_r = 2025-a_r$. 
%The normal distribution on $a_r$ is fairly diffuse, which generates a large number of unique values for the year of birth. 
For every record, we generate the race from a discrete uniform distribution over $\{1,\dots,5\}$. To generate last names, we use the \texttt{randomNames} package in \texttt{R}, and generate $n_A+n_B-n_{AB}$ unique last names (so that every record has a unique last name).

\paragraph{Linking variables in the low discriminatory power setting:}
In this setting, we generate four linking variables to represent an individual's gender, year of birth, race, and last name.
We generate gender for each record from a discrete uniform distribution over $\{1,2\}$.
We generate the year of birth and race in the same manner as in the high discriminatory power setting. We generate a set of last names using the \texttt{randomNames} package, with the number of unique values set to $n_A$. We assign a last name to every record by sampling with replacement from this set.

\paragraph{Variables exclusive to files $\mathbf{A}$ and $\mathbf{B}$:} 
\noindent We assume that the larger file ($\mathbf{B}$) includes the covariate $\mathbf{X}_B$ and the smaller file ($\mathbf{A}$) includes the outcome $\mathbf{Y}_A$. For the true links, we generate \begin{align} \label{eq:datagensup}
\begin{pmatrix}
Y_{Al} \\
X_{Bl}
\end{pmatrix}\overset{iid}\sim N_2\left(\begin{pmatrix}
0 \\
0
\end{pmatrix},\begin{pmatrix}
4 & 2\rho \\
2\rho & 1
\end{pmatrix}\right).
\end{align}
Under this setup, a linear regression of $\mathbf{Y}_{A}$ on $\mathbf{X}_{B}$ is of the form ${Y}_{Al} = \beta X_{Bl} + \epsilon_l$, where $\beta=2\rho$, and $\epsilon_l \overset{iid}\sim N(0,\sigma^2=4(1-\rho^2))$. The parameter $\rho$ controls the signal-to-noise ratio, which is defined as $\text{SNR} = ||\beta||_{2}^{2}/\sigma^2$. We consider $\rho \in \{0.9,0.4\}$. When $\rho=0.9$, SNR $\approx 4$, indicating a strong linear relationship between $\mathbf{Y}_{A}$ and $\mathbf{X}_{B}$, and a low noise variance. When $\rho=0.4$, SNR $\approx 0.2$, representing increased noise variance and a weak linear relationship between $\mathbf{Y}_{A}$ and $\mathbf{X}_{B}$. Among non-links, we generate $\mathbf{Y}_{A}$ and $\mathbf{X}_{B}$ independently as $Y_{Ai} \overset{iid}\sim N(0,1)$ and $X_{Bj} \overset{iid}\sim N(0,1)$. 

\paragraph{Generating measurement error:}
We induce measurement error in two linking variables: the last name and the year of birth. We consider two error levels ($e$): 40\% (high), and 10\% (low). To generate errors as per the LCAR mechanism, we randomly choose $e n_A$ records from file $\mathbf{A}$, and $ e n_B$ records from file $\mathbf{B}$, and randomly perturb the aforementioned linking variables in the chosen records. For the SNL, NL, WNL, and IL mechanisms, we model the propensity of each record having measurement error using logistic regression. Let $\zeta_{r}$ represent the probability of record $r$ having measurement error. 
We have
\begin{align} \label{propensity1}
\zeta_{r} &= \frac{1}{1+exp{(-(c_1+c_2 V_r))}},
\end{align}
where we set $c_2=1$, and choose $c_1$ using bisection to get the desired error level $e$. The variable $\mathbf{V} = \{V_r\}$ is chosen based on the linkage mechanism. Under SNL, we let $\mathbf{V}$ be a set of indicator variables, where $V_r = 1$, if the race for record $r$ is either category 4 or 5, and $V_r = 0$, otherwise. Under NL, we let $\mathbf{V} \equiv \mathbf{X}_B$. Under WNL, we let $\mathbf{V} \equiv \mathbf{Y}_A$. Under IL, the probability of measurement error depends on an extraneous unobserved variable associated with the analysis variables, $\mathbf{W}=-0.3+\exp(\mathbf{Y}_A)+0.5\mathbf{X}_B$. In this case, we set $\mathbf{V}\equiv\mathbf{W}$ in Equation \eqref{propensity1}. After computing $\zeta_r$, we generate an indicator variable $I_r \sim Bernoulli(\zeta_{r})$, and induce errors in records with $I_r=1$. \par
We generate errors in the year of birth by shuffling its values among the records containing error. We generate two types of errors in the last name: (i) deletion, where a randomly chosen character is deleted from the string; and (ii) transposition, where two characters from the string are randomly chosen, and their position is interchanged to form a new string. We generate each type of error with probability $e/2$.

\subsubsection{Implementation of record linkage algorithms}
\paragraph{Parameters in the \citet{steorts16} model:}
The entity resolution model of \citet{steorts16} entails specification of four parameters: (i) the maximum number of latent population entities; (ii) a ``steepness'' parameter, that governs how likely it is for a string field to be distorted to values that are far from the latent entity’s value; and (iii) two hyperparameters $(b_1, b_2)$ of the Beta prior on the probability that a record in a file is distorted on the $k^{th}$ linking variable. We set the maximum number of latent entities to be the total sample size of the concatenated, block-level file. Following \citet{steorts16}, we set the steepness parameter equal to 1. We set the hyperparameters $b_1=1$ and $b_2=99$, so that the prior in (iii) has a small mean, and is reasonably diffuse \citep{steorts16}.

\paragraph{Modifying the \citet{gutman2013} model:}
We modify the likelihood of \citet{gutman2013} to include comparisons between variables common to both files. Suppose that files $\mathbf{A}$ and $\mathbf{B}$ are divided into $H$ blocks, where each block in file $\mathbf{B}$ is larger than the corresponding block in file $\mathbf{A}$. Let $\boldsymbol{\theta}$ denote parameters governing the distribution of the linking variable comparisons, and let $\boldsymbol{\tau}=(\beta, \sigma^2)$ denote the linear regression parameters. Further, let $\mathbf{Z}_h$ denote the matching permutation for block $h \in \{1,\dots,H\}$. We modify the likelihood in Equations (2) and (5) of \citet{gutman2013} as
\begin{equation}
\mathcal{L}(\boldsymbol{\theta},\boldsymbol{\tau},\mathbf{Z}_h) = \prod_{h=1}^{H}  \big(\prod_{(i,j) \in \mathbf{M}_h} f_1(\mathbf{Y}_{Ai},\mathbf{X}_{Bj}|\boldsymbol{\tau},\mathbf{Z}_h) \text{  } f_2(\mathbf{L}_{Ai},\mathbf{L}_{Bj}|\boldsymbol{\theta},\mathbf{Z}_h) \big) \times \big(\prod_{j \in \mathbf{U}_h} f_3(\mathbf{X}_{Bj}) f_4(\mathbf{L}_{Bj}) \big ),
\end{equation}
where $\mathbf{M}_h$ denotes the set of linked record pairs in block $h$, and $\mathbf{U}_h$ denotes the set of unlinked records from File $\mathbf{B}$ in block $h$. For records in $\mathbf{M}_h$, we model the agreements on categorical linking variables as Bernoulli random variables. Formally, if $((i,j): i \in \mathbf{A},j \in \mathbf{B})$ is a link, for the $k^{th}$ categorical linking variable, we let $\mathbbm{1}(L_{Aik} = L_{Bjk}) \sim Bernoulli(\theta_k)$. We model the number of record pairs with Levenshtein distances in the ranges $(0,0.25)$, $(0.25, 0.5)$, and $(0.5,1)$ using a multinomial distribution with parameters $\alpha_L=(\alpha_{L1},\alpha_{L2},\alpha_{L3})$. It is straightforward to impose conjugate priors on $\boldsymbol{\theta} = (\{\theta_k\},\alpha_L)$, and sample from the posterior distribution as part of the data augmentation algorithm in \cite{gutman2013}. It is not essential to model the marginal distributions $f_3$ and $f_4$ due to the monotone pattern of missing records in every block.

\subsection{Scenario 2: Secondary analysis}
\subsubsection{Data Generation}
We generate a linked file of size $n_{AB}=600$, comprising three blocks of sizes $n_1=100$, $n_2=200$, and $n_3=300$. For all records, we generate $\mathbf{Y}_A$ and $\mathbf{X}_B$ using Equation \eqref{eq:datagensup}, with $\rho \in \{0.9,0.4\}$.
\paragraph{Generating linkage errors:} 
We assume that linkage errors are confined within the blocks. In the low linkage error scenario, the block-specific linkage error rates are $\{e_1, e_2, e_3\}=\{5\%, 10\%, 15\%\}$, and in the high linkage error scenario, they are $\{e_1, e_2, e_3\}=\{15\%, 20\%, 25\%\}$. For each error level, we generate linkage errors by shuffling values of the outcome. We choose erroneous record-pairs according to three error mechanisms, as described below.
\begin{enumerate}
\item Exchangeable linkage error (ELE) or LCAR: In block $h$, we generate errors so that for every record in one file, the probability of linking with the correct record from the other file is $1-e_h$, and the probability of linking with any incorrect record is $\frac{e_h}{n_h-1}, h = 1, \dots, 3$. 
%Algorithm 1 describes the steps that are necessary to generate errors as per the ELE mechanism.
%\begin{algorithm}
%\caption{Generating blockwise exchangeable linkage errors}
%\begin{algorithmic}[1]
%\State In block $b$, define an empty set of records $S_b$, and configuration matrix $\Delta_b$. 
%
%    \For {$r$ in 1 to $n_b$}
%    
%    \State Generate $U \sim Uniform(0,1)$
%    
%    \If{$U<1-e_b$}
%        \State Append record $r$ to set $S_b$.
%        
%        \State Set the $(r,r)^{th}$ diagonal element of $\Delta_b$, i.e., $\Delta_{brr}=1$.
%        
%    \EndIf
%\EndFor
%\State Define $S_{b1} = \{1,\dots,n_b\} \backslash S_b$.
%\For {$r_1 \text{ in } S_{b1}$}
%\State Define $S_{b2} = S_{b1} \backslash \{r_1\}$.
%\State Randomly sample a record from $S_{b2}$, say $r_2$.
%\State Set the $(r,r_2)^{th}$ element of $\Delta_b$, i.e., $\Delta_{brr_2}=1$.
%\EndFor
%\State Calculate the permuted outcome vector in block $b$ as $Y^*_b = \Delta_bY_b$. 
%\end{algorithmic}
%\end{algorithm}

\item NL, WNL, and IL mechanisms: In block $h$, let $\zeta_{rh}$ represent the probability of a linked
record pair $r$ being a false link. 
We let
\begin{align} \label{propensity}
\zeta_{rh} &= \frac{1}{1+exp({-(c_1+c_2 V_r)})},
\end{align}
where we set $c_2=1$, and choose $c_1$ using bisection to get the desired error level $e_h$.
The variable $\mathbf{V} \equiv \{V_r\}$ is chosen based on the linkage mechanism. Under NL, we let $\mathbf{V} \equiv \mathbf{X}_B$. Under WNL, we let $\mathbf{V}\equiv \mathbf{Y}_A$. Under IL, we set $\mathbf{V} \equiv \mathbf{W}$, where $\mathbf{W}=-0.3+\exp(\mathbf{Y}_A)+0.5\mathbf{X}_B$. Next, we simulate indicator variables $I_{rh} \sim Bernoulli(\zeta_{rh})$. For records with $I_{rh}=1$, we shuffle the values of the outcome.
\end{enumerate}

\section{Additional simulation results}
\subsection{Primary analysis}

\begin{table}
\caption[]{Bias, SE, and coverage probability for estimates of $\boldsymbol{\beta}$ in the primary analysis scenario with complete overlap. The measurement error level is high, the linking variables have high discriminatory power.\vspace{0.2cm}}
\centering
\begin{tabular}{llccc}
\hline
\makecell{Linkage\\Mechanism}  & Method   & Bias      & SE      & Coverage \\ \hline
\multirow{6}{*}{LCAR} &FS & -0.097 & 0.079 & 0.67 \\ 
 &SL & -0.008 & 0.060 & 0.94 \\ 
 &ST & -0.011 & 0.098 & 0.93 \\ 
 & GT  &-0.020   &0.072      &0.91    \\
 & KSG  &-0.029   &0.070     &0.91   \\ 
 &HLF & -0.352 & 0.076 & 0.62 \\ 
 &HL2 & -0.339 & 0.076 & 0.62 \\ 
 &HL1 & -0.327 & 0.076 & 0.62 \\ 
 &HZ & 0.120 & 0.141 & 0.84 \\ 
   \hline
\multirow{6}{*}{SNL}  &FS & -0.081 & 0.080 & 0.71 \\ 
 &SL & -0.008 & 0.060 & 0.94 \\ 
 &ST & -0.004 & 0.096 & 0.96 \\ 
 & GT  &-0.032   &0.075      &0.91          \\
 & KSG &-0.034   &0.076      &0.91\\  
 &HLF & -0.338 & 0.075 & 0.67 \\ 
 &HL2 & -0.323 & 0.075 & 0.67 \\ 
 &HL1 & -0.308 & 0.076 & 0.69 \\ 
 &HZ & 0.111 & 0.142 & 0.90 \\ 
   \hline
\multirow{6}{*}{NL}   &FS & -0.029 & 0.077 & 0.93 \\ 
 &SL & -0.013 & 0.061 & 0.94 \\ 
 &ST & 0.006 & 0.107 & 0.96 \\ 
& GT  &-0.039   &0.079      &0.93          \\
& KSG &-0.032   &0.080      &0.93 \\  
 &HLF & -0.312 & 0.074 & 0.63 \\ 
 &HL2 & -0.305 & 0.074 & 0.63 \\ 
 &HL1 & -0.289 & 0.074 & 0.64 \\ 
 &HZ & 0.105 & 0.143 & 0.88 \\ 
   \hline
\multirow{6}{*}{WNL} &FS & -0.097 & 0.075 & 0.83 \\ 
 &SL & -0.121 & 0.061 & 0.63 \\ 
 &ST & -0.276 & 0.078 & 0.04 \\ 
 & GT  &-0.112   &0.076      &0.82   \\
 & KSG &-0.106   &0.076      &0.82  \\  
 &HLF & -0.279 & 0.073 & 0.74 \\ 
 &HL2 & -0.274 & 0.074 & 0.74 \\ 
 &HL1 & -0.254 & 0.074 & 0.76 \\ 
 &HZ & -0.117 & 0.148 & 0.96 \\ 
   \hline
\multirow{6}{*}{IL} &FS & -0.121 & 0.073 & 0.58 \\ 
&SL & -0.280 & 0.062 & 0.53 \\ 
&ST & -0.339 & 0.078 & 0.02 \\ 
& GT  &-0.262   &0.081      &0.76        \\
& KSG &-0.266  &0.084     &0.76  \\  
&HLF & -0.278 & 0.073 & 0.75 \\ 
&HL2 & -0.274 & 0.073 & 0.75 \\ 
&HL1 & -0.244 & 0.073 & 0.78 \\ 
&HZ & -0.157 & 0.153 & 0.94 \\ 
   \hline
\end{tabular}
\end{table}
\FloatBarrier

\begin{table}
\caption[]{Bias, SE, and coverage probability for estimates of $\boldsymbol{\beta}$ in the primary analysis scenario with complete overlap. The measurement error level is low, the linking variables have low discriminatory power.\vspace{0.2cm}} 
\centering
\begin{tabular}{llccc}
\hline
\makecell{Linkage\\Mechanism}  & Method   & Bias      & SE      & Coverage \\ \hline
\multirow{6}{*}{LCAR} &FS & -0.028 & 0.065 & 0.88 \\ 
  &SL & -0.011 & 0.062 & 0.96 \\ 
  &ST & 0.000 & 0.066 & 0.95 \\ 
  & GT  & -0.018&0.069&0.93             \\ 
  & KSG    & -0.019&0.070&0.93            \\ 
  &HLF & -0.064 & 0.074 & 0.84 \\ 
  &HL2 & -0.063 & 0.074 & 0.84 \\ 
  &HL1 & -0.042 & 0.075 & 0.89 \\ 
  &HZ & 0.177 & 0.136 & 0.72 \\ \hline
\multirow{6}{*}{SNL}  &FS & -0.028 & 0.065 & 0.88 \\ 
 &SL & -0.010 & 0.062 & 0.96 \\ 
 &ST & -0.001 & 0.065 & 0.97 \\ 
 & GT   &-0.018&0.071&0.93                     \\ 
 & KSG     &-0.018&0.070&0.93          \\  
 &HLF & -0.058 & 0.074 & 0.87 \\ 
 &HL2 & -0.057 & 0.074 & 0.87 \\ 
 &HL1 & -0.038 & 0.074 & 0.91 \\ 
 &HZ & 0.178 & 0.136 & 0.70 \\ 
   \hline
\multirow{6}{*}{NL}  &FS & -0.030 & 0.069 & 0.89  \\ 
 &SL & -0.019 & 0.064 & 0.96 \\ 
 &ST & -0.078 & 0.058 & 0.72 \\ 
 & GT   &-0.020&0.068&0.94                         \\ 
& KSG   &-0.022&0.070&0.94             \\ 
 &HLF & -0.037 & 0.079 & 0.93 \\ 
 &HL2 & -0.036 & 0.079 & 0.94 \\ 
 &HL1 & -0.013 & 0.080 & 0.94 \\ 
 &HZ & 0.165 & 0.137 & 0.77 \\ 
   \hline
\multirow{6}{*}{WNL} &FS & -0.143 & 0.063 & 0.32 \\ 
&SL & -0.121 & 0.066 & 0.80 \\ 
&ST & -0.390 & 0.049 & 0.15 \\ 
& GT    &-0.093&0.069&0.83             \\
& KSG     &-0.090&0.069&0.83         \\  
&HLF & -0.116 & 0.088 & 0.79 \\ 
&HL2 & -0.115 & 0.088 & 0.79 \\ 
&HL1 & -0.089 & 0.090 & 0.85 \\ 
&HZ & 0.046 & 0.144 & 0.91 \\ 
   \hline
\multirow{6}{*}{IL} &FS & -0.179 & 0.061 & 0.16 \\ 
&SL & -0.127 & 0.069 & 0.60 \\ 
&ST & -0.410 & 0.047 & 0.03 \\ 
& GT     &-0.109&0.074&0.74                    \\
& KSG     &-0.104&0.077&0.74    \\  
&HLF & -0.129 & 0.090 & 0.70 \\ 
&HL2 & -0.128 & 0.090 & 0.70 \\ 
&HL1 & -0.097 & 0.092 & 0.79 \\ 
&HZ & 0.006 & 0.146 & 0.91 \\ 
   \hline
\end{tabular}
\end{table}

\FloatBarrier
\begin{table}
\caption[]{Bias, SE, and coverage probability for estimates of $\boldsymbol{\beta}$ in the primary analysis scenario with complete overlap. The measurement error level is high, the linking variables have low discriminatory power.\vspace{0.2cm}} 
\centering
\begin{tabular}{llccc}
\hline
\makecell{Linkage\\Mechanism}  & Method   & Bias      & SE      & Coverage \\ \hline
\multirow{6}{*}{LCAR} &FS & -0.219 & 0.094 & 0.45 \\ 
 &SL & -0.038 & 0.066 & 0.90 \\ 
 &ST & 0.002 & 0.098 & 0.94 \\ 
 &GT & 0.017 & 0.072 & 0.93 \\ 
 &KSG & 0.009 & 0.070 & 0.93 \\ 
 &HLF & -0.340 & 0.096 & 0.44 \\ 
 &HL2 & -0.334 & 0.096 & 0.45 \\ 
 &HL1 & -0.325 & 0.097 & 0.44 \\ 
 &HZ & 0.326 & 0.127 & 0.53 \\ 
   \hline
\multirow{6}{*}{SNL}  &FS & -0.227 & 0.092 & 0.45 \\ 
 &SL & -0.039 & 0.065 & 0.88 \\ 
 &ST & -0.258 & 0.111 & 0.71 \\ 
 &GT & 0.036 & 0.069 & 0.91 \\ 
 &KSG & 0.029 & 0.071 & 0.92 \\ 
 &HLF & -0.329 & 0.102 & 0.43 \\ 
 &HL2 & -0.324 & 0.102 & 0.43 \\ 
 &HL1 & -0.314 & 0.103 & 0.44 \\ 
 &HZ & 0.202 & 0.133 & 0.59 \\ 
   \hline
\multirow{6}{*}{NL}   &FS & -0.227 & 0.092 & 0.45 \\ 
 &SL & -0.060 & 0.071 & 0.84 \\ 
 &ST & 0.964 & 0.095 & 0.48 \\ 
 &GT & -0.060 & 0.081 & 0.89 \\ 
 &KSG & -0.063 & 0.080 & 0.88 \\ 
 &HLF & -0.238 & 0.091 & 0.53 \\ 
 &HL2 & -0.236 & 0.089 & 0.52 \\ 
 &HL1 & -0.219 & 0.089 & 0.54 \\ 
 &HZ & 0.298 & 0.129 & 0.57 \\ 
   \hline
\multirow{6}{*}{WNL} &FS & -0.186 & 0.093 & 0.49  \\ 
 &SL & -0.080 & 0.073 & 0.82 \\ 
 &ST & -0.487 & 0.063 & 0.02 \\ 
 &GT & -0.088 & 0.081 & 0.84 \\ 
 &KSG & -0.081 & 0.084 & 0.86 \\ 
 &HLF & -0.172 & 0.103 & 0.58 \\ 
 &HL2 & -0.170 & 0.103 & 0.58 \\ 
 &HL1 & -0.147 & 0.105 & 0.59 \\ 
 & HZ & 0.064 & 0.136 & 0.79 \\ 
   \hline
\multirow{6}{*}{IL} &FS & -0.225 & 0.083 & 0.31 \\ 
 &SL & -0.092 & 0.075 & 0.77 \\ 
 &ST & -0.503 & 0.064 & 0.01 \\ 
 &GT & -0.104 & 0.088 & 0.77\\ 
 &KSG & -0.120 & 0.092 & 0.77 \\ 
 &HLF & -0.172 & 0.104 & 0.58 \\ 
 &HL2 & -0.170 & 0.104 & 0.58 \\ 
 &HL1 & -0.145 & 0.105 & 0.59 \\ 
 &HZ & 0.017 & 0.140 & 0.82 \\ 
   \hline
\end{tabular}
\end{table}
\FloatBarrier

\FloatBarrier
\begin{table}[]
\caption[]{Bias, SE, and coverage probability for estimates of $\boldsymbol{\beta}$ in the primary analysis scenario with medium and low overlap. The measurement error level is high, and the linking variables have high discriminatory power.\vspace{0.2cm}}
\centering
\begin{tabular}{llcccccc}
\hline
                      &   &  \multicolumn{3}{c}{Medium Overlap} & \multicolumn{3}{c}{Low Overlap} \\ \cmidrule(r){3-5} \cmidrule(l){6-8}
\makecell{Linkage\\Mechanism}   &Method  & \makecell{Mean \\ Bias}        & SE     & Coverage       & \makecell{Mean \\ Bias}     & SE    & Coverage    \\ \hline
\multirow{2}{*}{LCAR} & FS & -0.236 & 0.111 & 0.46 & -0.595 & 0.146 & 0.21 \\ 
 &SL & -0.189 & 0.137 & 0.78 & -0.405 & 0.221 & 0.53 \\ 
 &ST & 0.020 & 0.140 & 0.94 & -0.009 & 0.227 & 0.93 \\ 
 &KSG & -0.169 & 0.141 & 0.82 & -0.319 & 0.227 & 0.60 \\ 
 &HZ & 0.089 & 0.159 & 0.88 & 0.020 & 0.188 & 0.88 \\ 
   \hline
\multirow{2}{*}{SNL} & FS & -0.246 & 0.109 & 0.48 & -0.565 & 0.146 & 0.26 \\ 
  &SL & -0.184 & 0.136 & 0.78 & -0.404 & 0.221 & 0.55 \\ 
  &ST & -0.018 & 0.136 & 0.92 & 0.011 & 0.223 & 0.94 \\ 
  &KSG & -0.166 & 0.142 & 0.83 & -0.319 & 0.227 & 0.60 \\ 
  &HZ & 0.061 & 0.162 & 0.91 & 0.057 & 0.179 & 0.87 \\ 
   \hline
                    
\multirow{2}{*}{NL} & FS & -0.131 & 0.106 & 0.62 & -0.432 & 0.143 & 0.30 \\ 
 &SL & -0.221 & 0.143 & 0.68 & -0.462 & 0.224 & 0.38 \\ 
 &ST & 0.005 & 0.151 & 0.95 & -0.013 & 0.248 & 0.95 \\ 
 &KSG & -0.189 & 0.145 & 0.72 & -0.361 & 0.229 & 0.50 \\ 
 &HZ & 0.071 & 0.162 & 0.90 & 0.022 & 0.183 & 0.89 \\ 
   \hline    
\multirow{2}{*}{WNL} & FS & -0.261 & 0.103 & 0.40 & -0.561 & 0.145 & 0.18 \\ 
  &SL & -0.194 & 0.135 & 0.76 & -0.416 & 0.224 & 0.53 \\ 
  &ST & -0.284 & 0.112 & 0.28 & -0.278 & 0.186 & 0.65 \\ 
  &KSG & -0.182 & 0.138 & 0.77 & -0.314 & 0.229 & 0.56 \\ 
  &HZ & -0.140 & 0.166 & 0.93 & -0.164 & 0.182 & 0.88 \\ 
   \hline          
\multirow{2}{*}{IL} & FS & -0.283 & 0.102 & 0.34 & -0.564 & 0.138 & 0.20 \\ 
 &SL & -0.207 & 0.138 & 0.72 & -0.441 & 0.225 & 0.46 \\ 
 &ST & -0.337 & 0.115 & 0.15 & -0.334 & 0.189 & 0.52 \\ 
 &KSG & -0.185 & 0.141 & 0.75 & -0.378 & 0.231 & 0.51 \\ 
 &HZ & -0.166 & 0.172 & 0.91 & -0.187 & 0.190 & 0.82 \\ 
   \hline
\end{tabular}
\end{table}

\begin{table}[]

\caption[]{Bias, SE, and coverage probability for estimates of $\boldsymbol{\beta}$ in the primary analysis scenario with medium and low overlap. The measurement error level is low, the linking variables have low discriminatory power. \vspace{0.2cm}} 
\centering
\begin{tabular}{llcccccc}
\hline
                      &   &  \multicolumn{3}{c}{Medium Overlap} & \multicolumn{3}{c}{Low Overlap}\\ \cmidrule(r){3-5} \cmidrule(l){6-8}
\makecell{Linkage\\Mechanism}   &Method  & \makecell{Mean \\ Bias}        & SE     & Coverage       & \makecell{Mean \\ Bias}     & SE    & Coverage    \\ \hline
\multirow{2}{*}{LCAR} & FS & -0.207 & 0.086 & 0.46 & -0.510 & 0.116 & 0.20 \\ 
 &SL & -0.246 & 0.146 & 0.61 & -0.541 & 0.216 & 0.24 \\ 
 &ST & -0.013 & 0.092 & 0.91 & -0.018 & 0.152 & 0.95 \\ 
 &KSG & -0.212 & 0.151 & 0.66 & -0.456 & 0.226 & 0.43 \\ 
 &HZ & 0.122 & 0.155 & 0.81 & 0.091 & 0.166 & 0.71 \\ 
   \hline
\multirow{2}{*}{SNL} & FS & -0.205 & 0.086 & 0.45 & -0.498 & 0.116 & 0.23  \\ 
  &SL & -0.246 & 0.146 & 0.61 & -0.545 & 0.216 & 0.26 \\ 
  &ST & -0.016 & 0.092 & 0.93 & -0.022 & 0.153 & 0.96 \\ 
  &KSG & -0.213 & 0.152 & 0.66 & -0.453 & 0.225 & 0.45 \\ 
  &HZ & 0.113 & 0.156 & 0.81 & 0.094 & 0.166 & 0.74 \\ 
   \hline
                    
\multirow{2}{*}{NL} & FS & -0.220 & 0.091 & 0.45 & -0.555 & 0.120 & 0.17  \\ 
  &SL & -0.288 & 0.158 & 0.53 & -0.628 & 0.231 & 0.23 \\ 
  &ST & -0.089 & 0.082 & 0.67 & -0.091 & 0.139 & 0.74 \\ 
  &KSG & -0.231 & 0.159 & 0.54 & -0.511 & 0.240 & 0.38 \\ 
  &HZ & 0.114 & 0.156 & 0.81 & 0.078 & 0.168 & 0.77 \\ 
   \hline
\multirow{2}{*}{WNL} & FS & -0.344 & 0.085 & 0.15 & -0.667 & 0.114 & 0.07 \\ 
  &SL & -0.352 & 0.156 & 0.30 & -0.700 & 0.225 & 0.08 \\ 
  &ST & -0.405 & 0.069 & 0.26 & -0.412 & 0.117 & 0.40 \\ 
  &KSG & -0.333 & 0.158 & 0.32 & -0.623 & 0.229 & 0.12 \\ 
  &HZ & -0.022 & 0.166 & 0.89 & -0.069 & 0.182 & 0.77 \\ 
   \hline
\multirow{2}{*}{IL} & FS & -0.411 & 0.084 & 0.09 & -0.730 & 0.113 & 0.03\\ 
  &SL & -0.420 & 0.160 & 0.17 & -0.786 & 0.229 & 0.03 \\ 
  &ST & -0.443 & 0.068 & 0.08 & -0.452 & 0.116 & 0.28 \\ 
  &KSG & -0.412 & 0.162 & 0.20 & -0.708 & 0.232 & 0.10 \\ 
  &HZ & -0.079 & 0.170 & 0.83 & -0.151 & 0.190 & 0.68 \\ 
   \hline
\end{tabular}
\end{table}
\FloatBarrier

\FloatBarrier
\begin{table}
\caption[]{Bias, SE, and coverage probability for estimates of $\boldsymbol{\beta}$ in the primary analysis scenario with medium and low overlap. The measurement error level is high, and the linking variables have low discriminatory power.\vspace{0.2cm}} 
\centering
\begin{tabular}{llcccccc}
\hline
                      &   &  \multicolumn{3}{c}{Medium Overlap} & \multicolumn{3}{c}{Low Overlap} \\ \cmidrule(r){3-5} \cmidrule(l){6-8}
\makecell{Linkage\\Mechanism}   &Method  & \makecell{Mean \\ Bias}        & SE     & Coverage       & \makecell{Mean \\ Bias}     & SE    & Coverage    \\ \hline
\multirow{2}{*}{LCAR} & FS & -0.511 & 0.111 & 0.19 & -0.895 & 0.114 & 0.03 \\ 
  &SL & -0.423 & 0.170 & 0.23 & -0.717 & 0.221 & 0.12 \\ 
  &ST & -0.029 & 0.143 & 0.95 & -0.105 & 0.238 & 0.91 \\ 
  &KSG & -0.389 & 0.173 & 0.38 & -0.607 & 0.229 & 0.20 \\ 
  &HZ & 0.231 & 0.145 & 0.58 & 0.163 & 0.159 & 0.56 \\ 
   \hline
\multirow{2}{*}{SNL} & FS & -0.504 & 0.110 & 0.17 & -0.895 & 0.112 & 0.03 \\ 
  &SL & -0.419 & 0.171 & 0.26 & -0.712 & 0.222 & 0.12 \\ 
  &ST & -0.285 & 0.161 & 0.70 & -0.374 & 0.270 & 0.69 \\ 
  &KSG & -0.347 & 0.174 & 0.33 & -0.604 & 0.230 & 0.23 \\ 
  &HZ & 0.102 & 0.154 & 0.40 & 0.081 & 0.166 & 0.41 \\ 
   \hline
                    
\multirow{2}{*}{NL} & FS & -0.458 & 0.111 & 0.21 & -0.871 & 0.115 & 0.02 \\ 
  &SL & -0.484 & 0.176 & 0.18 & -0.804 & 0.226 & 0.08 \\ 
  &ST & -0.110 & 0.129 & 0.70 & -0.178 & 0.233 & 0.74 \\ 
  &KSG & -0.400 & 0.179 & 0.22 & -0.712 & 0.228 & 0.15 \\ 
  &HZ & 0.212 & 0.146 & 0.63 & 0.119 & 0.165 & 0.63 \\ 
   \hline    
\multirow{2}{*}{WNL} & FS & -0.476 & 0.104 & 0.14 & -0.811 & 0.116 & 0.03 \\ 
 &SL & -0.456 & 0.180 & 0.18 & -0.819 & 0.242 & 0.04 \\ 
 &ST & -0.503 & 0.092 & 0.12 & -0.516 & 0.162 & 0.37 \\ 
 &KSG & -0.321 & 0.185 & 0.26 & -0.709 & 0.250 & 0.19 \\ 
 &HZ & -0.035 & 0.157 & 0.71 & -0.108 & 0.175 & 0.62 \\ 
   \hline
\multirow{2}{*}{IL} & FS & -0.492 & 0.102 & 0.12 & -0.854 & 0.112 & 0.01 \\ 
  &SL & -0.484 & 0.180 & 0.14 & -0.831 & 0.240 & 0.04 \\ 
  &ST & -0.517 & 0.096 & 0.07 & -0.548 & 0.170 & 0.32 \\ 
  &KSG & -0.412 & 0.187& 0.23 & -0.800 & 0.244 & 0.11 \\ 
  &HZ & -0.047 & 0.161 & 0.64 & -0.120 & 0.179 & 0.58 \\ 
   \hline
\end{tabular}
\end{table}
\begin{table}
\centering
\vspace{2cm}
\caption[]{Bias, SE, and coverage probability for estimates of $\boldsymbol{\beta}$ in the primary analysis scenario, when $n_A=600$ and $n_B=1800$. The error level is high, the linking variables are highly discriminatory, and the error mechanism is IL.\vspace{0.2cm}} 
\begin{tabular}{llccc}
\hline
Overlap                 & Method & Bias & SE                   & Coverage             \\ \hline
\multirow{8}{*}{Complete}& SL     & -0.327      &0.051                      &0.50                     \\
                        & ST     &-0.342      &0.047                     &0.00                      \\
                        & GT     &-0.264      & 0.079 &0.75 \\
                        & KSG    & -0.269      &0.079 & 0.75                                          \\
                        & HLF    &-0.278      & 0.067       & 0.70                     \\
                        & HL2    & -0.275      & 0.067& 0.70  \\
                        & HL1    & -0.241      & 0.069 & 0.72 \\
                        & HZ     &-0.165      &0.144  &0.88  \\ \hline
\multirow{4}{*}{Medium} & SL     &-0.212      &0.110                      &0.62                     \\
                        & ST     &-0.329      &0.102                      &0.09                    \\
                        & KSG    &-0.282      & 0.119                   &0.68                      \\
                        & HZ     &-0.167      &0.159                      & 0.84                    \\ \hline
\multirow{4}{*}{Low}    & SL     & -0.436      &0.196                     &0.39                      \\
                        & ST     &-0.330      &0.176                     &0.49                      \\
                        & KSG    &-0.370      &0.200                     &0.49                     \\
                        & HZ     &-0.187      &0.181                      &0.75   \\ \hline                 
\end{tabular}
\end{table}
\FloatBarrier

\begin{table}[]
\centering
\caption[]{Results under GT and KSG, when the record linkage models allow for dependence of $Y_A$ on the comparison vectors. The overlap is complete, the measurement error level is high, and the linking variables have high discriminatory power.\vspace{0.1cm}} \label{tab:gtksg}
\begin{tabular}{llccc}
\hline
Overlap                 & Method & Bias & SE                   & Coverage             \\ \hline
\multirow{2}{*}{LCAR}
                        & GT     &-0.032     & 0.074 &0.91 \\
                        & KSG    & -0.033      &0.074 & 0.91                                          \\ \hline
\multirow{2}{*}{SNL} 
                        & GT    &-0.033      & 0.075               &0.91                    \\
                        & KSG     &-0.036      &0.075                      & 0.91
                        \\ \hline
\multirow{2}{*}{NL}    
                        & GT    &-0.028      &0.076                     &0.93                     \\
                        & KSG     &-0.029      &0.077                      &0.93   \\ \hline
 \multirow{2}{*}{WNL}    
                        & GT    &-0.088     &0.077                     &0.89                     \\
                        & KSG     &-0.090      &0.079                      &0.88  \\ \hline
\multirow{2}{*}{IL}    
                        & GT    &-0.189      &0.101                     &0.82                   \\
                        & KSG     &-0.195      &0.099                      &0.77            \\
                        \hline               
\end{tabular}
\end{table}
\FloatBarrier

\subsection{Secondary analysis}

\FloatBarrier
\begin{table}[h]
\caption[]{Bias, estimated SE, and coverage probability for estimates of $\boldsymbol{\beta}$ in the secondary analysis scenario, when linkage error levels are low, and the SNR is low.\vspace{0.2cm}} 
\centering
\begin{tabular}{llccc}
Linkage Mechanism         & Method & Bias  & SE & Coverage \\ \hline
                          & Naive  &-0.102     &0.077    & 0.76        \\
\multirow{5}{*}{ELE (LCAR)} & ChR  & 0.002     &0.090    & 0.95       \\
                          & ChL    & 0.001     &0.090    & 0.94         \\
                          & ChB    & 0.001     &0.089    & 0.93         \\
                          & GT     & -0.089     &0.103    & 0.96         \\ 
                          & SLW     & -0.078     &0.103    & 0.96         \\ \hline
                          & Naive  & -0.088     &0.565    & 0.91         \\
\multirow{5}{*}{NL}       & ChR    & 0.024     &0.090    & 0.97         \\
                          & ChL    & 0.023     &0.089    & 0.98         \\
                          & ChB    & 0.025     &0.089    & 0.97        \\
                          & GT     & -0.064     &0.103    & 0.91          \\ 
                          & SLW     & -0.056     &0.548    & 0.94 \\ \hline
                          & Naive  & -0.065     &0.076    & 0.78         \\
\multirow{5}{*}{WNL}      & ChR    & 0.048     &0.090    & 0.89         \\
                          & ChL    &0.047      &0.089    & 0.87        \\
                          & ChB    &0.047     &0.089   & 0.88       \\
                          & GT     & -0.045     &0.097    & 0.94         \\ 
                         & SLW     & -0.034     &0.587    & 0.97 \\ \hline
                          & Naive  & -0.027    &0.075    &0.88      \\
\multirow{5}{*}{IL}       & ChR    & 0.093     &0.088    & 0.82         \\
                          & ChL    & 0.091     &0.088    & 0.84       \\
                          & ChB    & 0.090     &0.088    & 0.85        \\
                          & GT     & 0.017    &0.092    & 0.97         \\ 
                          & SLW     & -0.005     &0.487    & 1.00 \\\hline
\end{tabular}

\end{table}
\FloatBarrier

\begin{table}
\centering
\caption[]{Bias, estimated SE, and coverage probability for estimates of $\boldsymbol{\beta}$ in the secondary analysis scenario, when linkage error levels are high and the SNR is high.\vspace{0.2cm}}
\begin{tabular}{llccc}
Linkage Mechanism         & Method & Bias  & SE & Coverage \\ \hline
                          & Naive  &-0.387     &0.058    & 0.00       \\
\multirow{5}{*}{ELE (LCAR)} & ChR  & 0.006     &0.107    & 0.99       \\
                          & ChL    & 0.004     &0.106    & 0.99         \\
                          & ChB    & -0.003     &0.099    & 1.00         \\
                          & GT     & -0.016     &0.055    & 0.99         \\ 
                          & SLW     & 0.002     &0.050    & 0.99         \\\hline
                          & Naive  & -0.312     &0.055    & 0.00         \\
\multirow{5}{*}{NL}       & ChR    & 0.088     &0.106    & 0.97         \\
                          & ChL    & 0.086     &0.105    & 0.97         \\
                          & ChB    & 0.121     &0.098    & 0.84        \\
                          & GT     & -0.038     &0.061    & 0.90          \\ 
                          & SLW     & -0.015     &0.047    & 0.93         \\\hline
                          & Naive  & -0.213     &0.050    & 0.00         \\
\multirow{5}{*}{WNL}      & ChR    & 0.213     &0.113    & 0.57         \\
                          & ChL    &0.207      &0.111    & 0.57        \\
                          & ChB    &0.260     &0.104   & 0.15      \\
                          & GT     & -0.132     &0.059    & 0.61         \\ 
                          & SLW     & -0.081     &0.044    & 0.52        \\\hline
                          & Naive  & -0.104    &0.044    &0.26     \\
\multirow{5}{*}{IL}       & ChR    & 0.351     &0.121    & 0.03         \\
                          & ChL    & 0.346     &0.119    & 0.03       \\
                          & ChB    & 0.399     &0.113    & 0.01        \\
                          & GT     & -0.137    &0.073    & 0.49         \\ 
                          & SLW     & -0.080     &0.045    & 0.54         \\\hline
\end{tabular}

\end{table}

\FloatBarrier
\begin{table}[]
\centering
\caption[]{Bias, estimated SE, and coverage probability for estimates of $\boldsymbol{\beta}$ in the secondary analysis scenario, when linkage error levels are high and the SNR is low.\vspace{0.2cm}} 
\begin{tabular}{llccc}
Linkage Mechanism         & Method & Bias & SE & Coverage \\ \hline
                          & Naive  &-0.170   &0.078    & 0.44      \\
\multirow{5}{*}{ELE (LCAR)} & ChR  & 0.004   &0.106    & 0.96       \\
                          & ChL    & 0.002   &0.105    & 0.96         \\
                          & ChB    & 0.002   &0.105    & 0.97        \\
                          & GT     & -0.165   &0.103    & 0.96         \\ 
                          & SLW     & -0.152     &0.847    & 0.82         \\\hline
                          & Naive  & -0.135   &0.078    & 0.58         \\
\multirow{5}{*}{NL}       & ChR    & 0.046   &0.105   & 0.97         \\
                          & ChL    & 0.044   &0.104    & 0.96         \\
                          & ChB    & 0.048   &0.104    & 0.96        \\
                          & GT     & -0.064   &0.103    & 0.91          \\ 
                          & SLW     & -0.121     &0.704    &0.89        \\\hline
                          & Naive  & -0.374   &0.042    & 0.59         \\
\multirow{5}{*}{WNL}      & ChR    & -0.181   &0.070    & 0.76         \\
                          & ChL    &-0.183   &0.070   & 0.76        \\
                          & ChB    &-0.183  &0.069 & 0.73       \\
                          & GT     & -0.045   &0.097    & 0.94         \\ 
                          & SLW     & -0.063     &0.980    &0.93         \\\hline
                          & Naive & -0.029  &0.076    &0.95      \\
\multirow{5}{*}{IL}       & ChR    & 0.176 &0.104  & 0.61         \\
                          & ChL    & 0.173   &0.104    & 0.63       \\
                          & ChB    & 0.174   &0.103    & 0.63        \\
                          & GT     & 0.017  &0.092    & 0.97         \\ 
                          & SLW     &-0.012     &0.493    & 1.00         \\\hline
\end{tabular}
\end{table}
\FloatBarrier

\end{document}